\documentclass[a4paper,11pt]{article}

\usepackage{amsmath,amssymb,amsthm,amsfonts}    
\usepackage{mathrsfs,bm}                        
\usepackage{dsfont}
\usepackage{bm}
\usepackage{eurosym}                            
\usepackage{xcolor}                             
\usepackage{url,hyperref}                       
\usepackage{float}                              
\usepackage[small,bf]{caption}                  
\usepackage{graphicx}
\usepackage{epstopdf}                           
\usepackage{subfigure}
\graphicspath{{./Figures/}}                     
\usepackage{psfrag}                             
\usepackage{epsfig}                             
\usepackage{booktabs,multirow,longtable}        
\usepackage[titletoc,page]{appendix}
\usepackage[english]{babel}
\usepackage{authblk}                            
\usepackage{geometry}
\makeatletter \@addtoreset{equation}{section} \makeatother

\begin{document}
%

\newcommand{\ie}{i.e.~} 	                    
\newcommand{\eg}{e.g.~} 	                    
\newcommand{\rhs}{r.h.s.~} 	                    
\newcommand{\wrt}{w.r.t.~} 	                    
\newcommand{\old}[1]{\textcolor{blue}{#1}}      
\newcommand{\new}[1]{\textcolor{red}{#1}}       
\newcommand{\comment}[1]{\textcolor{green}{#1}} 
\newcommand{\textbox}[1]{\mbox{\textit{#1}}} 

\newtheorem{theorem}{Theorem}[section]
\newtheorem{definition}[theorem]{Definition}
\newtheorem{definitions}[theorem]{Definition}
\newtheorem{proposition}[theorem]{Proposition}
\newtheorem{remark}[theorem]{Remark}
\newtheorem{corollary}[theorem]{Corollary}
\newtheorem{example}[theorem]{Example}
\newtheorem{lemma}[theorem]{Lemma}

\newenvironment{system}{\left\lbrace\begin{array}{@{}l@{}}}{\end{array}\right.}

\newcommand{\Parenthesis}[1]{\left( #1 \right)}         
\newcommand{\Brack}[1]{\left\lbrack #1 \right\rbrack}   
\newcommand{\Brace}[1]{\left\lbrace #1 \right\rbrace}   
\newcommand{\Abs}[1]{\left\lvert #1 \right\rvert}       
\newcommand{\Norm}[1]{\left\lVert #1 \right\rVert}      
\newcommand{\Mean}[1]{\left\langle #1 \right\rangle}    
\newcommand{\QuoteDouble}[1]{``#1''}                    
\newcommand{\QuoteSingle}[1]{`#1'}                      

\newcommand{\beq}{\begin{equation}}         
\newcommand{\eeq}{\end{equation}}           
\newcommand{\Max}{\mbox{Max}}                                   
\newcommand{\Min}{\mbox{Min}}                                   
\newcommand{\Derp}[2]{\frac{\partial #1}{\partial #2}}          
\newcommand{\DerpXX}[2]{\frac{\partial^2 #1}{\partial #2 ^2}}   
\newcommand{\DerpXY}[3]{\frac{\partial^2 #1}{\partial #2 \partial #3}}    
\newcommand{\half}[0]{\frac{1}{2}}                                                  
\newcommand{\ind}[1]{\mathbf{1}_{#1}} 	                                            
\newcommand{\E}[1]{\mathbb{E}\left[#1\right]}                                       
\newcommand{\Econd}[2]{\mathbb{E}\left[#1\;\middle \vert\;\mathcal{F}_{#2}\right]}  

\newcommand{\Nset}{\mathbb{N}}      
\newcommand{\Zset}{\mathbb{Z}}      
\newcommand{\Qset}{\mathbb{Q}}      
\newcommand{\Rset}{\mathbb{R}}      

\newcommand{\Black}{\mbox{Black}}				    

\newcommand{\Depo}{\mbox{\textbf{Depo}}}			
\newcommand{\FRA}{\mbox{\textbf{FRA}}}			    
\newcommand{\Futures}{\mbox{\textbf{Futures}}}      
\newcommand{\ZCB}{\mbox{ZCB}}					    
\newcommand{\Swap}{\mbox{\textbf{Swap}}}			
\newcommand{\Swaplet}{\mbox{\textbf{Swaplet}}}		
\newcommand{\IRS}{\mbox{\textbf{IRS}}}			    
\newcommand{\IRSlet}{\mbox{\textbf{IRSlet}}}        
\newcommand{\OIS}{\mbox{\textbf{OIS}}}              
\newcommand{\OISlet}{\mbox{\textbf{OISlet}}}        
\newcommand{\BSwap}{\mbox{\textbf{BSwap}}}          
\newcommand{\BSwaplet}{\mbox{\textbf{BSwaplet}}}    
\newcommand{\IRBS}{\mbox{\textbf{IRBS}}}            
\newcommand{\IRBSlet}{\mbox{\textbf{IRBSlet}}}      
\newcommand{\CCS}{\mbox{\textbf{CCS}}}              
\newcommand{\CCSwap}{\mbox{\textbf{CCSwap}}}        
\newcommand{\CCSwaplet}{\mbox{\textbf{CCSwaplet}}}  
\newcommand{\CCSlet}{\mbox{\textbf{CCSlet}}}        
\newcommand{\Caplet}{\mbox{\textbf{Caplet}}}        
\newcommand{\Floorlet}{\mbox{\textbf{Floorlet}}}    
\newcommand{\cf}{\mbox{\textbf{cf}}}                
\newcommand{\CAP}{\mbox{\textbf{Cap}}}              
\newcommand{\Floor}{\mbox{\textbf{Floor}}}          
\newcommand{\CF}{\mbox{\textbf{CF}}}                
\newcommand{\Swaption}{\mbox{\textbf{Swaption}}}    
\newcommand{\CMSlet}{\mbox{\textbf{CMSlet}}}        
\newcommand{\CMS}{\mbox{\textbf{CMS}}}              
\newcommand{\CMScf}{\mbox{\textbf{CMScf}}}          
\newcommand{\FXFwd}{\mbox{\textbf{FXFwd}}}          

\title{\textbf{Application of Quasi Monte Carlo and\\ Global Sensitivity Analysis\\ to Option Pricing and Greeks}}

\author[1]{Stefano Scoleri}
\author[2,3,*]{Marco Bianchetti}
\author[4]{Sergei Kucherenko}

\affil[1]{Iason Ltd, Italy,\texttt{stefano.scoleri@gmail.com}}
\affil[2]{Market Risk Management, Banca Intesa Sanpaolo, Milan, Italy}
\affil[3]{Department of Statistical Sciences "Paolo Fortunati", University of Bologna, Italy}
\affil[4]{Imperial College, London, UK, \texttt{s.kucherenko@imperial.ac.uk}}
\affil[*]{Corresponding author, \texttt{marco.bianchetti@unibo.it}}

\date{5 February 2017}

\maketitle

\begin{abstract}
\noindent
Quasi Monte Carlo (QMC) and Global Sensitivity Analysis (GSA) techniques are applied for pricing and hedging representative financial instruments of increasing complexity. We compare standard Monte Carlo (MC) vs QMC results using Sobol' low discrepancy sequences, different sampling strategies, and various analyses of performance. \\
We find that QMC outperforms MC in most cases, including the highest-dimensional simulations, showing faster and more stable convergence. Regarding greeks computation, we compare standard approaches, based on finite differences (FD) approximations, with adjoint methods (AAD) providing evidences that, when the number of greeks is small, the FD approach combined with QMC can lead to the same accuracy as AAD, thanks to increased convergence rate and stability, thus saving a lot of implementation effort while keeping low computational cost. Using GSA, we are able to fully explain our findings in terms of reduced effective dimension of QMC simulation,
allowed in most cases, but not always, by Brownian bridge discretization or PCA construction. \\
We conclude that, beyond pricing, QMC is a very efficient technique also for computing risk measures, greeks in particular, as it allows to reduce the computational effort of high-dimensional Monte Carlo simulations typical of modern risk management.
\end{abstract}

\newpage
\tableofcontents

\vspace{2cm} \noindent \textbf{JEL classifications}: C63, G12, G13.

\vspace{1cm} \noindent \textbf{Keywords}: derivative, option, European, Asian, barrier, knock-out, cliquet, greeks, Monte Carlo, Quasi Monte Carlo, random, pseudo random, quasi random, low discrepancy, Sobol', convergence, speed-up, Brownian
bridge, global sensitivity analysis, principal component analysis, adjoint algorithmic differentiation.

\vspace{1cm} \noindent \textbf{Acknowledgements}: M.B. acknowledges fruitful discussions with many colleagues at international conferences, in Risk Management of Intesa Sanpaolo, and in Financial Engineering and Trading Desks of Banca IMI.

\vspace{1cm} \noindent \textbf{Disclaimer}: the views expressed here are those of the authors and do not represent the opinions of their employers. They are not responsible for any use that may be made of these contents.

\vspace{1cm}
\noindent \textbf{Availability}: this paper was published in Wilmott, Volume 2021, Issue 116, November 2021, pages 66-83 \url{https://doi.org/10.1002/wilm.10972}.
\newpage

\section{Introduction}
\label{SecIntro}
 \noindent
Monte Carlo simulation is a common way to tackle problems in finance.
However, this method is rather time consuming and the convergence
rate is slow, since the root mean
square error decays as $N^{-1/2}$, where $N$ is the number of
sampled points. Various ``variance reduction'' techniques exist,
which can improve the efficiency of the simulation, but they don't
modify the convergence rate \cite{Jac01,Gla03}.
\par
Quasi Monte Carlo (QMC) represents a very efficient alternative to
standard MC, capable to achieve, in many cases, a faster
convergence rate and, hence, higher accuracy
\cite{Jac01,Gla03,MonFer1999,SobKuc05a,SobKuc05b,Wan09,KucBal11,SobAso12}.
QMC methods are based on using instead of
pseudo-random numbers (PRN), low discrepancy sequences (LDS), also
known as quasi-random numbers for sampling points. LDS are
designed in such a way that the integration domain is covered as
uniformly as possible, while PRN are known
 to statistically form clusters and holes of points. On the contrary,
LDS ``know'' about the positions of
previously sampled points and fill the gaps between them. Among
several known LDS, Sobol' sequences have been proven to show
better performance than others and for this reason they are widely
used in Finance \cite{Jac01,Gla03}. However, it is also known
that construction of efficient high-dimensional Sobol' sequences heavily depends on
the so-called initial numbers and therefore very few Sobol'
sequence generators show good efficiency in practical tests (see \eg
\cite{SobAso12}).
\par
Compared to MC, QMC techniques also have
some disadvantages. Firstly, there is no ``in sample'' estimation
of errors: since LDS are deterministic, there is not a notion of
probabilistic error. There are some techniques,
known under the name of randomized QMC, which
introduce appropriate randomizations in the construction of LDS,
opening up the possibility of measuring errors through a
confidence interval while preserving the convergence rate of Quasi
Monte Carlo (see \eg \cite{Gla03}). Secondly,
effectiveness of QMC depends on the integrand
function,  and, most importantly, the convergence rate can depend
on the dimensionality of the problem. However, it's been shown that in
many financial applications
QMC outperforms standard Monte Carlo even in the
presence of very high dimensions
\cite{SobAso12, PasTrau1995, PapPas1999, CafMor1997, KreMer1998a, KreMer1998b, KucSha07, BiaKuc15}.
This fact is usually explained by a reduced effective dimension of
the problem, with respect to its nominal dimension. The concept of
effective dimensions was introduced in \cite{CafMor1997}. It was
suggested that QMC is superior to MC if the effective dimension of
an integrand is not too large. The notion is based on the ANalysis
Of VAriances (ANOVA). In \cite{Lemieux2000} it was shown how the
ANOVA components are linked to the effectiveness of QMC
integration methods. It is important to measure the effective
dimension in order to predict the efficiency of a Quasi Monte
Carlo algorithm. Moreover, various techniques can be used to
reduce effective dimension and, thus, improve efficiency: this is
possible because the effective dimension\footnote{Actually, the
effective dimension in the truncation sense can be reduced in this
way. See Section \ref{SecGSA} for the formal definition of
effective dimensions.} can vary by changing the order in which the
variables are sampled. The optimal way to achieve this can be a
hard task, it could depend on the specific model and a general
solution is not known at present. One popular choice in the
financial literature on path-dependent option pricing
\cite{CafMor1997,KucSha07} is to apply the Brownian bridge
discretization to the simulation of the underlying stochastic
process, which is based on the use of conditional distributions.
Unlike the standard discretization, which generates values of the
Brownian motion sequentially along the time horizon, the Brownian
bridge discretization first generates the Brownian motion value at
the terminal point, then it fills a midpoint using the value
already found at the terminal point and then subsequent values at
the successive midpoints using points already simulated at
previous steps. In terms of QMC sampling, this simulation scheme
means that the first coordinate of the QMC vector is used to
simulate the terminal value of the Brownian motion, while
subsequent coordinates are used to generate intermediate points.
There are many studies which show that superior performance of the
QMC approach with the Brownian bridge discretization in comparison
with the standard discretization using MC or QMC sampling, in
application \eg to Asian options \cite{CafMor1997,KucSha07}.
However, it was pointed in \cite{Pap01} that, in some cases, the
Brownian bridge can perform worse than the standard discretization
in QMC simulation. The big question is how to know with certainty
which numerical scheme will provide superior efficiency in QMC
simulation. Global Sensitivity Analysis (GSA) can provide this answer.
\par
GSA is a very powerful tool in the analysis of complex models as
it offers a comprehensive approach to model analysis.
Traditional sensitivity analysis, within the present context,
is "local" since, given a function $f(x_0)$ and a point $x_0$ in the function domain,
computes the derivative $\Derp{f}{x}$ at $x = x_0$. GSA instead does not require to
specify a particular point $x_0$ in the domain, since it explores
the whole domain (hence the name \textit{global}). It also
quantifies the effect of varying a given input (or set of inputs)
while all other inputs are varied as well, providing a measure of
interactions among variables. GSA is used to identify key
parameters whose uncertainty most affects the output. This
information can be used to rank variables, fix unessential
variables and decrease problem dimensionality. Reviews of GSA can
be found in \cite{SobKuc05b} and \cite{Sal10}. The
variance-based method of global sensitivity indices developed by
\cite{Sob01} became very popular among practitioners due to its
efficiency and easiness of interpretation. There are two types of
Sobol' sensitivity indices: the main effect indices, which
estimate the individual contribution of each input parameter to
the output variance, and the total sensitivity indices, which
measure the total contribution of a single input factor or a group
of inputs.
\par
For modelling and complexity reduction purposes, it is important
to distinguish between the model \textit{nominal} dimension and
its \textit{effective} dimension. The notions of effective
dimension in the truncation and superposition sense were
introduced by \cite{CafMor1997}. Further, \cite{LiuOwe06} added
the notion of ``average dimension'' which is more practical from
the computational point of view. Definitions and evaluations of
effective dimensions are based on the knowledge of Sobol'
sensitivity indices. Quite often complex mathematical models have
effective dimensions much lower than their nominal dimensions. In the
context of quantitative finance, GSA can be used to estimate
effective dimensions of a given problem. In particular, it can
assess the efficiency of a particular numerical scheme (such as
the Brownian bridge, Principal Component Analysis or standard
discretizations along the time direction or Cholesky factorization
or Principal Component Analysis in the risk factor space).
\par
``Greeks'' represent the sensitivity of the price of a financial instrument with respect to specific risk factors and are formally defined as partial derivatives of the price function. They are very important quantities which need to be computed besides price both on the Front Office side (for hedging purposes) and on Risk Management (to monitor the risk of a portfolio w.r.t. individual risk factors). If a simulation approach is used to price the instrument, standard techniques (Finite Differences, FD) require bumping each risk factor and re-pricing the instrument on each MC path. The computational cost of computing all the greeks, therefore, increases linearly with the number of underlying risk factors and becomes particularly expensive \eg for options on multiple underlyings. One popular and faster alternative to finite differences is Adjoint Algorithmic Differentiation (AAD) \cite{Gri08,Nau12,GilGla06,LecLia09,CapGil10,CapGil11,Cap11,CapLee12}. It is based on the ``Pathwise Derivative'' method: unbiased estimators of the Greeks are obtained by differentiating the discounted payoff along each MC path, see \cite{BroGla1996,Gla03}. If we want to compute the gradient of a single output w.r.t. many variables (as in the case of Greeks of multi-asset options), the adjoint mode of algorithmic differentiation can be employed to dramatically increase the efficiency of pathwise differentiation. In particular, it can be proven that the computational cost of evaluating a function and its gradient with AAD is less then approximately four times the cost of evaluating the function alone, independently of the number of derivatives to compute. So, the relative computational cost of computing all the greeks with this approach is constant (and this constant is a small number, say $\leq 4$) making AAD favourable in presence of many risk factors, such as in the case of multi-asset options. In the present work we apply adjoints to simple test cases in multi-asset option pricing with both MC and QMC, and we measure its efficiency w.r.t. finite differences, taking into account the accuracy of the computation.
\par
The paper is organized as follows: Section \ref{SecMC} contains
a brief review on QMC methodology and on Low Discrepancy Sequences,
with particular emphasis on financial applications. Section \ref{SecGSA} introduces GSA and
the notions of effective dimensions, establishing a link with QMC efficiency. In Section \ref{SecRes} we present the
results of prices and sensitivities (greeks) computation for selected payoffs: both GSA and convergence analysis are
performed, with the purpose to compare MC and QMC efficiencies via a thorough error analysis.
Finally, conclusions and directions of future work are given in Section \ref{SecConclusions}.
In particular we propose to apply our methodology to risk management issues, where a faster and smoother convergence would represent a great advantage in terms of both computational effort and budget.
A brief review of the AAD methodology is contained in the Appendix.

\section{Monte Carlo and quasi Monte Carlo methods in finance}
\label{SecMC}
\subsection{General motivation}
\label{SecGM}\noindent
Let's consider a generic financial instrument written on $N_{rf}$
assets $\bm{S}=(S_1,\ldots,S_{N_{rf}})$ with a single payment date $T$. We denote the
instrument's payoff at time $T$ as $\mathcal{P}(\bm{S}(t),\bm{\theta})$,
where $\bm{S}(t)$ is the underlying assets' value at time $t\in [0,T]$,
and $\bm{\theta}$ is a set of relevant parameters, including
\emph{instrument} parameters, such as strikes, barriers, fixing
dates of the underlyings, callable dates, payment dates, etc.,
described in the contract, and \emph{pricing} parameters, such as
interest rates, volatilities, correlations, etc., associated with
the pricing model.
\par
Using standard no-arbitrage pricing theory, see \eg \cite{Duf01},
the price of the instrument at time $t=0$ is given by
\begin{gather}\label{opt:price}
V_0(\bm{\theta})=\mathbb{E}^Q[D(0,T)\mathcal{P}(\bm{S}(t),\bm{\theta})|\mathcal{F}_0],
\end{gather}
where $\Parenthesis{\Omega,\mathcal{F},Q}$ is a probability space with risk-neutral probability measure $Q$ and filtration $\mathcal{F}_t$ at time $t$, $\mathbb{E}^Q[\,\cdot\,]$ is the expectation with respect to $Q$, $D(0,T)= \exp{\Parenthesis{-\int_0^T r(t)dt}}$ is the stochastic discount factor, and $r(t)$ is the risk-neutral short spot interest rate.
Notice that the values of $\bm{S}$ at intermediate times $t$ before final payment date $T$ may enter into the definition of the payoff $\mathcal{P}$. \QuoteDouble{Greeks} are derivatives of the price
$V_0(\bm{\theta})$ \wrt specific parameters $\bm{\theta}$. In the present work, we will
consider in particular the following component greeks:
\begin{equation}
\label{Greeks}
\Delta_i = \frac{\partial V_0}{\partial S_i(0)},\quad
\mathcal{V}_i = \frac{\partial V_0}{\partial\sigma_i},\quad
\Gamma_{ij} = \frac{\partial^2 V_0}{\partial S_i(0)\partial S_j(0)},
\end{equation}
called delta, vega and gamma respectively. Notice that \emph{model} parameters $\bm{\sigma}$ denote the volatilities of assets $\bm{S}$ and are assumed to be constant in a Black-Scholes framework.
\par
\par
The underlying assets dynamics is described by a set of stochastic differential equations (SDE). A generic Wiener diffusion model in $N_{rf}$ dimensions is generally considered and it is characterized by the following dynamics:
\begin{equation}
\label{SDE}
d\bm{S}(t) = \bm{\mu}(t,\bm{S})dt + \Sigma(t,\bm{S})\, d\bm{X}^P(t),
\end{equation}
with initial conditions $\bm{S}_0$, where $P$ is the real-world probability measure, $\bm{\mu}$ is the real-world drift, $\Sigma$ is the $N_{rf}\times N_{rf}$ volatility matrix, and $\bm{X}^P(t)$ is a $N_{rf}$-dimensional Brownian motion under $P$ with correlation matrix $R$\footnote{A correlated $N_{rf}$-dimensional Brownian motion with correlation matrix $R$ satisfies $\mathbb{E}[dX_i(t)dX_k(t)]=\rho_{ik}dt$, where $i,k=1,\ldots,N_{rf}$ and $\rho_{ik}$ are the entries of $R$, which is a symmetric, positive (semi)definite matrix with diagonal terms equal to 1.}.
In particular, in the Black-Scholes model the underlying assets $\bm{S}(t)$ follow a $N_{rf}$-dimensional geometric Brownian motion, \ie in (\ref{SDE}) $\mu_i(t,\bm{S})=\mu_iS_i(t)$ and $[\Sigma(t,\bm{S})\, d\bm{X}^P(t)]_i=\sigma_iS_i(t)\, d\bm{X}_i^P(t)$ for $i=1,\ldots,N_{rf}$, with constant drift and volatility parameters, $\bm{\mu}=(\mu_1,\ldots,\mu_{N_{rf}})$ and $\bm{\sigma}=(\sigma_1,\ldots,\sigma_{N{rf}})$ respectively. The covariance matrix is defined as
\begin{equation}\label{CovMat}
\Sigma=\tilde{D}R\tilde{D}
\end{equation}
where $\tilde{D}=\text{diag}(\sigma_1,\ldots,\sigma_{N_{rf}})$. Geometric Brownian motion dynamics can be reformulated in terms of independent Brownian motions $\bm{W}^P(t)$:
\begin{equation}\label{BS}
dS_i(t) = \mu_i\,S_i(t)\,dt + S_i(t)\, \sum_{k=1}^{N_{rf}}A_{ik}\,dW_k^P(t),
\end{equation}
where $A$ is a square root of $\Sigma$, \ie any matrix such that $AA^T=\Sigma$. The solution to equation (\ref{BS}) in a risk-neutral world (under the risk-neutral
probability measure $Q$) is given by\footnote{We assume a constant
interest rate $r$ for simplicity. See \eg \cite{BriMer06},
appendix B, for a generalization to stochastic interest rates.}
\begin{equation}
\label{BSsol}
S_i(t)=S_i(0)\,\exp{\Brack{\Parenthesis{r-\frac{1}{2}\sigma_i^2}t+\sum_{k=1}^{N_{rf}}A_{ik}\,W_k^Q(t)}}.
\end{equation}
We notice that $\bm{Y}(t)=\tilde{D}\bm{X}(t)=A\bm{W}(t)$, appearing in (\ref{BS}) and (\ref{BSsol}), is a $N_{rf}$-dimensional Brownian motion with covariance $\Sigma$. In this work, we use two different methods in order to find such matrix $A$, for any fixed $t\in[0,T]$. The first one is the Cholesky method, which yields a triangular matrix thus reducing the number of elementary operations subsequently needed to compute the Brownian motion. The second one is the Principal Component Analysis (PCA) construction, which requires a diagonalization of $\Sigma$. Let $\lambda_i$ and $\bm{v}_i$ be the $N_{rf}$ eigenvalues and an orthonormal set of corresponding eigenvectors of the covariance matrix, respectively\footnote{Since $\Sigma$ is symmetric and positive semidefinite, it has $N_{rf}$ real non-negative eigenvalues and, by the spectral theorem, an orthonormal set of eigenvectors.}. Then, the covariance matrix can be written as
\begin{equation}
\Sigma = V\,\Lambda\,V^T\, ,
\end{equation}
where $\Lambda = \text{diag}(\lambda_1,\ldots,\lambda_{N_{rf}})$ and $V=(\bm{v}_1|\cdots|\bm{v}_{N_{rf}})$. It follows that
\begin{equation}\label{PCA}
A=V\Lambda^{1/2}
\end{equation}
is a square root of $\Sigma$. Even though the PCA factorization isn't faster than the Cholesky method, it is optimal in the sense that, if the eigenvalues are ordered so that $\lambda_1\ge\lambda_2\ge\cdots\ge\lambda_{N_{rf}}$, most of the variance of the Brownian motion $\bm{Y}$ is explained by the first few principal components: formally, if $Z_1,\ldots,Z_K$ ($K\le N_{rf}$) are independent standard normals, then the error $\mathbb{E}\left[\|\bm{Y}-\sum_{i=1}^K\bm{a}_iZ_i\|^2\right]$ is minimized taking $\bm{a}_i$ as the columns of $A$ as given in (\ref{PCA}) and $Z_i=\bm{v}_i^T\bm{Y}/\sqrt{\lambda_i}$, called the $i$th principal component of $\bm{Y}$. This optimality turns out to be relevant in QMC applications.
\par
The solution to the pricing equation (\ref{opt:price}) requires
the knowledge of the values of the underlying assets $\bm{S}$ at the
relevant contract dates $\Brace{T_1,\ldots,T_n}$. Such values may
be obtained by solving the SDE (\ref{SDE}). If the SDE
cannot be solved explicitly, we must resort to a discretization
scheme, computing the  values of $\bm{S}$ on a time grid
$\Brace{t_1,\ldots,t_{N_{ts}}}$, where $t_1<t_2<\cdots<t_{N_{ts}}$, and $N_{ts}$ is
the number of time steps. Notice that the contract dates must be
included in the time grid,
$\Brace{T_1,\ldots,T_n}\subset\Brace{t_1,\ldots,t_{N_{ts}}}$. For
example, the Euler discretization scheme consists of approximating
the SDE (\ref{SDE}) by
\begin{equation}
\label{SDEdiscrete}
S_i(t_j) = S_i(t_{j-1}) + \mu_i\Parenthesis{t_{j-1},S_i(t_{j-1})}\Delta t_j + [\Sigma\Parenthesis{t_{j-1},\bm{S}(t_{j-1})}\Delta \bm{X}(t_j)]_i,\quad j=1,\ldots,N_{ts},
\end{equation}
where $\Delta t_j = t_j - t_{j-1}$, $\Delta X_i(t_j) = X_i(t_j)-X_i(t_{j-1})$ and $t_0 =0$.
In particular, the discretization of Black-Scholes solution (\ref{BSsol}) leads to
\begin{equation}\label{BSsolDiscr}
S_i(t_j)=S_i(t_{j-1})\exp{\Brack{\Parenthesis{r-\frac{\sigma_i^2}{2}}\Delta t_j+\sum_{k=1}^{N_{rf}}A_{ik}\,\Delta W_k(t_j)}},\quad i=1,\ldots,N_{rf}\, ,\;j=1,\ldots,N_{ts}.
\end{equation}
where $\Delta W_i(t_j) = W_i(t_j)-W_i(t_{j-1})$.
\par
Clearly, the price in eq. (\ref{opt:price}) will depend on the
discretization scheme adopted. We consider three discretization schemes in eq. (\ref{BSsolDiscr}): standard discretization (SD), Brownian bridge discretization (BBD) and Principal Component Analysis (PCA).
In the SD scheme the Brownian motion is discretized as follows:
\begin{equation}
\label{std}
\Delta W_i(t_j) = \sqrt{\Delta t_j}Z_{ij},\quad i=1,\ldots,N_{rf}\, ,\;j=1,\ldots,N_{ts},
\end{equation}
where $Z_{ij}$ are $N_{ts}N_{rf}$ independent standard normal variates\footnote{Gaussian numbers $Z_{ij}$ are usually sampled from a $N_{ts} N_{rf}$-dimensional gaussian vector using PRN or QRN generators. As will be discussed in the following sections, the sampling order chosen to fill $\bm{Z}$ has a relevant impact when QRN such as Sobol' sequences are used.}.

In the BBD scheme the first variate is used to generate the terminal value of the Brownian motion, while subsequent variates are used to generate intermediate points, conditioned to points already simulated at earlier and later time steps, according to the following formula,
\begin{equation}
\label{BB}
\begin{split}
&W_i(t_0)=0,\\
&W_i(t_{N_{ts}})=\sqrt{\Delta t_{N_{ts}0}} Z_{i1},\\
&W_i(t_j)=\frac{\Delta t_{kj}}{\Delta t_{kh}}\,W_i(t_h) + \frac{\Delta
t_{jh}}{\Delta t_{kh}}\,W_i(t_k)+\sqrt{\frac{\Delta t_{kj}\Delta
t_{jh}}{\Delta t_{kh}}} Z_{il}, \quad t_h<t_j<t_k\,
,\;\;\;l=2,\ldots,N_{ts},
\end{split}
\end{equation}
where $\Delta t_{ab}=t_a-t_b$.
Unlike the SD scheme, which generates the Brownian motion sequentially across time steps, the BBD scheme uses different orderings: as a result, the variance in the stochastic part of (\ref{BB}) is smaller than that in (\ref{std}) for the same time steps, so that the first few points contain most of the variance. It follows that, with the BBD, much of the shape of the Brownian motions are determined by the first few coordinates of $\bm{Z}$. However, in this way, the points of the Brownian motion where to concentrate the variance are somewhat a priori determined.

The PCA discretization scheme optimally samples from the gaussian vector so that most of the variance of the Brownian path is explained by the first few coordinates of $\bm{Z}$. It is based on the PCA factorization of the covariance matrix $C$ of the (discretized) Brownian motion vector $\left(W_i(t_1),\ldots,W_i(t_{N_{ts}})\right)$. In the case of a multi-dimensional Brownian motion, when the covariance matrix $\Sigma$ of the underlying assets is also factorized by PCA, the optimality of principal components would be reduced if an independent PCA time discretization were applied to each component. Therefore, one has to work with the full covariance matrix $C\otimes\Sigma$ of the (discretized) $N_{ts}N_{rf}$-dimensional Brownian motion
\begin{equation}
\left(Y_1(t_1),\ldots,Y_{N_{rf}}(t_1),\ldots,Y_1(t_{N_{ts}}),\ldots,Y_{N_{rf}}(t_{N_{ts}})\right)
\end{equation}
and then apply a single PCA directly to it. This also reduces the computational effort of the diagonalization, \cite{Gla03}. As we will discuss in the following sections, QMC sampling shows different efficiencies for SD, BBD and PCA.
\par
Throughout this work, we will focus on the \textit{relative}
effects of the dimension $D$ and of the discretization schemes on
the MC and QMC simulations. Thus, we will assume a simple
Black-Scholes underlying dynamics for simplicity.

\subsection{Pseudo Random Numbers and Low Discrepancy Sequences}
\label{SecRN} \noindent
The nominal dimension of the computational problem of finding option prices and greeks is $D=N_{ts} N_{rf}$, \ie the product of the number of time steps required in the discretization of the SDE (\ref{SDEdiscrete}) and the number of risk factors (the underlying assets): indeed, the expectation value in (\ref{opt:price}) is formally an integral of the payoff, regarded as a function of $D$ standard normal\footnote{Independent standard gaussian numbers $Z_j$ are computed using transformation of uniform variates $x_j\sim\ i.i.d. \;U(0,1)$, $Z_j = \Phi^{-1}(x_j)$, $j=1,\ldots,D$ where $\Phi^{-1}$ is the inverse cumulative distribution function of the standard normal distribution.} variables $Z_1,\ldots,Z_D$.
Hence the pricing problem (\ref{opt:price}) can be reduced to the evaluation of integrals of the following generic form
\begin{equation}\label{integral}
V=\int_{H^D}f(\bm{x})d^D\bm{x},
\end{equation}
where $H^D=[0,1]^D$ is the $D-$dimensional unit hypercube. This motivates the use of Monte Carlo techniques.
The standard Monte Carlo estimator of (\ref{integral}) has the form
\begin{equation}
\label{estint}
V_N\simeq \frac{1}{N}\sum_{k=1}^Nf(\bm{x}_k),
\end{equation}
where $\Brace{\bm{x}_k}_{k=1}^N$ is a sequence of $N$ random
points in $H^D$. Consider an integration error
\begin{equation}\label{interr}
\varepsilon=|V-V_N|.
\end{equation}
By the Central Limit Theorem the root mean square error of the Monte Carlo method is
\begin{equation}\label{MC:err}
\varepsilon_{MC}=\Brack{ \mathbb{E}(\varepsilon^2)}^{1/2} =
\frac{\sigma_f}{\sqrt{N}}\, ,
\end{equation}
where $\sigma_f$ is the standard deviation of $f(x)$. Although
$\varepsilon_{MC}$ does not depend on the dimension $D$, as in the
case of lattice integration on a regular grid, it decreases slowly
with increasing $N$. Variance reduction techniques, such as
antithetic variables \cite{Jac01,Gla03}, only affect the
numerator in (\ref{MC:err}).
\par
In order to increase the rate of convergence one has to
resort to Low Discrepancy Sequences (LDS), also called Quasi
Random Numbers (QRNs), instead of PRNs. The discrepancy of a
sequence $\Brace{\bm{x}_k}_{k=1}^N$ is a measure of how
inhomogeneously the sequence is distributed inside the unit
hypercube $H^D$ and is denoted by $\mathcal{D}^D_N(\bm{x}_1,\ldots,\bm{x}_N)$. A Low Discrepancy Sequence is a sequence
$\Brace{\bm{x}_k}_{k=1}^N$ in $H^D$ such that, for any $N>1$, the
first $N$ points ${\bm{x}_1,\ldots,\bm{x}_N}$ satisfy inequality
\begin{equation}
\mathcal{D}^D_N(\bm{x}_1,\ldots,\bm{x}_N) \leq c(D)\frac{\ln^D
N}{N}\, ,
\end{equation}
for some constant $c(D)$ depending only on $D$, see \cite{Nid88} for details.
\par
A QMC estimator of the integral (\ref{integral}) is of the form (\ref{estint}) with the only
difference that the sequence $\Brace{\bm{x}_k}_{k=1}^N$ is sampled using LDS instead of PRNs. An upper bound for the QMC integration error is given by the Koksma-Hlawka inequality
\begin{equation}\label{QMC:errbound}
\varepsilon_{QMC}\le
V(f)\mathcal{D}^D_N=\mathcal{O}\left(\frac{\ln^D N}{N}\right)\, ,
\end{equation}
where $V(f)$ is the variation of the integrand function in the
sense of Hardy and Krause (\cite{Nid88}), which is finite for functions of
bounded variation, see \cite{KucBal11}. The convergence rate of
(\ref{QMC:errbound}) is asymptotically faster than (\ref{MC:err}),
but it depends on
the dimensionality $D$. However, eq. (\ref{QMC:errbound}) is just
an upper bound: what is observed in most numerical tests
\cite{KucBal11,CafMor1997} is a power law
\begin{equation}
\label{err:QMC}
\varepsilon_{QMC}\sim \frac{c}{N^\alpha}\, ,
\end{equation}
where the value of $\alpha$ depends on the model function.
When $\alpha>0.5$ the QMC method outperforms standard MC: this situation turns out to be quite common in financial problems.
We will measure $\alpha$ for some representative financial instruments, showing that its value can be very close to 1 when the \emph{effective} dimension of $f$ is low, irrespective of the nominal dimension $D$. The concept of effective dimension, and the methodology to compute it, will be introduced in the following sections.
\par
Since LDS are deterministic, there are no
statistical measures like variances associated with them. Hence,
the constant $c$ in (\ref{err:QMC}) is not a variance and
(\ref{err:QMC}) does not have a probabilistic interpretation as
for standard MC. To overcome this limitation, \cite{Owe93}
suggested to introduce randomization into LDS at the same time
preserving their superiority to PRN uniformity properties. Such
LDS became known as \emph{scrambled} (see also \cite{Gla03}). In
practice, the root mean square error (RMSE) for both MC and QMC methods for
any fixed $N$ can be estimated by computing the following error
averaged over $L$ independent runs:
\begin{equation}\label{error}
\varepsilon_N=\sqrt{\frac{1}{L}\,\sum_{\ell=1}^L\left(V-V_N^{(\ell)}\right)^2},
\end{equation}
where $V$ is the exact, or estimated at a very large extreme value
of $N\rightarrow\infty$, value of the integral and $V_N^{(\ell)}$
is the simulated value for the $\ell$th run, performed using $N$
PRNs, LDS, or scrambled LDS. For MC and QMC based on scrambled
LDS, runs based on different seed points are statistically
independent. In the case of QMC, different runs are obtained using
non overlapping sections of the LDS. Actually, scrambling LDSs
weakens the smoothness and stability properties of the Monte Carlo
convergence, as we will see in Section \ref{SecStability}. Hence,
in this paper we will use the approach based on non-overlapping
LDSs, as in \cite{SobKuc05a}.
\par
The most known LDS are Halton, Faure, Niederreiter and Sobol'
sequences. Sobol' sequences, also called $LP\tau$ sequences or
$(t, s)$ sequences in base 2, see \cite{Nid88}, became the most
known and widely used LDS in finance due to their efficiency
\cite{Jac01,Gla03}. As explained \eg in
\cite{Sob1967}, Sobol' sequences were constructed under the
following requirements:
\begin{enumerate}
    \item Best uniformity of distribution as $N\to\infty$.
    \item Good distribution for fairly small initial sets.
    \item A very fast computational algorithm.
\end{enumerate}
The efficiency of Sobol' LDS depend on the so-called
initialisation numbers. In this work we used \texttt{SobolSeq8192}
generator provided by \cite{BRODA}. \texttt{SobolSeq} is an
implementation of the 8192-dimensional Sobol' sequences with
modified initialisation numbers. Sobol' sequences produced by
\texttt{SobolSeq8192} can be up to and including dimension
$2^{13}$, and satisfy additional uniformity properties: Property A
for all dimensions and Property A' for adjacent dimensions (see
\cite{SobAso12} for details\footnote{The most recent generator released by BRODA is \texttt{SobolSeq65536} which, not only has the highest dimensionality available and employs the super fast generation algorithm, but also the generated Sobol' sequences satisfy Property A in all dimensions and property A' for the adjacent dimensions.}). It has been
found in \cite{SobAso12} that \texttt{SobolSeq} generator
outperforms all other known LDS generators both in speed and
accuracy.

\section{Global sensitivity analysis and effective dimensions}
\label{SecGSA} \noindent
 Effective dimension is the key to explain the superior efficiency of QMC \wrt MC.
Hence, it is crucial to develop techniques to estimate the effective dimension
and to find the most important variables in a MC simulation.
\par
The variance-based method of global sensitivity indices developed
by Sobol' became very popular among practitioners due to its
efficiency and easiness of interpretation \cite{SobKuc05b,Sal10}.
There are two types of Sobol' sensitivity indices: the main effect
indices, which estimate the individual contribution of each input
parameter to the output variance, and the total sensitivity
indices, which measure the total contribution of a single input
factor or a group of inputs. Sobol' indices can be used to rank
variables in order of importance,  to identify non-important
variables, which can then be fixed at their nominal values to
reduce model complexity, and to analyze the efficiency of various
numerical schemes.
\par
Consider a mathematical model described by an integrable function $f(x)$, where the input $x=(x_1,\ldots,x_D)$ is taken in a $D$-dimensional domain $\Omega$ and the output is a scalar.
Without loss of generality, we choose $\Omega$ to be the unit hypercube $H^D$. The input variables $x_1,\ldots,x_D$ can, then, be regarded as independent uniform random variables each defined in the unit interval $[0,1]$.
The starting point of global sensitivity analysis (GSA) is the analysis of variance (ANOVA) decomposition of the model function,
\begin{equation}\label{ANOVA1}
f(x) = f_0 + \sum_i f_i(x_i) + \sum_{i<j}f_{ij}(x_i,x_j) + \ldots
+ f_{1\,2\cdots D}(x_1,\ldots,x_D)\, .
\end{equation}
The expansion (\ref{ANOVA1}) is unique, provided that
\begin{equation}\label{ANOVA2}
\int_0^1f_{i_1\cdots i_s}(x_{i_1},\ldots,x_{i_s})dx_{i_k}=0\,
,\;\;\;\forall k=1,\ldots,s\, .
\end{equation}
The ANOVA decomposition expands the function $f$ into a sum of terms, each depending on an increasing number of variables: a generic component $f_{i_1\cdots i_s}(x_{i_1},\ldots,x_{i_s})$, depending on $s$ variables, is called an $s$-order term. It follows from (\ref{ANOVA2}) that the ANOVA decomposition is orthogonal and that its terms can be explicitly found as follows,
\begin{equation}
\begin{split}
f_0=&\int_{H^D}f(x)d^Dx,\\
f_i(x_i)=&\int_{H^{D-1}}f(x)\prod_{k\neq i}dx_k\, -f_0,\\
f_{ij}(x_i,x_j)=&\int_{H^{D-2}}f(x)\prod_{k\neq i,j}dx_k\, -f_0-f_i(x_i)-f_j(x_j),
\end{split}
\end{equation}
and so on. If $f$ is square-integrable, its variance decomposes into a sum of partial variances:
\begin{equation}\label{totvar}
\sigma^2 = \sum_i\sigma^2_i+\sum_{i<j}\sigma^2_{ij}+\ldots+\sigma^2_{12\cdots D},
\end{equation}
where
\begin{equation}
\label{var}
\sigma^2_{i_1\cdots i_s} =\int_0^1f^2_{i_1\cdots i_s}(x_{i_1},\ldots,x_{i_s})dx_{i_1}\cdots dx_{i_s}.
\end{equation}
\par
Sobol' sensitivity indices are defined as
\begin{equation}
\label{SobSensIndex}
S_{i_1\cdots i_s}=\frac{\sigma^2_{i_1\cdots i_s}}{\sigma^2}
\end{equation}
and measure the fraction of total variance accounted by each
$f_{i_1\cdots i_s}$ term of the ANOVA decomposition. From
(\ref{totvar}) it follows that all Sobol' indices are non negative
and normalized to 1. First order Sobol' indices $S_i$ measure the
effect of single variables $x_i$ on the output function; second
order Sobol' indices $S_{ij}$ measure the interactions between
pairs of variables, \ie the fraction of total variance due to
variables $x_i$ and $x_j$ which cannot be explained by a sum of
effects of single variables; higher order Sobol' indices
$S_{i_1\cdots i_s}$, with $s>2$, measure the interactions among
multiple variables, \ie the fraction of total variance due to
variables $x_{i_1},\ldots,x_{i_s}$ which cannot be explained by a
sum of effects of single variables or lower order interactions.
\par
The calculation of Sobol' sensitivity indices in eq. (\ref{SobSensIndex})
requires, in principle, $2^D$ valuations of the multi-dimensional integrals in eq. (\ref{var}), which is unfeasible if $D$ is large.
However, for practical purposes it is not actually necessary to know all the possible Sobol' indexes,
but just an appropriate selection of them.
It is very useful to introduce Sobol' indices for subsets of variables and total Sobol' indices.
Let $y=\Brace{x_{i_1},\ldots,x_{i_m}}\subseteq x, 1\le i_1\le\ldots,\le i_m\le D$, be a subset of $x$, and $z=y^\complement\subseteq x$ its complementary subset, and define
\begin{equation}
\begin{split}
&S_y = \sum_{s=1}^D\,\sum_{(i_1<\cdots<i_s)\in K}S_{i_1\cdots i_s},\\
&S_y^{tot} = \,1-S_z\,,
\end{split}
\end{equation}
where $K=\{i_1,\ldots,i_m\}$. Notice that $0\le S_y\le
S_y^{tot}\le 1$. The quantity $S_y^{tot}-S_y$ accounts for all the
interactions between the variables in subsets $y$ and $z$. It
turns out that there exist efficient formulas which allow to avoid
the knowledge of ANOVA components and to compute Sobol' indices
directly from the values of function $f$ \cite{Sob01}. These
formulas are based on the following integrals,
\begin{equation}
\label{SI}
\begin{split}
S_y=&\frac{1}{\sigma^2}\,\int_0^1[f(y',z')-f_0][f(y',z)-f(y,z)]dy\,dz\,dy'dz'\, ,\\
S_y^{tot}=&\frac{1}{2\sigma^2}\,\int_0^1[f(y,z)-f(y',z)]^2 dy\,dz\,dy'\, ,\\
\sigma^2=&\int_0^1f^2(y,z)dy\,dz\,-f_0^2\, ,\\
f_0=&\int_0^1f(y,z)dy\,dz\,,
\end{split}
\end{equation}
where the integration variables are the components of the vectors
$y,z,y',z'$, such that $x=y\cup z$, and the first two integrals
depend on the choice of $y$. Such integrals can be evaluated, in
general, via MC/QMC techniques \cite{KucBal11,Sal02}.
\par
Furthermore, usually enough information is already given by the first order indices $S_i$ and by corresponding total effect indices $S_i^{tot}$, linked to a single variable $y=\Brace{x_i}$.
For these Sobol' indices, it's easy to see that
\begin{itemize}
    \item $S_i^{tot}=0$: the output function does not depend on
    $x_i$,
    \item $S_i=1$: the output function depends only on $x_i$,
    \item $S_i=S_i^{tot}$: there is no interaction between $x_i$ and other variables.
\end{itemize}
Notice that just $D+2$ function evaluations for each MC trial are necessary to compute all $S_i$ and $S_i^{tot}$ indices in eqs. (\ref{SI}): one function evaluation at point $x=\Brace{y,z}$, one at point $x'=\Brace{y',z'}$, and $D$ evaluations at points  $x''=\Brace{y',z},\forall\; y' = \Brace{x_i},\;i = 1,\ldots,D$.
\par
Approach presented above is applicable only to
the case of independent input variables, which admits a unique
ANOVA decomposition. In the case of dependent (correlated) input
variables, the computation of variance-based global sensitivity
indices is more involved. A generalization of GSA to dependent
variables can be found in \cite{KucTar12}.
\par
We finally come to the notion of effective dimensions, firstly
introduced in \cite{CafMor1997}. Let $|y|$ be the cardinality of
a subset of variables $y$. The effective dimension in the
\emph{superposition sense}, for a function $f$ of $D$ variables,
is the smallest integer $d_S$ such that
\begin{equation}
\sum_{0<|y|<d_S }S_y\ge 1-\varepsilon
\end{equation}
for some threshold $\varepsilon$ (arbitrary and usually chosen to be less than $5\%$). If a function has an effective dimension $d_S$ in the superposition sense, it can be approximated by a sum of $d_S$-dimensional terms, with an approximation error below $\varepsilon$.
\par
The effective dimension in the \emph{truncation sense} is the smallest integer $d_T$ such that
\begin{equation}
\sum_{y\subseteq \{1,2,\ldots,d_T\}}S_y\ge 1-\varepsilon.
\end{equation}
The effective dimension $d_S$ does not depend on the order of sampling of variables, while $d_T$ does. In general, the following inequality holds,
\begin{equation}
\label{eqDimensions}
d_S\le d_T\le D.
\end{equation}
Effective dimensions can be estimated solely from indices $S_i$
and $S_i^{tot}$ using eqs. (\ref{SI}) with $y=i$, as described in
\cite{KucBal11}, where relationships among such indices are used
to classify functions in three categories according to their
dependence on variables. For the so-called type A functions,
variables are not all equally important and the effective
dimension in the truncation sense $d_T$ is small, such that
$d_S\leq d_T\ll D$. They are characterized by the following
relationship
\begin{equation}
\label{eqTypeA}
\frac{S^{tot}_y}{|y|}\gg \frac{S_z^{tot}}{|z|},
\end{equation}
where $y\subseteq x$ is a leading subset of variables, $z=y^\complement\subseteq x$ its complementary subset.
Functions with equally important variables have $d_T\simeq D$ and they can be further divided in two groups: type B and C functions. Type B functions have dominant low-order interactions and small effective dimension in the superposition sense $d_S$, such that $d_S\ll d_T\simeq D$. For such functions, Sobol' indices satisfy the following relationships:
\begin{equation}
\label{eqTypeB}
S_i\simeq S_i^{tot},\quad \forall\;i=1,\ldots,D, \quad\sum_{i=1}^DS_i\simeq 1.
\end{equation}
Type C functions have dominant higher-order interactions
\begin{equation}
\label{eqTypeC}
S_i\ll
S_i^{tot}\, ,\;\sum_{i=1}^DS_i\ll 1
\end{equation}
and effective dimensions $d_S\simeq d_T\simeq D$. This classification is summarized in Table \ref{tab:effdim}.
\begin{table}\small
\centering
\subtable{
\begin{tabular}{cccc}
  \toprule
  \textbf{Type} & \textbf{Description} & \textbf{Relationship between SI} & \textbf{Eff. dimensions} \\
  \midrule
   A & Few important variables & $S_y^{tot}/|y|\gg S_z^{tot}/|z|$ & $d_S\leq d_T\ll D$ \\
   B & Low-order interactions & $S_i\simeq S_j,S_i\simeq S_i^{tot}\,, \forall\; i,j$  & $d_S\ll d_T\simeq D$ \\
   C & High-order interactions & $S_i\simeq S_j,S_i\ll S_i^{tot}\,, \forall\; i,j$ & $d_S\simeq d_T\simeq D$ \\
  \bottomrule
\end{tabular}
}
\caption{Classification of functions \wrt their dependence on variables, based on GSA.}
\label{tab:effdim}
\end{table}
\par
\cite{Owe03} introduced the notion of the average dimension
$d_A$, which can assume fractional values, defined as
\begin{equation} d_A := \sum_{0<|y|<D}|y|\,S_y\,,
\end{equation}
and showed that it can be rather straightforwardly computed as
\begin{equation}\label{d_A}
d_A=\sum_{i=1}^DS_i^{tot}\, .
\end{equation}
It has been suggested in \cite{SobShu14} that QMC should
outperform MC when $d_A\lesssim 3$. This is confirmed in our
findings, see Section \ref{SecGSAresults}.
\par
It has been proved in many works \cite{KucBal11,CafMor1997,
Owe03} that QMC outperforms MC regardless of the nominal dimension
whenever the effective dimension is low in one or more senses.
Hence, in the case of type A and type B functions (we assume that
functions are sufficiently smooth), QMC always outperforms MC,
while for type C functions the two methods are expected to have
similar efficiency. Actually, type A and B functions are very
common in financial problems. We also note that the performance of
the QMC method for Type A functions sometimes, but not always, can
be greatly improved by using effective dimension reduction
techniques, such as Brownian bridge, which will be demonstrated in
the following section.

\section{Numerical results}
\label{SecRes} \noindent
In this Section we apply MC and QMC techniques to high-dimensional pricing problems. Our aim is to test the efficiency of QMC with respect to standard MC in computing prices and greeks (delta, gamma, vega) for selected option payoffs $\mathcal{P}$ with increasing degree of complexity and path-dependency. We will consider both single asset and multi-asset test cases.

\subsection{Selected payoffs and test set-up}
\label{SecPayoffs} \noindent
We selected the following instruments as test cases: the first three payoffs depend on a single underlying $S$, while the remaining two payoffs depend on multiple underlyings $\bm{S}$.
\begin{enumerate}
    \item \textbf{Asian call}:
        \beq\mathcal{P} = \max(\bar{S}-K,0),\quad \bar{S}=\frac{1}{N_{ts}}\sum_{j=1}^{N_{ts}} S(t_j)
        \label{eqPayoffAsian}.\eeq
    \item \textbf{Double knock-out call}:
        \beq\mathcal{P} = \max(S(T)-K,0)\,\mathds{1}_{\{\min_{j=1,\ldots,N_{ts}}{S(t_j)}>B_l,\;\max_{j=1,\ldots,N_{ts}}{S(t_j)}<B_u\}}
        \label{eqPayoffDKO}.\eeq
    \item \textbf{Cliquet}:
        \beq\mathcal{P} = \max\left\{\sum_{j=1}^{N_{ts}} \max\left[0,\min\left(C,\frac{S(t_j)-S(t_{j-1})}{S(t_{j-1})}\right)\right],F\right\}.\label{eqPayoffCliquet}\eeq
    \item \textbf{European  basket call}:
        \beq\mathcal{P} = \max(B(T)-K,0),\quad B(t) = \sum_{i=1}^{N_{rf}}w_i\,S_i(t)\label{eqPayoffBasketCall}.\eeq
    \item \textbf{Asian basket call}:
        \beq\mathcal{P} = \max(\bar{B}-K,0),\quad \bar{B}=\frac{1}{N_{ts}}\sum_{j=1}^{N_{ts}} B(t_j)\label{eqPayoffBasketAsian}.\eeq
\end{enumerate}
In the above definitions, $K$ denotes the strike price, $T$ is the maturity, $B_l$ and $B_u$ are the values of the lower and upper barrier, respectively, $C$ is a local cap, $F$ is a global floor and $w_i$ are the weights for a basket $B$ composed of $N_{rf}>1$ underlyings. Furthermore, payoffs (\ref{eqPayoffAsian}, \ref{eqPayoffDKO}, \ref{eqPayoffCliquet}, \ref{eqPayoffBasketAsian}) are defined on a schedule $(t_1,\ldots,t_{N_{ts}})$ of fixing dates.

In all test cases we use the following payoff parameters:
\begin{itemize}
    \item maturity: $T=1$,
    \item strike: $K=100$,
    \item lower barrier: $B_l = 0.5\,S_0$,
    \item upper barrier: $B_u = 1.5\, S_0$,
    \item global floor: $F=0.16$,
    \item local cap: $C=0.08$,
    \item number of basket components: $N_{rf} = 5$,
    \item basket weights: $w_i=0.2\quad\forall i = 1,\ldots,N_{rf}$.
\end{itemize}
Such selection guarantees an increasing level of complexity and
path-dependency. The Asian call with
arithmetic average is the simplest and most diffused non-European
payoff. The double barrier is another very diffused
payoff with stronger path-dependency. The cliquet option
is a typical strongly path-dependent payoff based on the
performance of the underlying stock \cite{Wil06}. Finally, basket options allow to test, in addition, the contribution of correlation among different stocks. Clearly, many other possible
payoffs could be added to the test (\eg autocallable), but we
think that such selection should be complete enough to cover most
of the path-dependency and correlation characteristics relevant in the Monte Carlo
simulation.

We assume that the stochastic processes $\bm{S}(t)$, which govern the evolution of the underlying assets, follow a $N_{rf}$-dimensional geometric Brownian motion as described in Section \ref{SecGM}, with the
following model parameters:
\begin{enumerate}
 \item Payoffs (\ref{eqPayoffAsian}, \ref{eqPayoffDKO}, \ref{eqPayoffCliquet}):
    \begin{itemize}
    \item spot price: $S_0=100$,
    \item volatility: $\sigma = 0.3$.
\end{itemize}
  \item Payoffs (\ref{eqPayoffBasketCall}, \ref{eqPayoffBasketAsian})
     \begin{itemize}
        \item spot prices: $\bm{S}_0=(80,90,100,110,120)$,\label{OrderSpot}
        \item volatilities: $\bm{\sigma} = (0.5, 0.4, 0.2, 0.3, 0.6)$,\label{OrderVola}
        \item correlations: $\rho_{ij}=\rho,\;i\neq j\, ,\;\rho=0,0.3,0.6,0.9$.
    \end{itemize}
\end{enumerate}
Therefore, we assume for simplicity that all assets have the same (constant) correlation $\rho$ and we let $\rho$ vary from 0 to 1. We notice that the choice of the values of $\bm{S}_0$ and $\bm{\sigma}$ ensures that the variables associated to different assets have different importance, which is a realistic case and is crucial in the application of QMC techniques, as will be discussed. The processes $\bm{S}(t)$ are discretized across $N_{ts}$ time steps
$\Brace{t_1<\cdots<t_j<\cdots<t_{N_{ts}}}$, where $N_{ts}=32$ for the single asset cases and $N_{ts}=16$ for the multi-asset cases (also for the European basket in order to have the same dimensionality as for the Asian one). The dimension of the simulation is therefore $D=N_{ts}=32$ for the single asset cases and $D=N_{ts}N_{rf}=80$ for the multi-asset cases.

The numerical computations are performed in \texttt{Matlab} using different combinations of sampling techniques\footnote{Notice that, in contrast to \cite{BiaKuc15}, here we use crude MC simulations instead of MC + antithetic variables.}, both in time dimension and in risk-factors space (in parenthesis the factorization method of the covariance matrix of the underlying assets, relevant only for the multi-asset case, is indicated):
\begin{itemize}
    \item MC + SD (+ CHOL),
    \item QMC + SD (+ CHOL),
    \item QMC + SD (+ PCA),
    \item QMC + BBD (+ CHOL),
    \item QMC + BBD (+ PCA),
    \item QMC + PCA (+ CHOL),
    \item QMC + PCA (+ PCA).
\end{itemize}
Regarding simulation parameters, we denote by $N$ the number of simulated paths for the underlyings and by $L$ the number of independent runs. Following the specifics of Sobol' sequences, we take $N=2^p$, where $p$ is an integer, since this guarantees the lowest discrepancy properties. For the MC case, we use Mersenne Twister generator, while in the case of QMC we use \texttt{SobolSeq8192} generator.
\par
Simulation errors $\varepsilon_N$ are analyzed by computing the root mean square error (RMSE) as defined by (\ref{error}), where $V$ is a reference value of prices or greeks simulated with a large number of MC scenarios ($N\approx 10^8\,\div\,10^9$ depending on the payoff, according to the necessary accuracy). To assess and compare performance of MC and QMC methods with different schemes, we compute the scaling of the RMSE as a function of $N$ by fitting the function $\varepsilon_N$ with a power law $c\,N^{-\alpha}$ (\ref{err:QMC}). RMSE is obtained with $L=100$ independent runs as discussed in Section \ref{SecRN}.
\par
Finally, greeks for the payoffs above are computed with the following alternative approaches:
\begin{itemize}
     \item FD: central finite differences with path recycling and with shift parameter  $\epsilon=10^{-3}$ (see appendix \ref{App:GreekErr} for details).
     \item AAD: hand coded adjoints as described in appendix \ref{SecAAD}, only for multi-asset options and for first order sensitivities.
\end{itemize}
Notice that the analysis of the RMSE for FD greeks is, in general,
more complex than that for prices, since the variance of the MC
simulation mixes with the bias due to the approximation of
derivatives with finite differences. This issue is discussed in Appendix \ref{App:GreekErr}.

\subsection{Global sensitivity analysis for prices and greeks}
\label{SecGSAresults} \noindent
Sobol' indices $S_i$ and $S_i^{tot}$ are
computed for different sampling techniques
using eqs. (\ref{SI}), where $f$ is the relevant model function
(the instrument payoff or a greek with finite differences\footnote{Notice that only FD estimators of greeks are considered, since AAD is discussed, here, only with standard MC, for which GSA is not necessary.}) and
$y=\Brace{x_i}$, $y'=\Brace{x'_i}$,
$z=\Brace{x_1,\ldots,x_{i-1},x_{i+1},\ldots,x_D}$,
$z'=\Brace{x'_1,\ldots,x'_{i-1},x'_{i+1},\ldots,x'_D}$. Here
${x_i}$ are the uniform variates used to construct gaussian numbers (see Section \ref{SecRN}). The integrals in eqs. (\ref{SI}) are computed
using QMC simulation.
\par
\begin{figure}[ht]
\centering
\subfigure[Price]{\includegraphics[width=3.1in,height=2.4in,keepaspectratio=false]{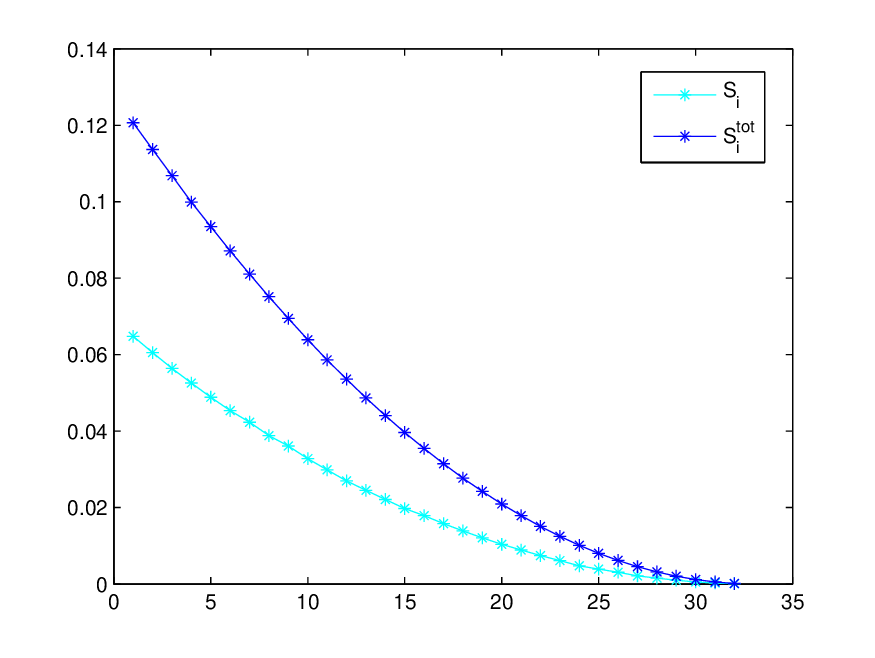}}
\subfigure[Delta]{\includegraphics[width=3.1in,height=2.4in,keepaspectratio=false]{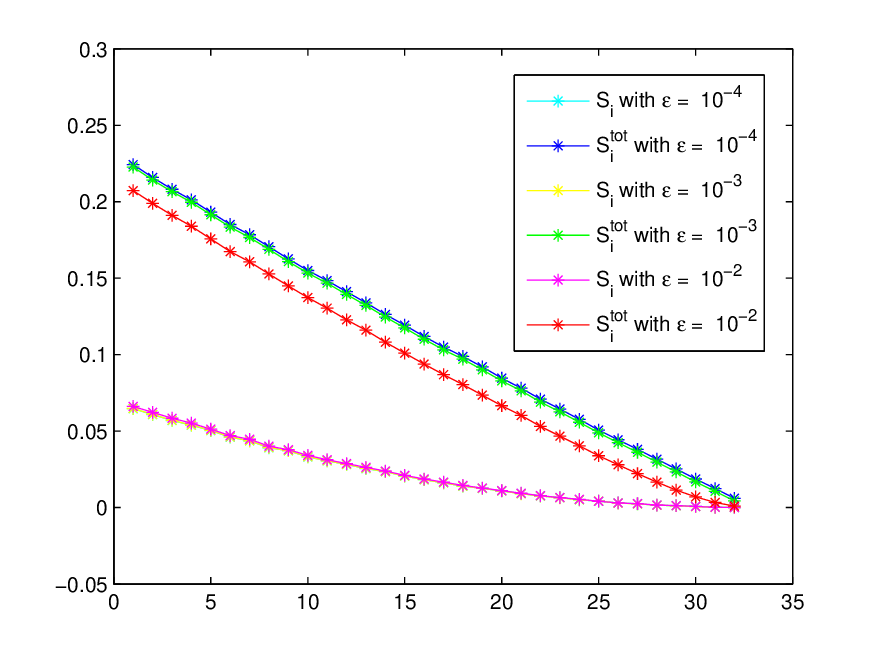}}
\subfigure[Gamma]{\includegraphics[width=3.1in,height=2.4in,keepaspectratio=false]{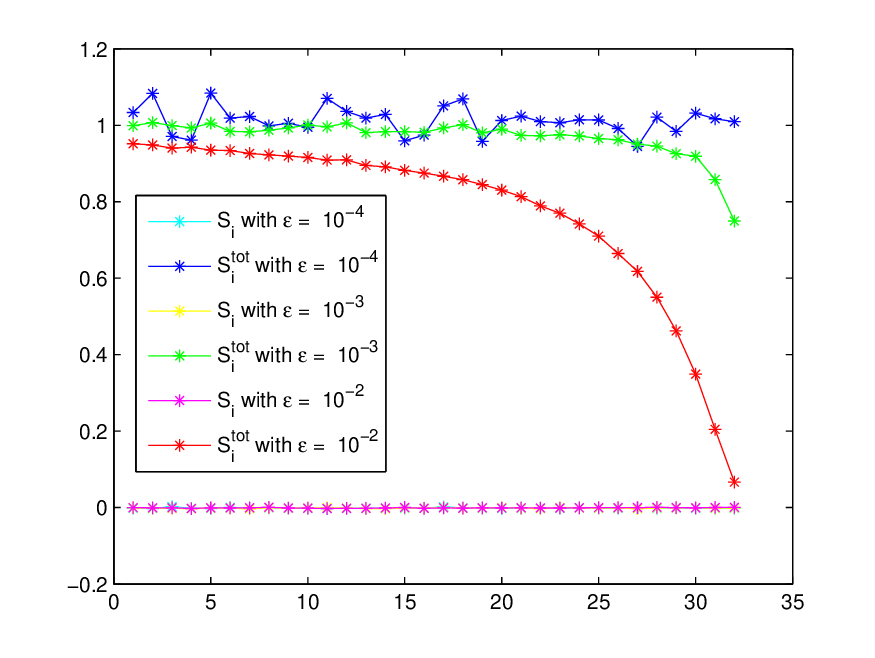}}
\subfigure[Vega] {\includegraphics[width=3.1in,height=2.4in,keepaspectratio=false]{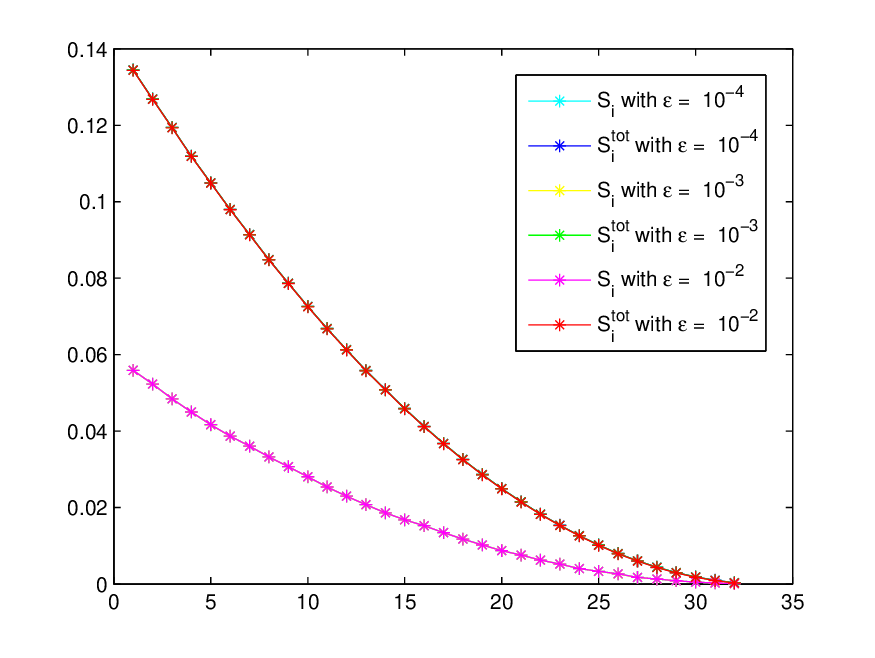}}
\caption{Asian call option price $(a)$ and greeks $(b),(c),(d)$, SD, $D=32$. First order Sobol' indices $S_{i}$ and total sensitivity indices $S_{i}^{tot}$ versus time step $i$.}
\label{fig:2}
\end{figure}
\begin{figure}[ht]
\centering
\subfigure[Price]{\includegraphics[width=3.1in,height=2.4in,keepaspectratio=false]{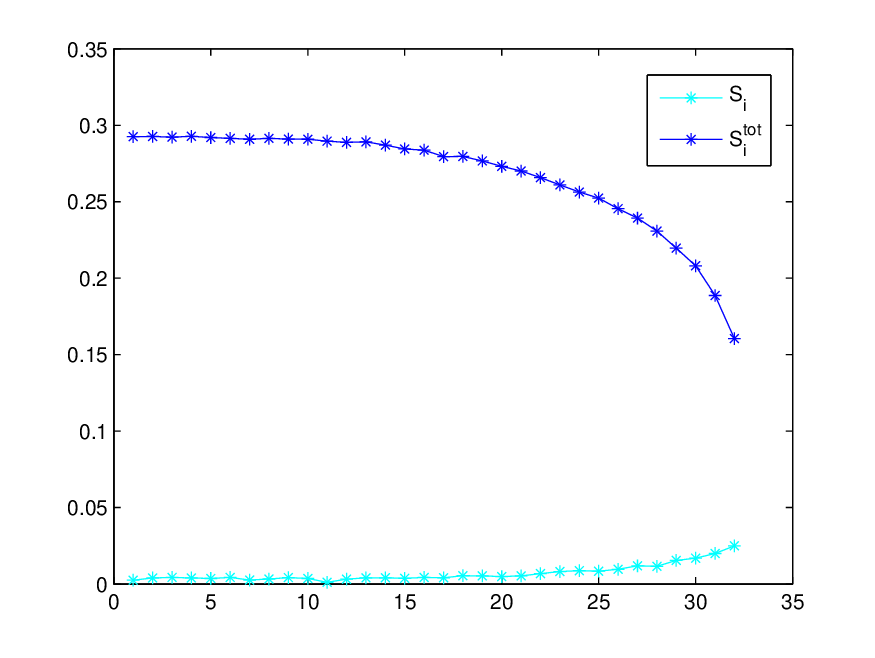}}
\subfigure[Delta]{\includegraphics[width=3.1in,height=2.4in,keepaspectratio=false]{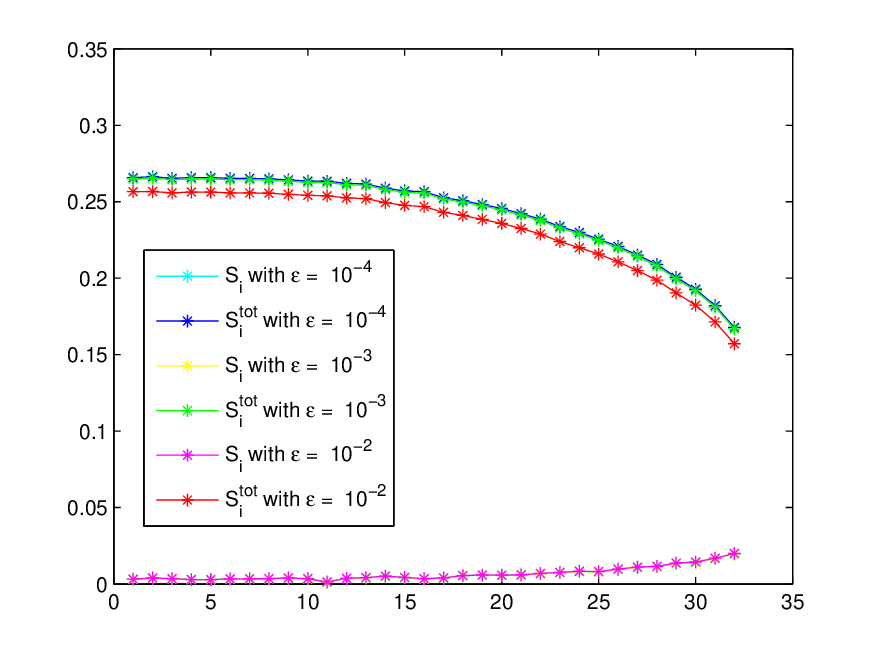}}
\subfigure[Gamma]{\includegraphics[width=3.1in,height=2.4in,keepaspectratio=false]{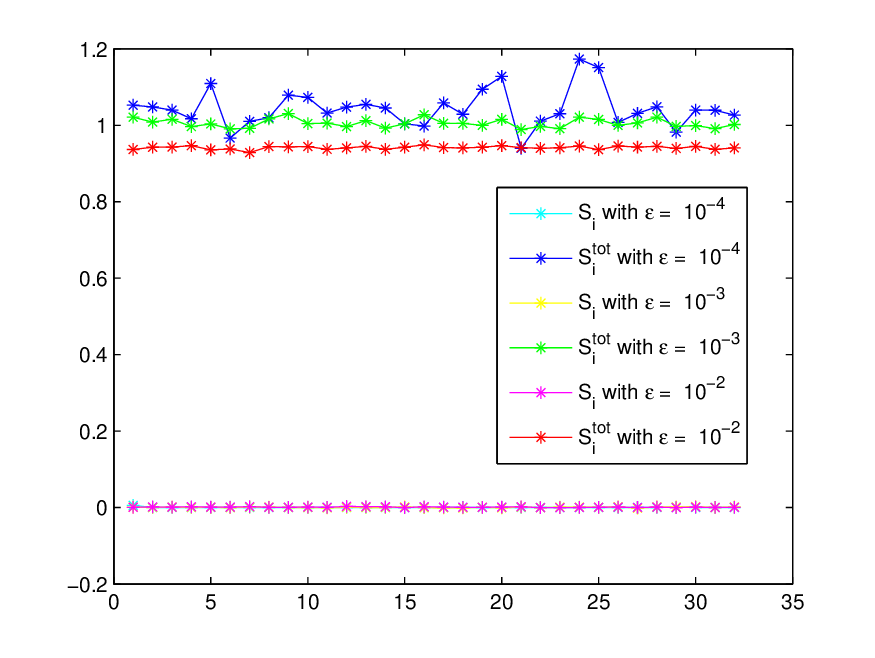}}
\subfigure[Vega] {\includegraphics[width=3.1in,height=2.4in,keepaspectratio=false]{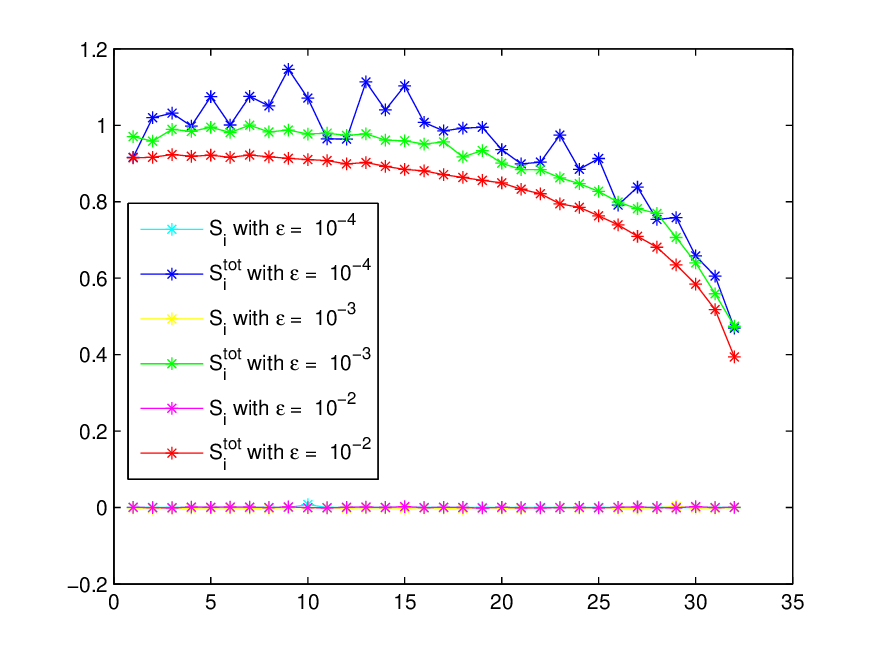}}
\caption{Double Knock-out call option. Parameters as in Fig.
\ref{fig:2}.} \label{fig:3}
\end{figure}
\begin{figure}[ht]
\centering
\subfigure[Price]{\includegraphics[width=3.1in,height=2.4in,keepaspectratio=false]{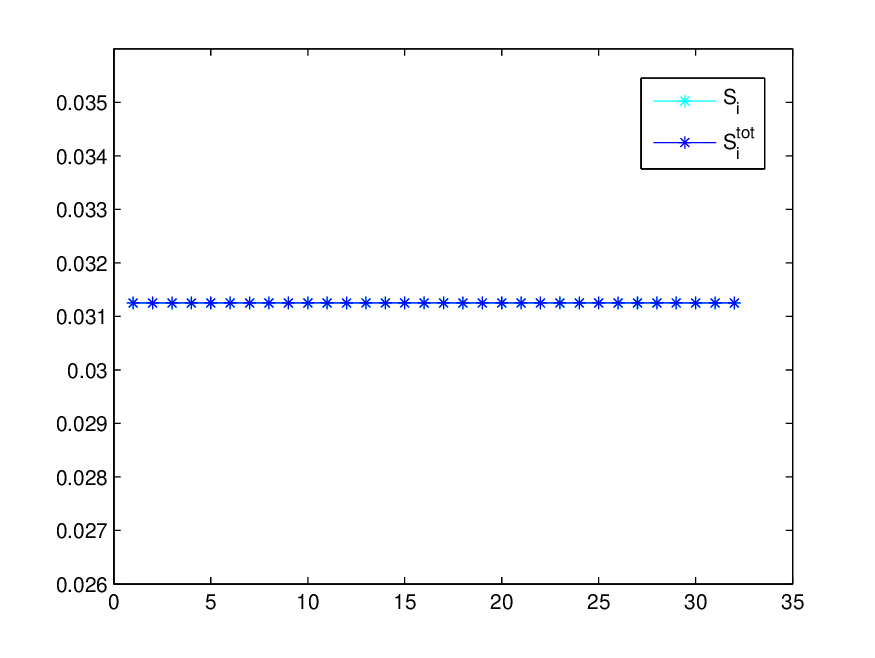}}
\subfigure[Vega] {\includegraphics[width=3.1in,height=2.4in,keepaspectratio=false]{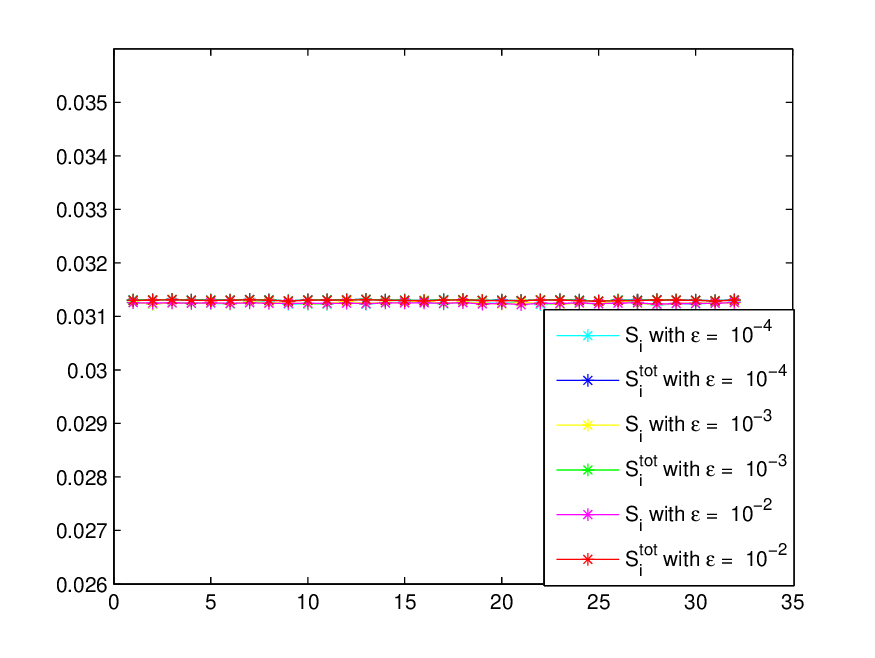}}
\caption{Cliquet option. Parameters as in Fig. \ref{fig:2}.
Delta and gamma are null for Cliquet options.} \label{fig:4}
\end{figure}

Effective dimensions are estimated in the following way:
\begin{itemize}
  \item The effective dimension in the truncation sense $d_T$ is computed using inequality (\ref{eqTypeA}), looking for a minimal set of variables $y=\Brace{x_1,\ldots,x_{d_T}}$ such that the quantity $S_z^{tot}|y|/S_y^{tot}|z|$ is smaller than 1\%. Since the calculation of $d_T$ depends on the order of sampling variables, the result depends on the sampling techniques.
  \item The effective dimension in the superposition sense $d_S$ is estimated using dimension $d_T$ as an upper bound according to inequality (\ref{eqDimensions}). In order to distinguish between Type B and Type C functions, we look at ratios $S_i/S_i^{tot}$ and $\sum_i S_i$ according to eqs. (\ref{eqTypeB}), (\ref{eqTypeC}).
  \item The effective average dimension $d_A$ is computed using eq. (\ref{d_A}).
\end{itemize}
Let's consider the single-asset payoffs (\ref{eqPayoffAsian}), (\ref{eqPayoffDKO}), (\ref{eqPayoffCliquet}) first.
The results of GSA for the SD are shown in Figs. \ref{fig:2}-\ref{fig:4}.
Measures based on Sobol' indices are provided in Table \ref{tab:1}. These measures are used to compute effective
dimensions and to classify integrands in (\ref{integral}) corresponding to prices and greeks according to Table
\ref{tab:effdim}.
\begin{table}\small
\centering
\subtable{%
\begin{tabular}{c c c c c c c c}
  \toprule
  \textbf{Payoff}& \textbf{Function} & $\mathbf{S_i/S_i^{tot}}$ & $\mathbf{\sum_iS_i}$ & $\mathbf{d_T}$ & $\mathbf{d_S}$ & $\mathbf{d_A}$ & \textbf{Effect of} $\mathbf{\epsilon}$ \\
  \midrule
  Asian    & Price & 0.54$\to$0.43 & 0.714 & $< 32$ & $< 32$ & 1.38 & - \\
                   & Delta & 0.32$\to 10^{-2}$ & 0.71$\to$0.74 & 32 & $< 32$ & 3.5 & small \\
                   & Gamma & $10^{-4}\to 10^{-2}$ & $10^{-4}\to 10^{-2}$ & 32 & $ 32$ & $31\to 25$ & high \\
                   & Vega & 0.42$\to$0.01 & 0.611 & $< 32$ & $< 32$ & 1.57 & no \\
  \hline
  Double KO& Price & 0.01$\to$0.15 & 0.22 & 32 & $< 32$ & 8.5 & - \\
                   & Delta & 0.01$\to 0.12$ & 0.22 & 32 & $< 32$ & 7.6 & no \\
                   & Gamma & $10^{-5}\to 10^{-7}$ & $10^{-4}\to 10^{-2}$ & 32 & $ 32$ & $31.2\to 29.8$ & high \\
                   & Vega & $10^{-5}\to 10^{-8}$ & $10^{-4}\to 10^{-2}$ & 32 & $ 32$ & 28 & high \\
  \hline
  Cliquet  & Price & 1 & 1 & 32 & 1 & 1 & - \\
                   & Vega & 1 & 1 & 32 & 1 & 1 & no \\
  \bottomrule
\end{tabular}}
\caption{Summary of GSA metrics and effective dimensions of prices
and greeks for SD scheme. Arrow ``$\to$'' in the column for
${S_i/S_i^{tot}}$ denotes the change in the value with the
increase of index $i$ and/or with the increase of shift parameter
$\epsilon=10^{-4},10^{-3},10^{-2}$ (see appendix \ref{App:GreekErr}); in the column for ${\sum_iS_i}$ it denotes the change
in the value with the increase of shift parameter $\epsilon$. The
numerical computations were obtained evaluating the $D+2$ integrals (\ref{SI}) on $2^{17}$ QMC scenarios for each function (price or greeks). We show significant digits only.} \label{tab:1}
\end{table}

From these results we draw the following conclusions.
\begin{enumerate}
\item Asian option (Figure \ref{fig:2}): price, and vega are type B functions, while delta and gamma are type C function.
\item Double KO option (Figure \ref{fig:3}): price and all greeks are type C functions.
\item Cliquet option (Figure \ref{fig:4}): price and vega are type B functions with $d_S=1$ (delta and gamma for a Cliquet option are null). We recall that $d_S=1$ means that there are no interactions among variables.
\end{enumerate}
\par
\begin{figure}[ht]
\centering
\subfigure[Price]{\includegraphics[width=3.1in,height=2.4in,keepaspectratio=false]{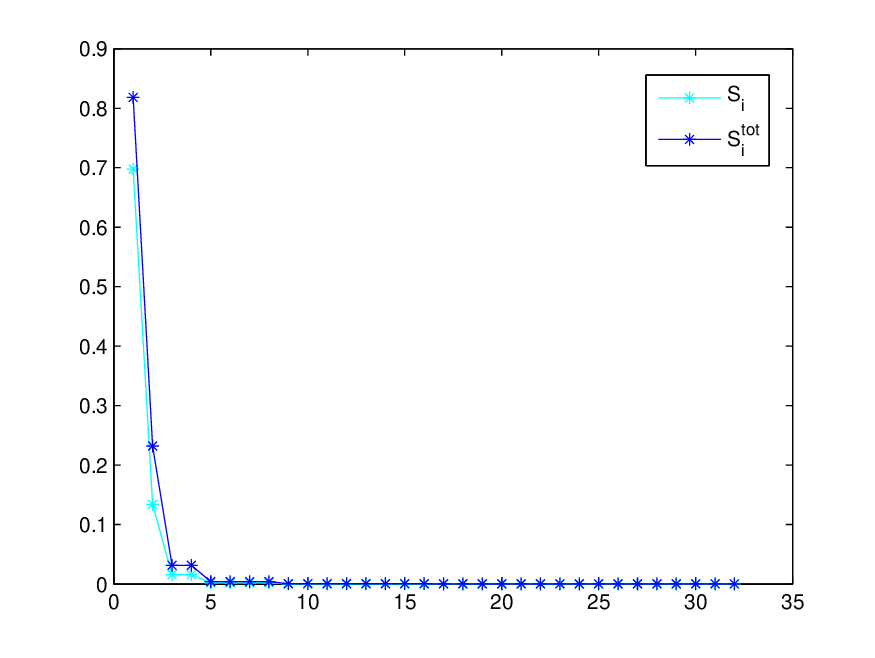}}
\subfigure[Delta]{\includegraphics[width=3.1in,height=2.4in,keepaspectratio=false]{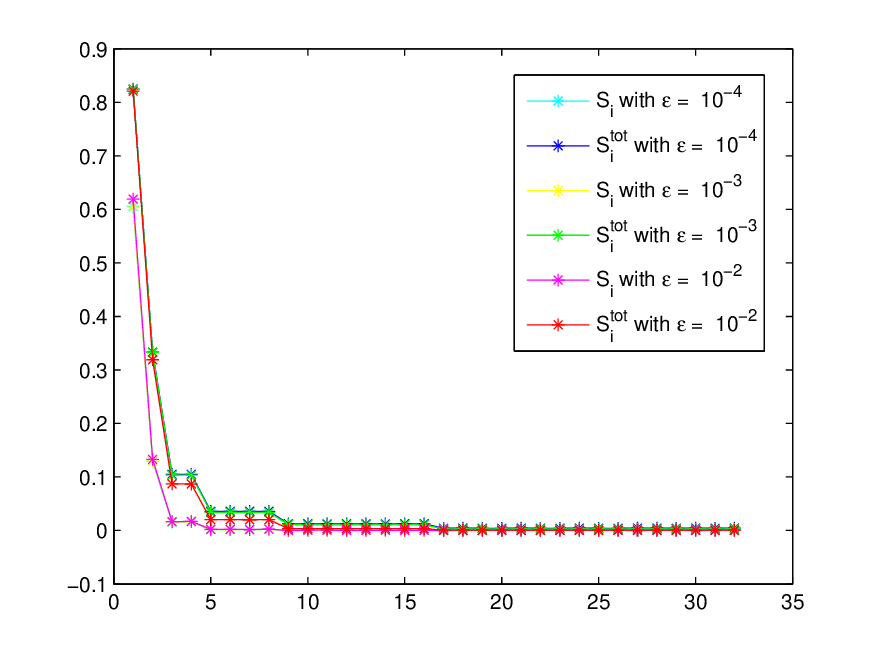}}
\subfigure[Gamma]{\includegraphics[width=3.1in,height=2.4in,keepaspectratio=false]{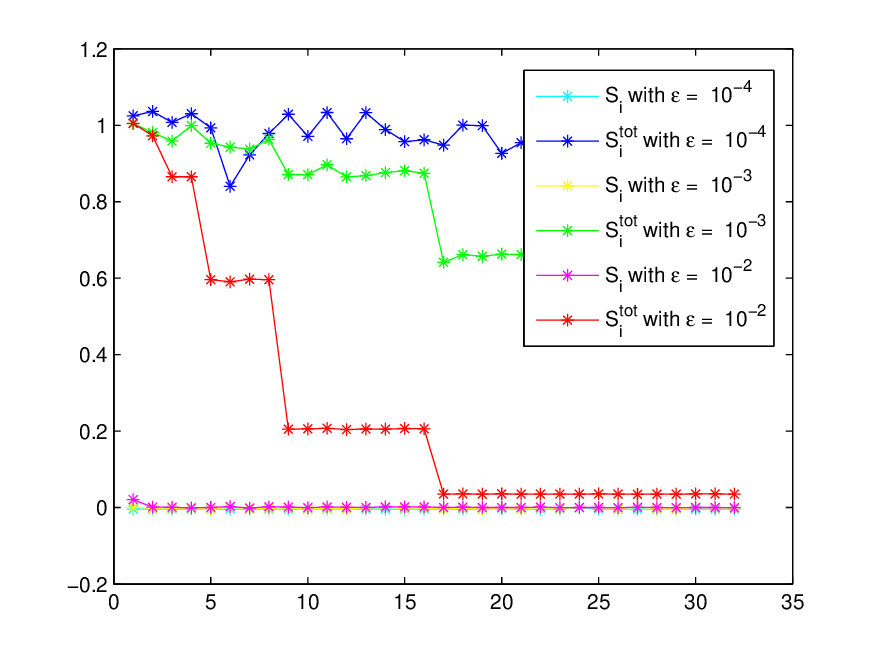}}
\subfigure[Vega] {\includegraphics[width=3.1in,height=2.4in,keepaspectratio=false]{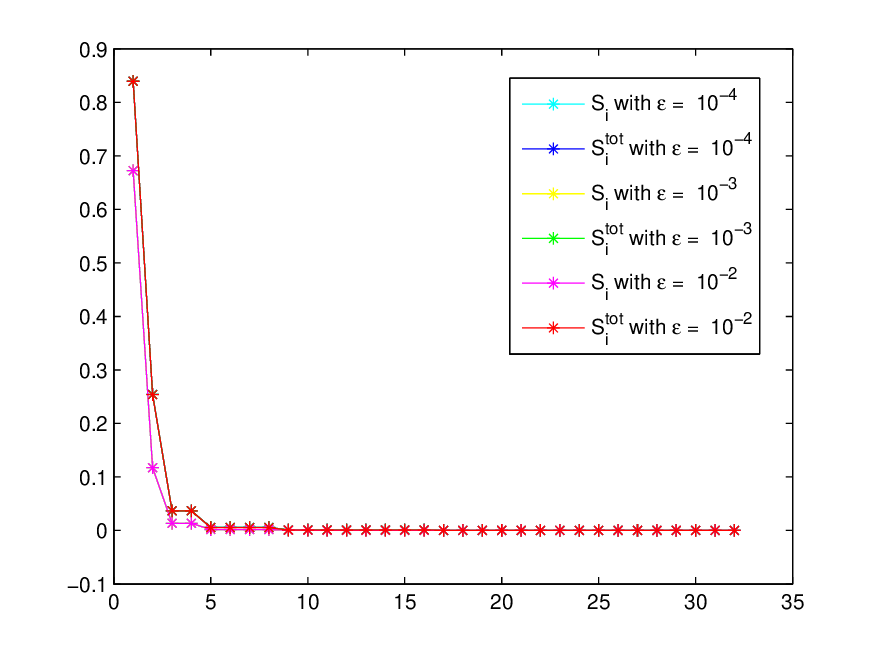}}
\caption{Asian call option price $(a)$ and greeks
$(b),(c),(d)$, BBD, $D=32$. First order Sobol' indices $S_{i}$ and
total sensitivity indices $S_{i}^{tot}$ versus time step $i$.}
\label{fig:6}
\end{figure}
\begin{figure}[ht]
\centering
\subfigure[Price]{\includegraphics[width=3.1in,height=2.4in,keepaspectratio=false]{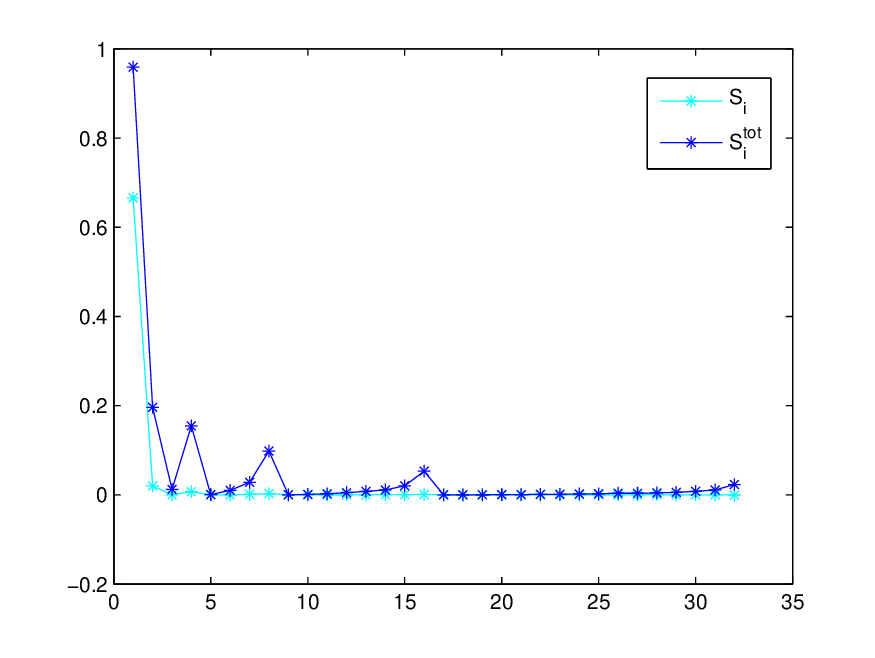}}
\subfigure[Delta]{\includegraphics[width=3.1in,height=2.4in,keepaspectratio=false]{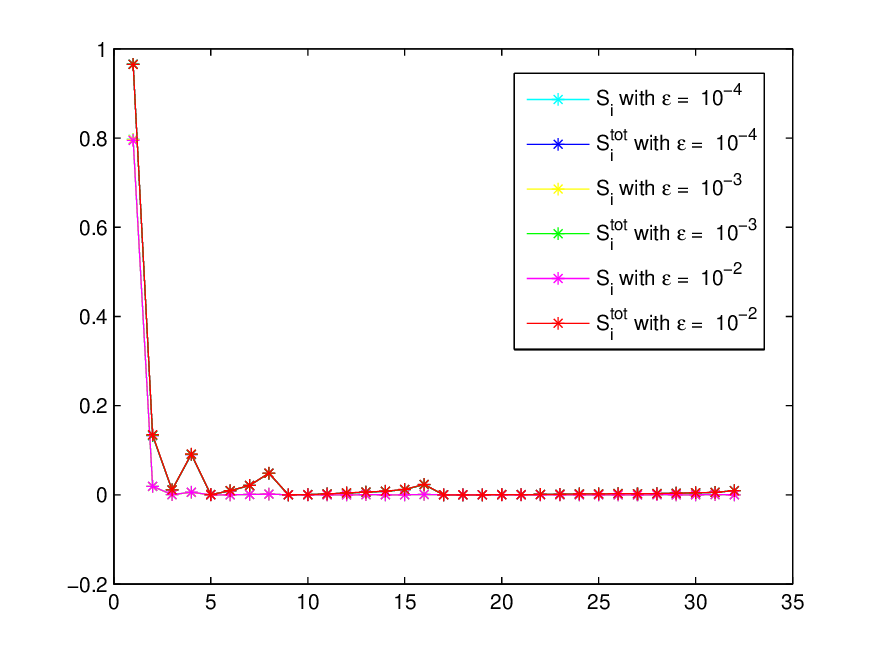}}
\subfigure[Gamma]{\includegraphics[width=3.1in,height=2.4in,keepaspectratio=false]{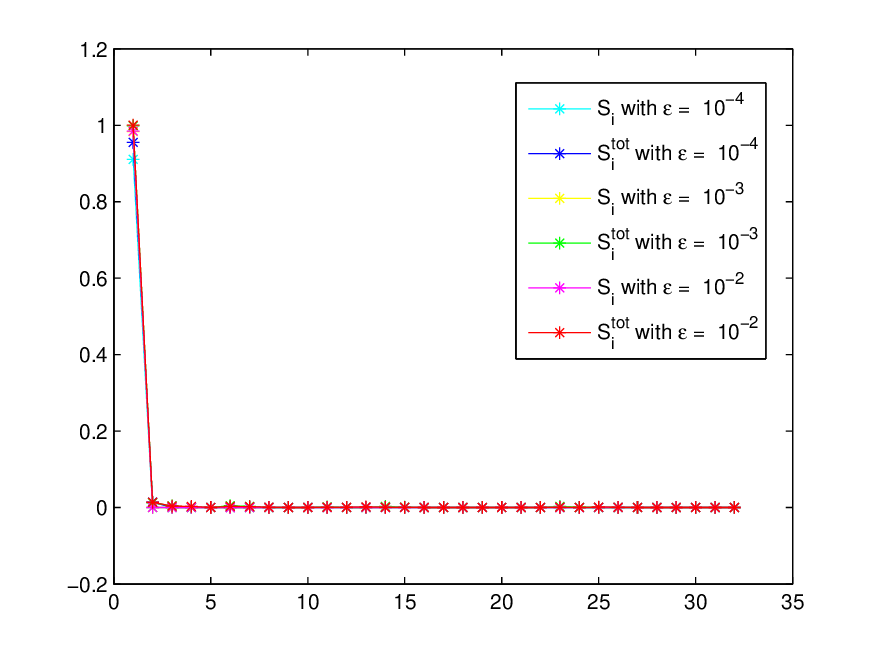}}
\subfigure[Vega] {\includegraphics[width=3.1in,height=2.4in,keepaspectratio=false]{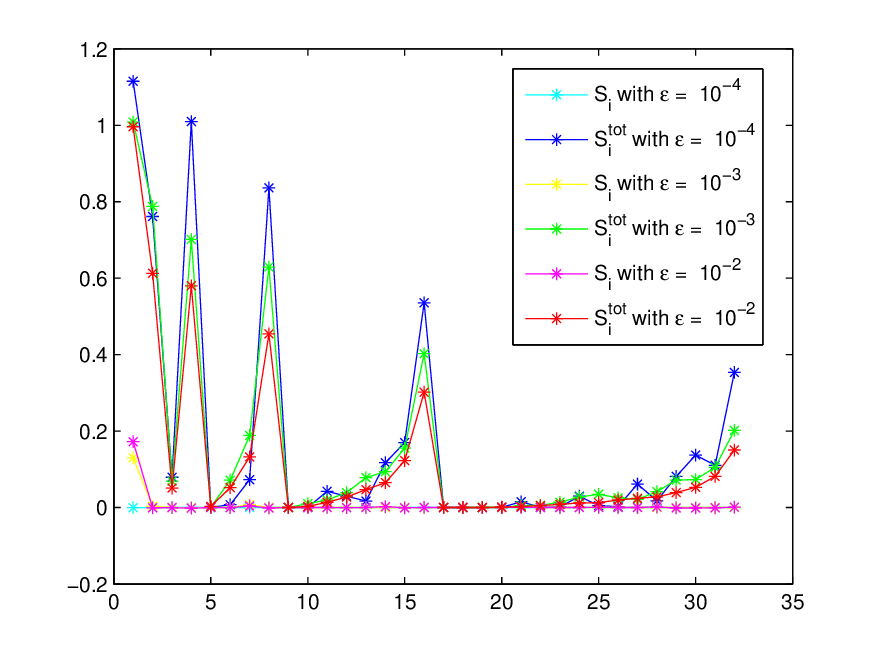}}
\caption{Double Knock-out call option. Details as in Fig.
\ref{fig:6}.} \label{fig:7}
\end{figure}
\begin{figure}[ht]
\centering
\subfigure[Price]{\includegraphics[width=3.1in,height=2.4in,keepaspectratio=false]{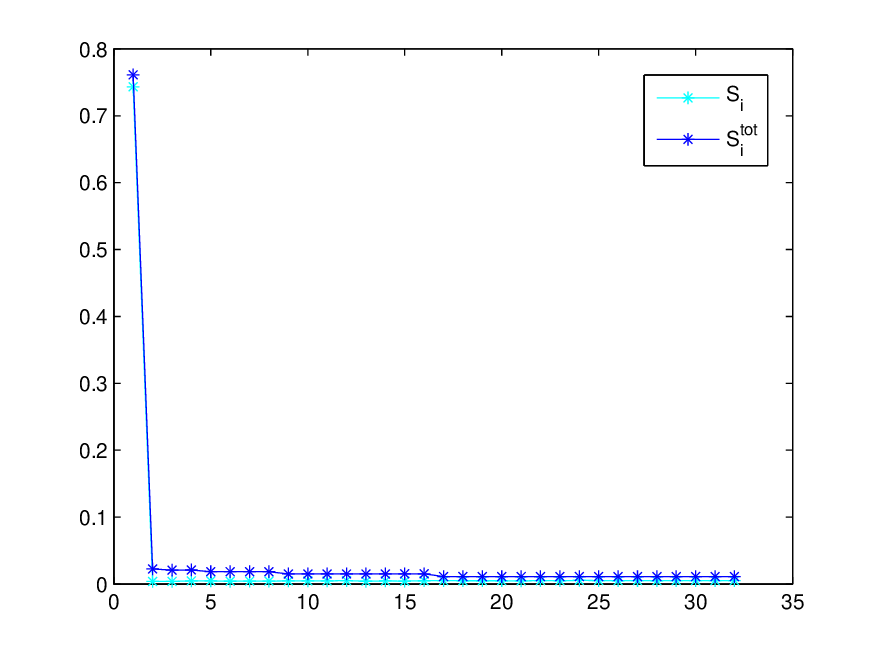}}
\subfigure[Vega] {\includegraphics[width=3.1in,height=2.4in,keepaspectratio=false]{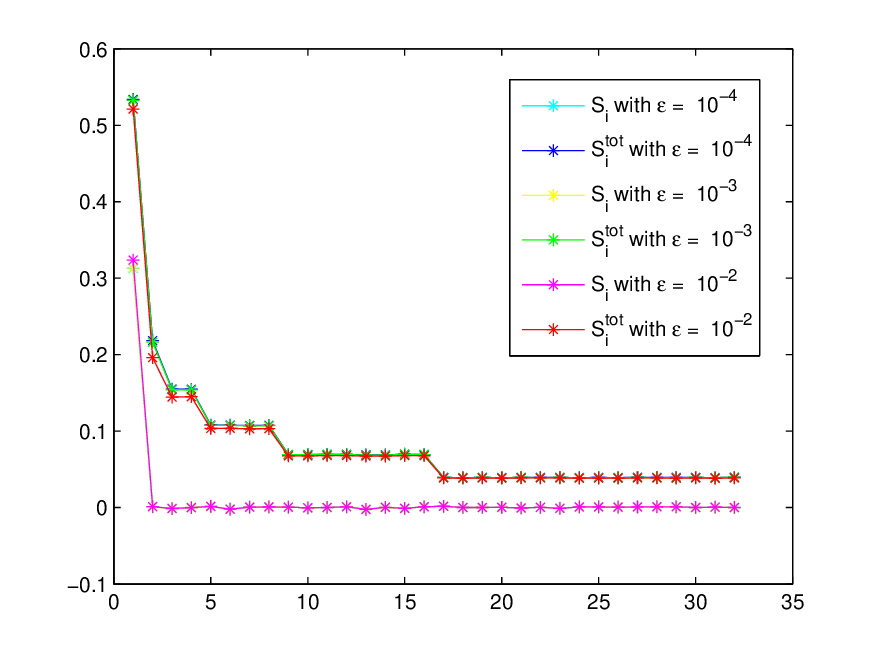}}
\caption{Cliquet option. Details as in Fig. \ref{fig:6}.}
\label{fig:8}
\end{figure}

The analogous results of GSA for BBD are shown in Fig.s \ref{fig:6}-\ref{fig:8} and in Table \ref{tab:2}.
\begin{table}[h]\small
\centering
\subtable{%
\begin{tabular}{c c c c c c c c}
  \toprule
  \textbf{Payoff}& \textbf{Function} & $\mathbf{S_i/S_i^{tot}}$ & $\mathbf{\sum_iS_i}$ & $\mathbf{d_T}$ & $\mathbf{d_S}$ & $\mathbf{d_A}$ & \textbf{Effect of} $\mathbf{\epsilon}$ \\
  \midrule
  Asian    & Price & 0.853$\to$0.4 & 0.875 & 2 & $\leq$2 & 1.13 & - \\
           & Delta & 0.733$\to$0.01 & 0.778 & 4 & $\leq$4 & $1.68\to 1.43$ & small \\
           & Gamma & $10^{-2}\to 10^{-4}$ & $0.022\to 10^{-4}$ & 32 & 32 & $31\to 8$ & high \\
           & Vega & 0.802$\to$0.03 & 0.827 & 2 & $\leq$2 & 1.20 & no \\
  \hline
  Double KO& Price & 0.70$\to$0.01 & 0.70 & $\simeq 2$ & $\leq$2 & 1.63 & - \\
           & Delta & 0.83$\to$0.01 & 0.83 & 2 & $\leq$2 & 1.37 & no \\
           & Gamma & 1 & $1\to 0.95$ & 1 & 1 & 1.0 & small \\
           & Vega & $10^{-4}\to 0.2$ & $10^{-6}\to 10^{-4}$ & 32 & 32 & $4.8\to 3.9$ & high \\
  \hline
  Cliquet  & Price & 0.978$\to$0.2 & 0.892 & $\simeq 2$ & $\leq 2$ & 1.19 & - \\
           & Vega & 0.595$\to$0.001 & 0.32 & $\simeq 32$ & $\leq 32$ & 2.6 & no \\
  \bottomrule
\end{tabular}}
\caption{Summary of GSA metrics and effective dimensions of prices
and greeks for BBD scheme. Details as in Table \ref{tab:1}.}
\label{tab:2}
\end{table}

\begin{figure}[ht]
\centering
\subfigure[Price]{\includegraphics[width=3.1in,height=2.4in,keepaspectratio=false]{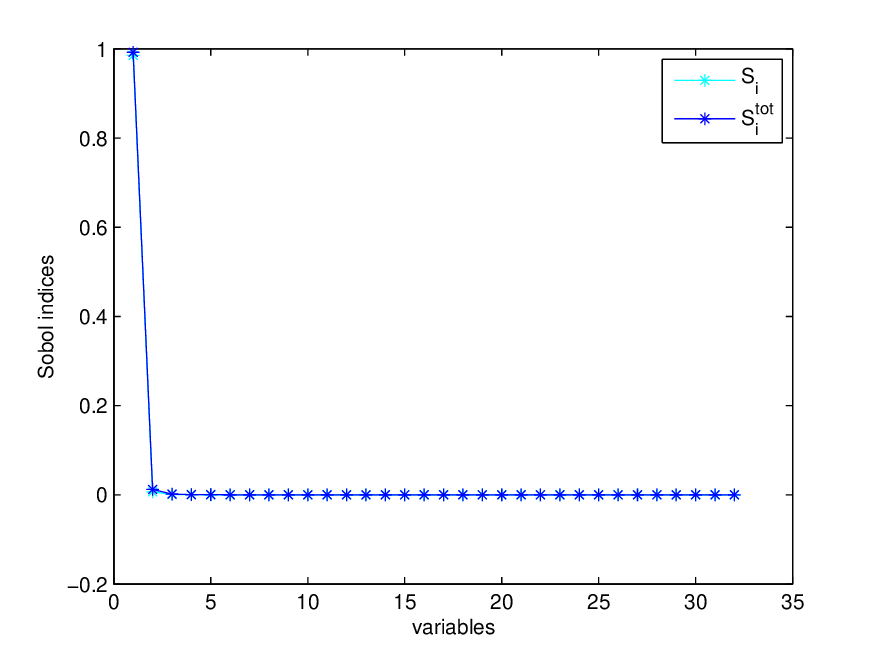}}
\subfigure[Delta]{\includegraphics[width=3.1in,height=2.4in,keepaspectratio=false]{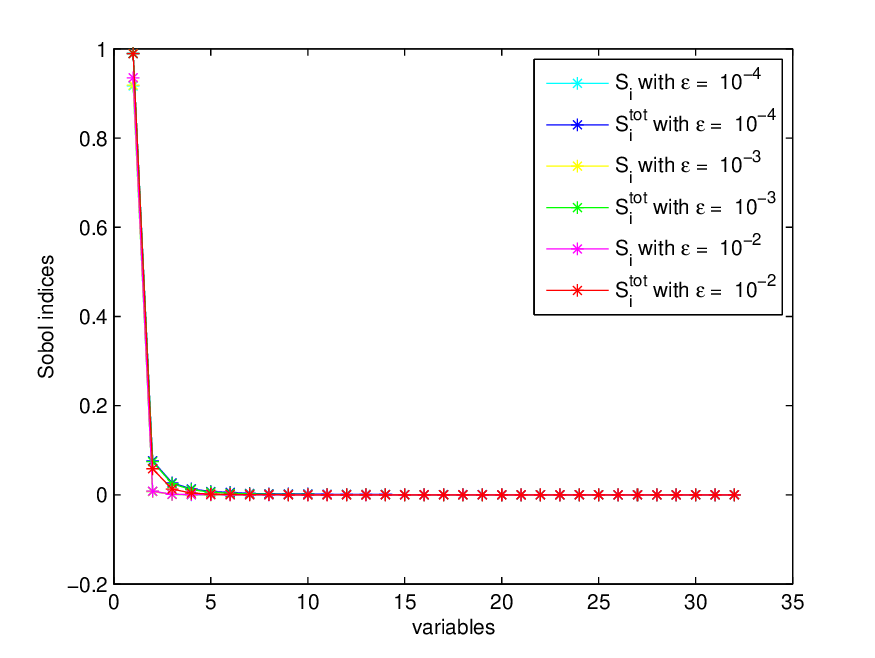}}
\subfigure[Gamma]{\includegraphics[width=3.1in,height=2.4in,keepaspectratio=false]{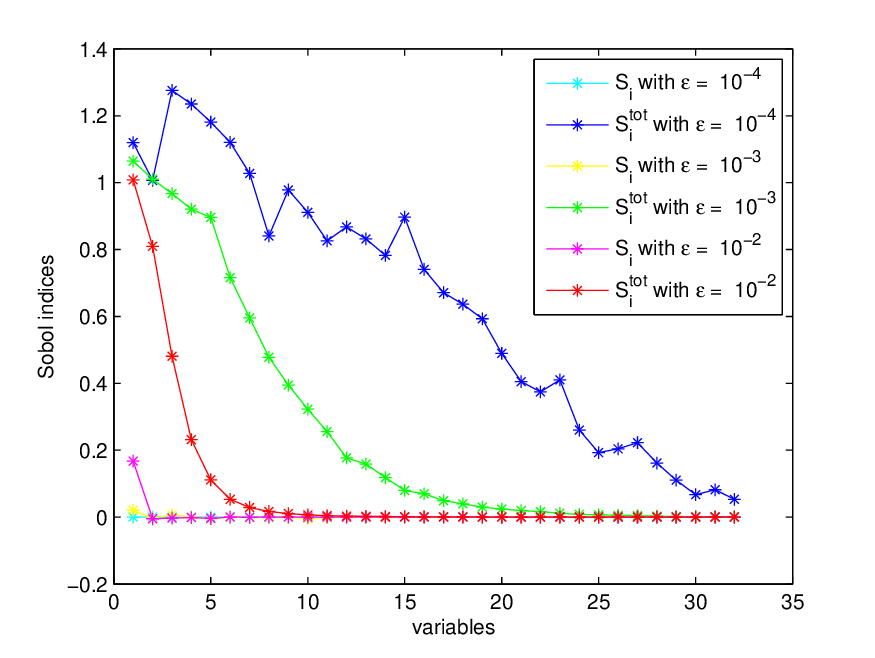}}
\subfigure[Vega] {\includegraphics[width=3.1in,height=2.4in,keepaspectratio=false]{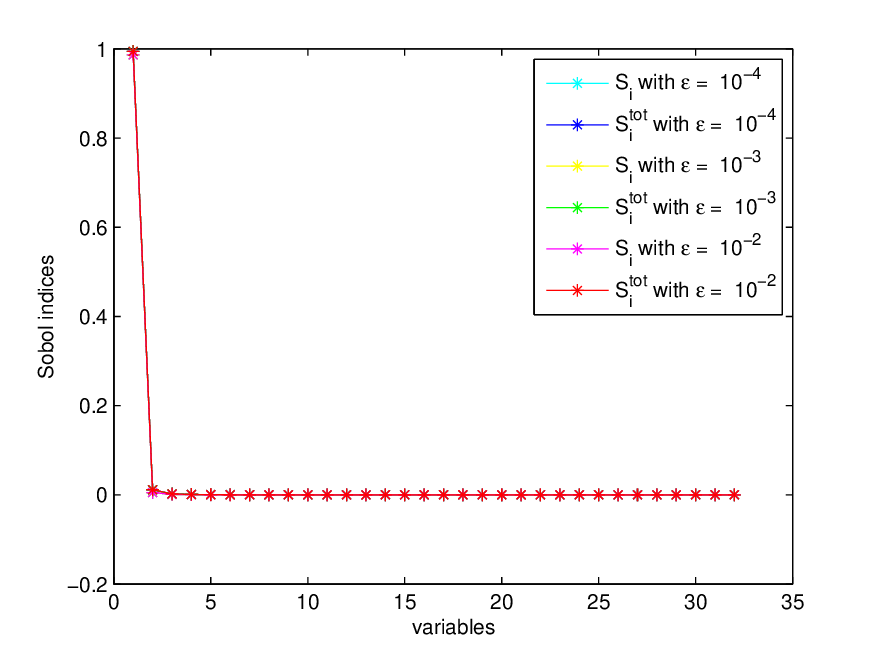}}
\caption{Asian call option price $(a)$ and greeks
$(b),(c),(d)$, PCA, $D=32$. First order Sobol' indices $S_{i}$ and
total sensitivity indices $S_{i}^{tot}$ versus variate $i$.}
\label{fig:6a}
\end{figure}

\begin{figure}[ht]
\centering
\subfigure[Price]{\includegraphics[width=3.1in,height=2.4in,keepaspectratio=false]{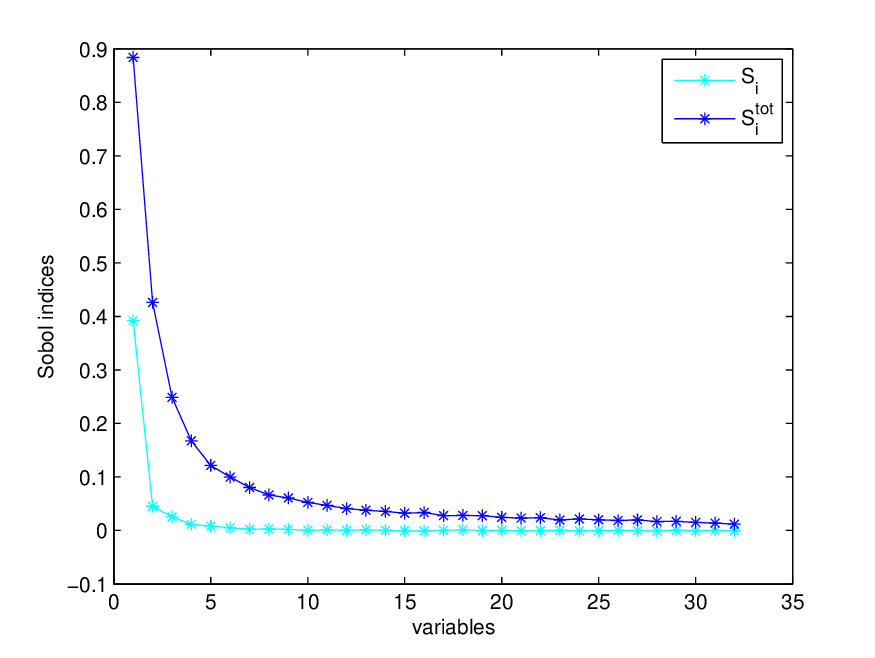}}
\subfigure[Delta]{\includegraphics[width=3.1in,height=2.4in,keepaspectratio=false]{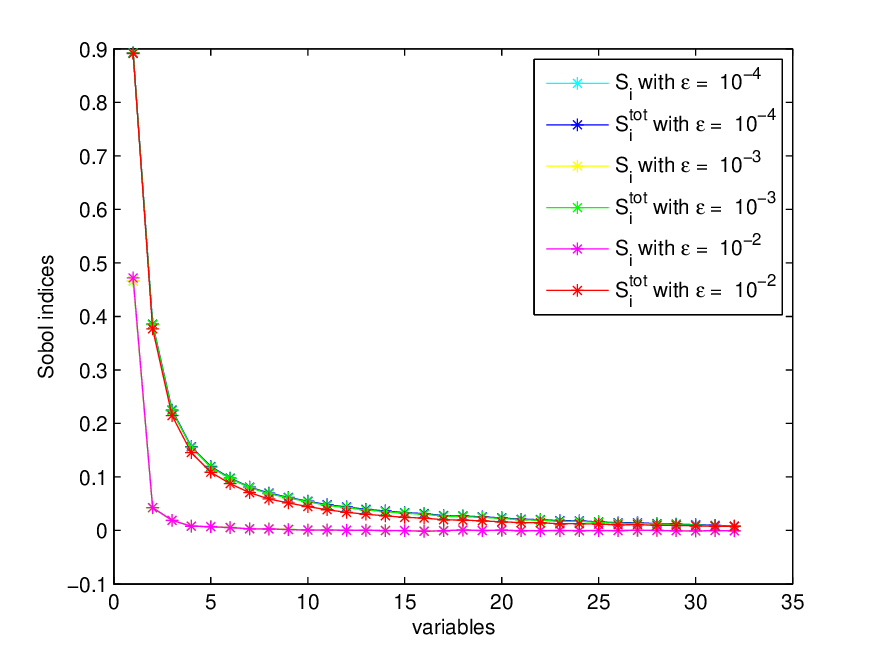}}
\subfigure[Gamma]{\includegraphics[width=3.1in,height=2.4in,keepaspectratio=false]{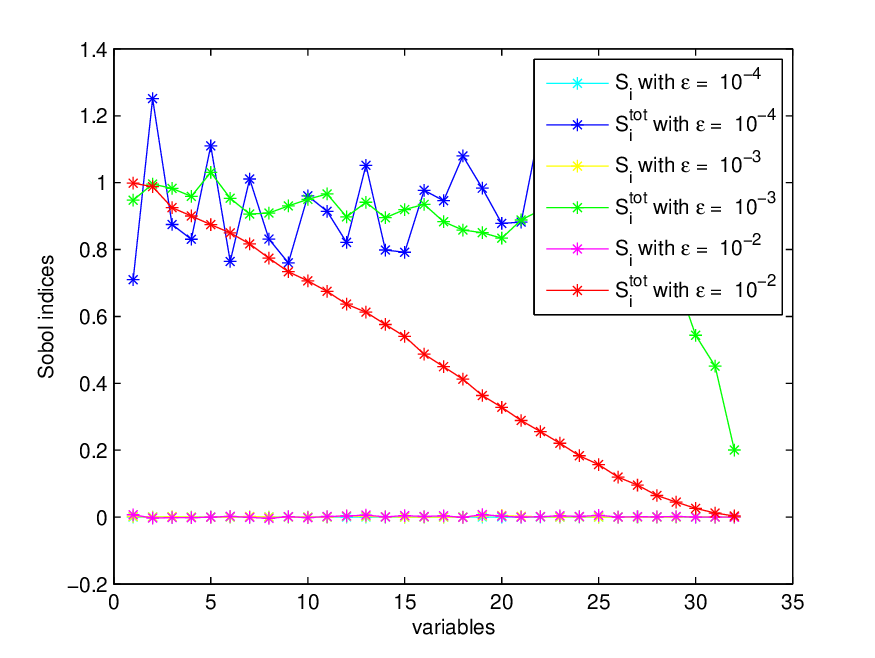}}
\subfigure[Vega] {\includegraphics[width=3.1in,height=2.4in,keepaspectratio=false]{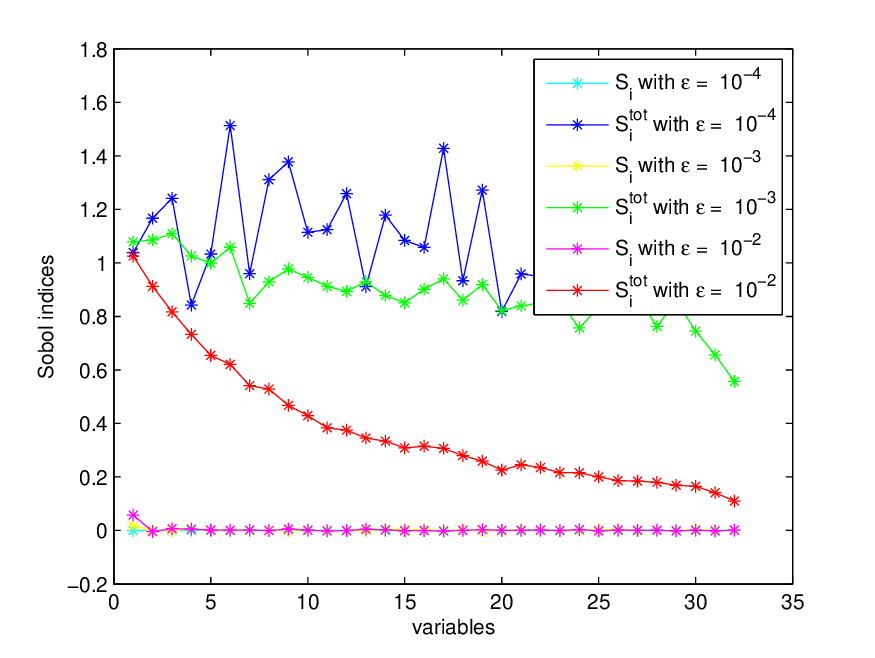}}
\caption{Double Knock-out call option. Details as in Fig.
\ref{fig:6a}.} 
\label{fig:7a}
\end{figure}

\begin{figure}[ht]
\centering
\subfigure[Price]{\includegraphics[width=3.1in,height=2.4in,keepaspectratio=false]{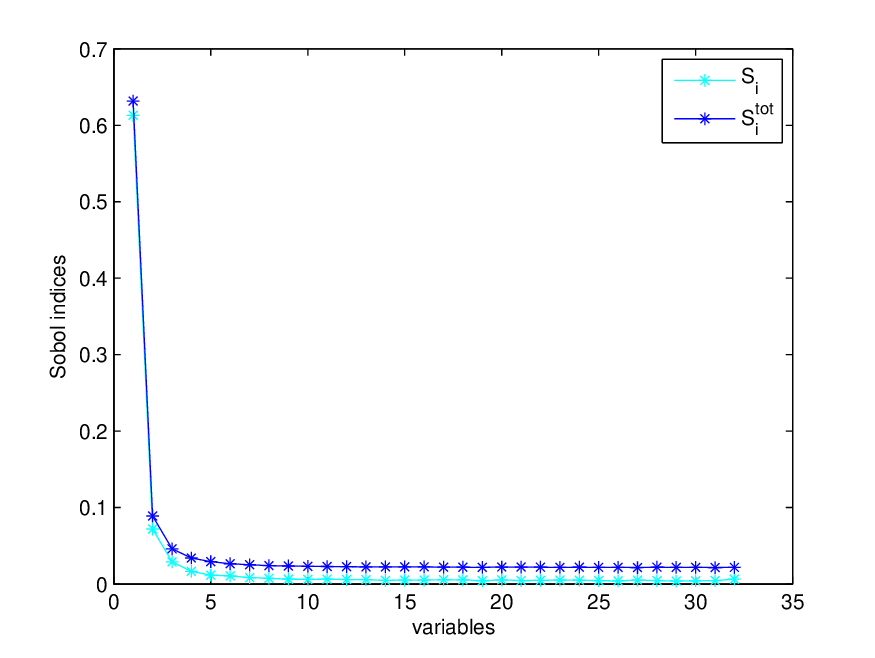}}
\subfigure[Vega]{\includegraphics[width=3.1in,height=2.4in,keepaspectratio=false]{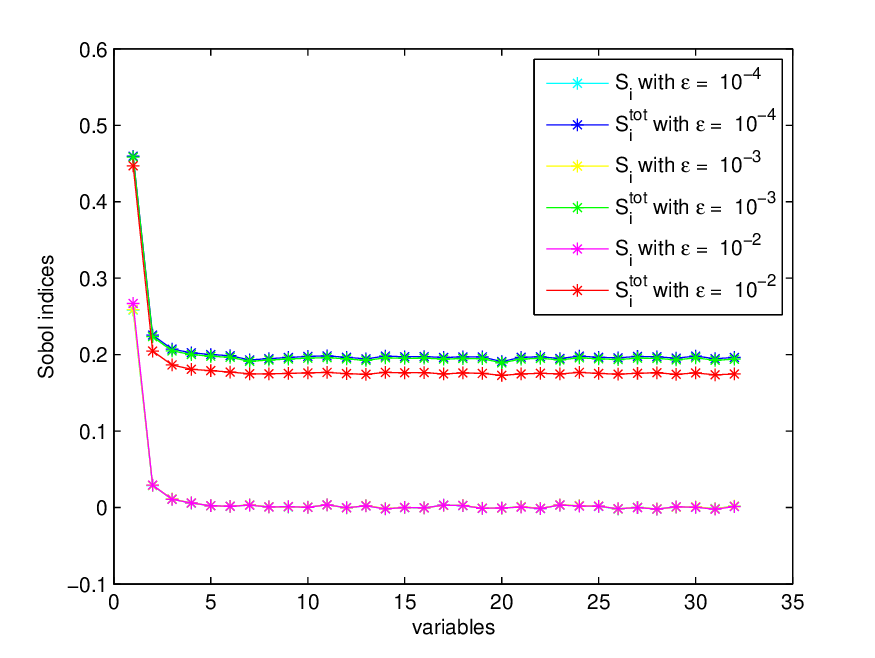}}
\caption{Cliquet option. Details as in Fig. \ref{fig:6a}.}
\label{fig:8a}
\end{figure}

\clearpage
From these results we draw the following conclusions.
\begin{enumerate}
\item Asian option (Figure \ref{fig:6}): price, delta and vega are type A functions with $d_S=1$. The value of sensitivity indexes for the first input, corresponding to the terminal value $t=T$, is $\simeq 1$, while the following variables have sensitivity indexes $\simeq 0$. Gamma remains a type C function as for SD.
\item Double KO option (Figure \ref{fig:7}): price, delta and gamma are type A functions. Comments as for the Asian option above. Vega remains a type C function as for SD.
\item Cliquet option (Figure \ref{fig:8}): price is a type A function. Similarly to the European option, the value of sensitivity indexes for the first input, corresponding to the terminal value $t = T$, is $\simeq 1$, while the following values of $S_i$ are $\simeq 0$. Vega is a type C function, since the ratio $S_i/S_i^{tot}$ reaches small values revealing interacting variables. Thus in this case BBD is much less efficient than SD.
\end{enumerate}
Finally, the results of GSA for PCA are shown in in Table \ref{tab:2a} and Figs. \ref{fig:6a}, \ref{fig:7a}, \ref{fig:8a}.
\begin{table}[h]\small
\centering
\subtable{%
\begin{tabular}{c c c c c c c c}
  \toprule
  \textbf{Payoff}& \textbf{Function} & $\mathbf{S_i/S_i^{tot}}$ & $\mathbf{\sum_iS_i}$ & $\mathbf{d_T}$ & $\mathbf{d_S}$ & $\mathbf{d_A}$ & \textbf{Effect of} $\mathbf{\epsilon}$ \\
  \midrule
  Asian    & Price & 0.99$\to$0.3 & 0.993 & 1 & 1 & 1.01 & - \\
           & Delta & 0.93$\to$0.001 & 0.925 & 1 & 1 & 1.12 & small \\
           & Gamma & $10^{-2}\to 10^{-4}$ & $0.1\to 10^{-3}$ & 32 & 32 & $20\to 3$ & high \\
           & Vega & 0.99$\to$0.05 & 0.992 & 1 & 1 & 1.01 & no \\
  \hline
  Double KO& Price & 0.44$\to 10^{-3}$ & 0.49 & 32 & $<32$ & 2.76 & - \\
           & Delta & 0.52$\to 10^{-3}$ & 0.56 & 32 & $<32$ & 2.6 & small \\
           & Gamma & $10^{-3}\to 10^{-4}$ & $0.04\to 10^{-3}$ & 32 & 32 & 27 & high \\
           & Vega & $0.02\to 10^{-4}$ & $0.08\to 10^{-3}$ & 32 & 32 & $36\to 12$ & high \\
  \hline
  Cliquet  & Price & 0.98$\to$0.2 & 0.888 & $\simeq 2$ & $\leq 2$ & 1.43 & - \\
           & Vega & 0.565$\to$0.001 & 0.33 & 32 & $\leq 32$ & 6.5 & small \\
  \bottomrule
\end{tabular}}
\caption{Summary of GSA metrics and effective dimensions of prices
and greeks for PCA scheme. Details as in Table \ref{tab:1}.}
\label{tab:2a}
\end{table}

\begin{figure}[ht]
\centering
\subfigure[SD+CHOL]{\includegraphics[width=3.1in,height=2.4in,keepaspectratio=false]{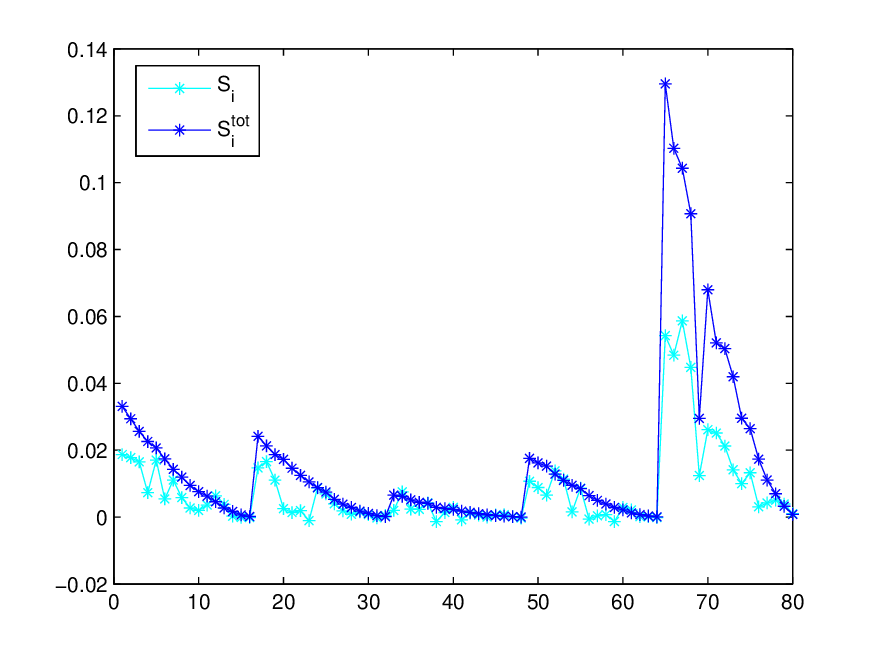}}
\subfigure[BB+PCA] {\includegraphics[width=3.1in,height=2.4in,keepaspectratio=false]{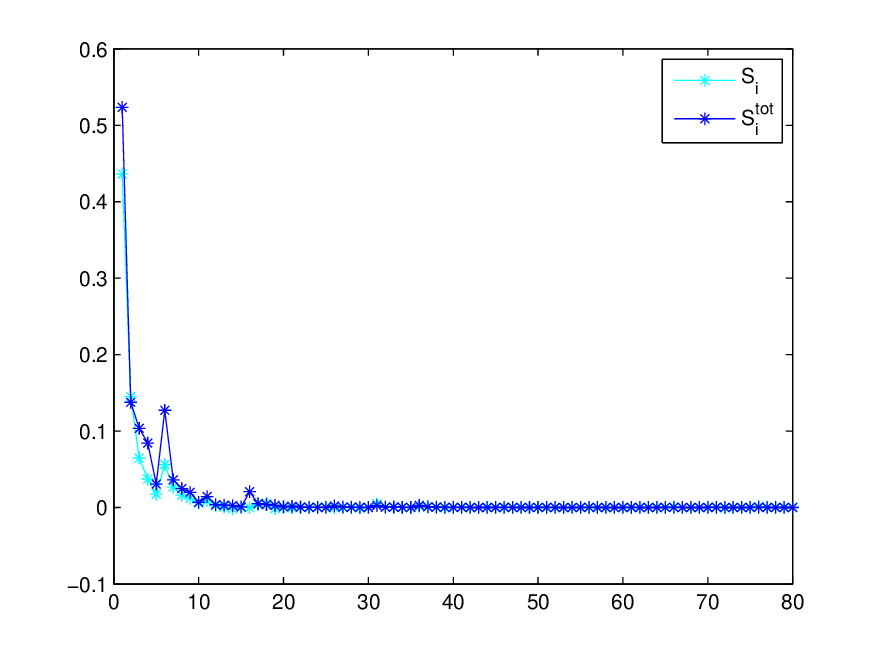}}
\subfigure[PCA+PCA]{\includegraphics[width=3.1in,height=2.4in,keepaspectratio=false]{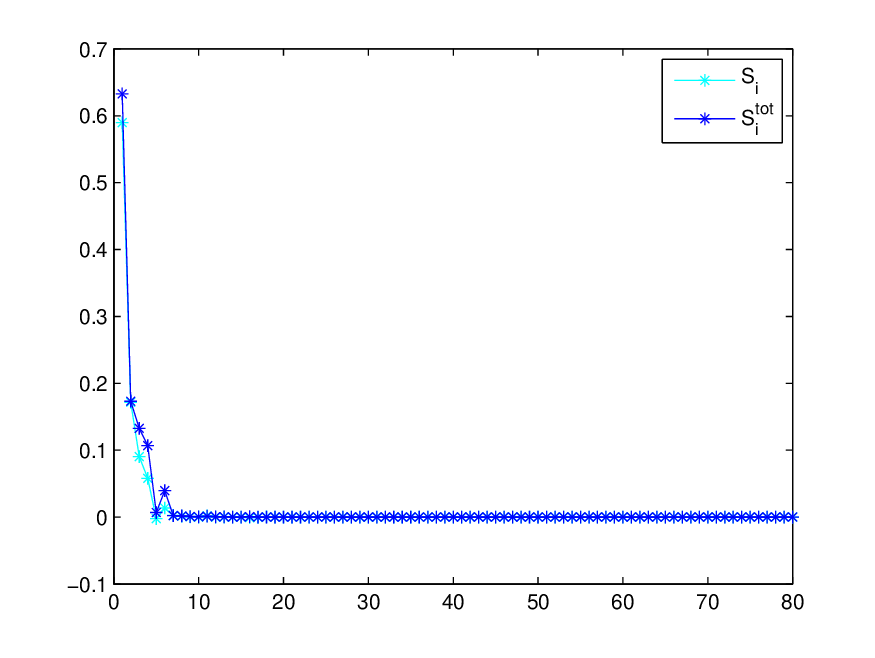}}
\caption{Asian basket call option price with different sampling strategies along time and risk fator directions: standard discretization and Cholesky factorization $(a)$, Brownian bridge and PCA
$(b)$, PCA on both directions $(c)$. Nominal dimension is $D=80$. First order Sobol' indices $S_{i}$ and
total sensitivity indices $S_{i}^{tot}$ versus variate $i$.}
\label{Fig:GSAmult}
\end{figure}

From these results we draw the following conclusions.
\begin{enumerate}
\item Asian option (Figure \ref{fig:6a}): price, delta and vega are type A functions. Comments as for the BBD case (actually, PCA results to be more efficient than BBD in this case). Gamma remains a type C function as for SD.
\item Double KO option (Figure \ref{fig:7a}): price and delta are type B functions. Gamma and Vega are type C functions.
\item Cliquet option (Figure \ref{fig:8a}): price is a type A function. The value of sensitivity indexes for the first input, corresponding to the terminal value $t = T$, is $\simeq 1$, while the following values of $S_i$ are $\simeq 0$. Vega is a type C function, since the ratio $S_i/S_i^{tot}$ reaches small values revealing interacting variables. Thus in this case, PCA is much less efficient than SD (as for BBD).
\end{enumerate}
Let's now move to the multi-asset payoffs (\ref{eqPayoffBasketCall}) and (\ref{eqPayoffBasketAsian}).
We have performed the same analysis as for the single-asset case, computing Sobol' indices for all possible combinations of the sampling strategies described in Section \ref{SecGM} and with zero correlation\footnote{This is due to the fact that the GSA approach described in this work assumes that the input variables are independent. However, based on the numerical results of Section \ref{SecPerformance}, we argue that the conclusions should be similar in the case of non zero correlation. In \cite{KucTar12} a possible extension of GSA to models with dependent variables was discussed, but the interpretation of Sobol' indexes is less transparent in that approach.}: we observed that, when BBD or PCA time discretizations and PCA factorization of the covariance matrix are used and input variables are appropriately re-arranged, a significant reduction in effective dimensions is obtained. In Fig. \ref{Fig:GSAmult} and Table \ref{Tab:GSAmult} we compare the results for the price of an Asian basket option with SD+CHOL, BB+PCA and PCA+PCA sampling strategies. In order to highlight the sensitivity of Sobol' indexes to the sampling order of the variables, we also re-arranged the inputs in the following way:
\begin{itemize}
  \item SD+CHOL: the underlying assets are not sorted, so that spots and volatilities are read in the original (and, thus, not optimal) order. The input variables $x_i$ are used to construct the Brownian motion in the following order (by risk factors):\\ $\left(W_1(t_1),\ldots,W_1(t_{Nts}),\ldots,W_{Nrf}(t_1),\ldots,W_{Nrf}(t_{Nts})\right)$.
  \item BB+PCA: the input variables $x_i$ are used to construct the Brownian motion in the following order (by time steps): $\left(W_{Nrf}(t_{Nts}),\ldots,W_1(t_{Nts}),\ldots\right)$.
  \item PCA+PCA: no particular ordering is applied, since the algorithm does the job by itself.
\end{itemize}

\begin{table}[h]\small
\centering
\subtable{%
\begin{tabular}{c c c c c c c c}
  \toprule
  \textbf{Payoff}& \textbf{Function} & \textbf{Method} & $\mathbf{S_i/S_i^{tot}}$ & $\mathbf{\sum_iS_i}$ & $\mathbf{d_T}$ & $\mathbf{d_S}$ & $\mathbf{d_A}$\\
  \midrule
  Asian basket  & Price & SD+CHOL & 0.7$\to$0.1 & 0.62 & $\simeq 80$ & $< 80$ & 1.3\\
                & Price & BB+PCA & 0.83$\to$0.1 & 0.85 & 2 & $\leq 2$ & 1.2\\
                & Price & PCA+PCA & 0.99$\to$0.34 & 0.93 & 1 & 1 & 1.1\\
  \bottomrule
\end{tabular}}
\caption{Summary of GSA metrics and effective dimensions for the price of the Asian basket option with SD+CHOL, BB+PCA and PCA+PCA schemes. In the first case the variables are ordered by risk factors, while in the second case by time steps. Nominal dimension is $D=80$. The
numerical computations were obtained evaluating the $D+2$ integrals (\ref{SI}) on $2^{13}$ QMC scenarios. We show significant digits only.}
\label{Tab:GSAmult}
\end{table}

From the results it is clear that, with BB+PCA or PCA+PCA, the price of the Asian basket option is a Type A function, while with SD+CHOL it is a Type B function. In the first case, the most important variables are in the first places, while in the latter they are placed in the end of the sampling vector.

In conclusion, prices and FD greeks are predominantly Type
A or Type B functions when BBD or PCA are employed, so that the
effective dimension in the truncation sense $d_T$ is reduced.
\par
The different efficiency of QMC with BBD or PCA is completely
explained by the properties of Sobol' low discrepancy sequences.
The initial coordinates of Sobol' LDS are much better distributed
than the later high dimensional coordinates
\cite{Gla03,CafMor1997}. BBD and PCA change the order in which
inputs (linked with time steps) are sampled. As follows from GSA,
in most cases for BBD and PCA the low index variables are much more important than
higher index variables. The BBD uses lower index, well distributed
coordinates from each $D$-dimensional LDS vector to determine most
of the structure of a path, and reserves other coordinates to fill
in finer details. That is, well distributed coordinates are used
for important variables and other not so well distributed
coordinates are used for far less important variables. Similar considerations hold for the PCA construction. This
results in a significantly improved accuracy of QMC integration.
However, this technique does not always improve the efficiency of
the QMC method as \eg for Cliquet options: in this case GSA
reveals that for SD all inputs are equally important and,
moreover, there are no interactions among them, which is an ideal
case for application of Sobol' low discrepancy sequences; BBD and PCA,
on the other hand, favoring higher index variables destroys
independence of inputs introducing interactions, which leads to
higher values of $d_S$ and $d_A$: as a result, we observe
degradation in performance of the corresponding QMC methods.

\subsection{Convergence analysis}
\label{SecPerformance} \noindent
In this Section we compare the relative performances of MC and QMC techniques. This analysis is crucial to establish if QMC outperforms MC, and in what sense.
\par
\begin{figure}[ht]
\centering
\subfigure[Price]{\includegraphics[width=3.1in,height=2.4in,keepaspectratio=false]{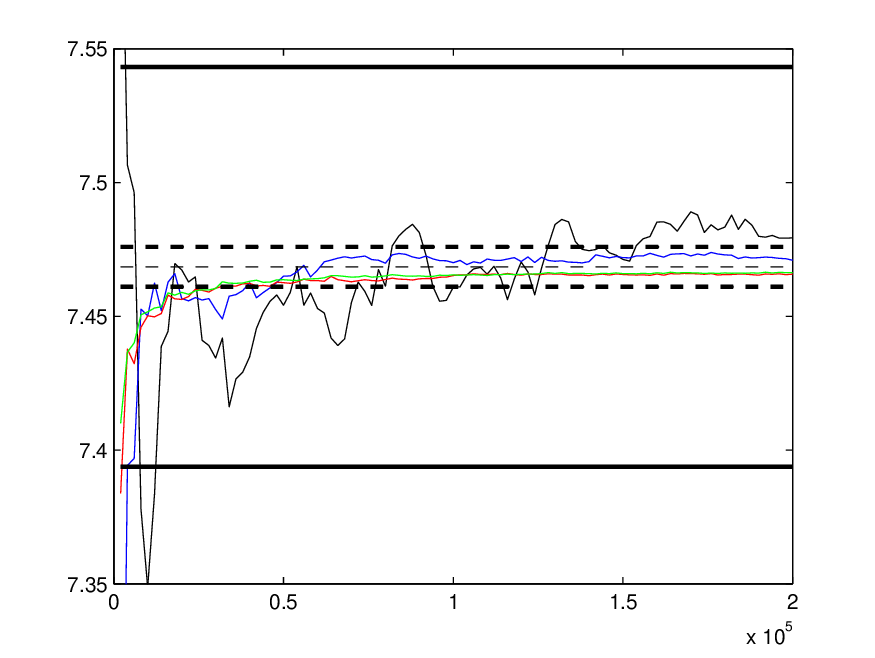}}
\subfigure[Delta]{\includegraphics[width=3.1in,height=2.4in,keepaspectratio=false]{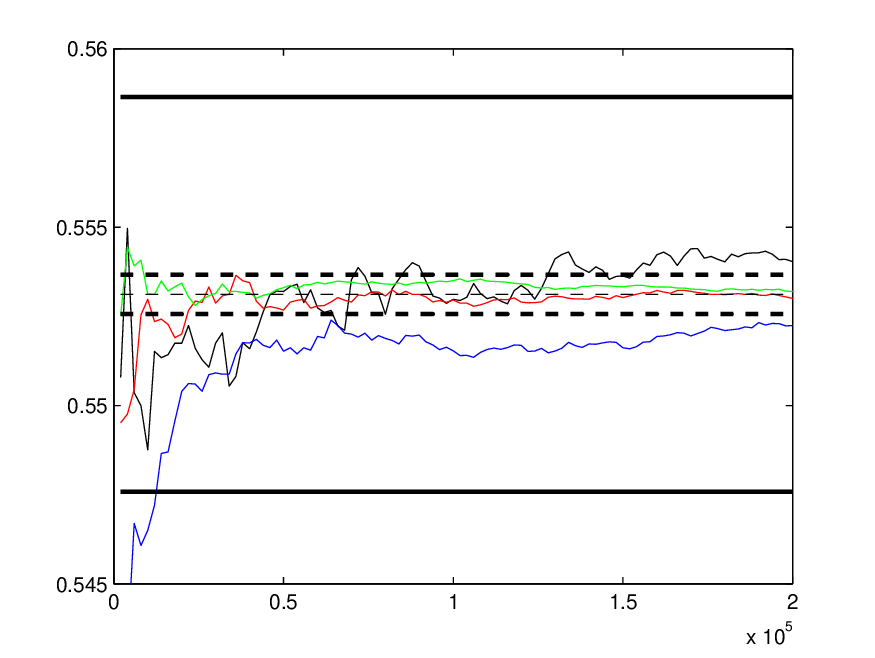}}
\subfigure[Gamma]{\includegraphics[width=3.1in,height=2.4in,keepaspectratio=false]{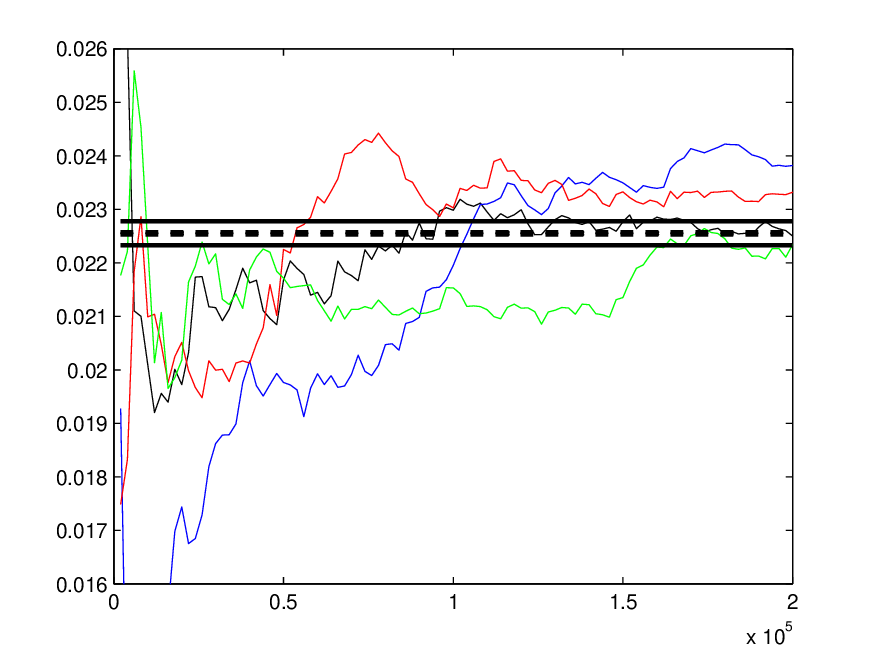}}
\subfigure[Vega] {\includegraphics[width=3.1in,height=2.4in,keepaspectratio=false]{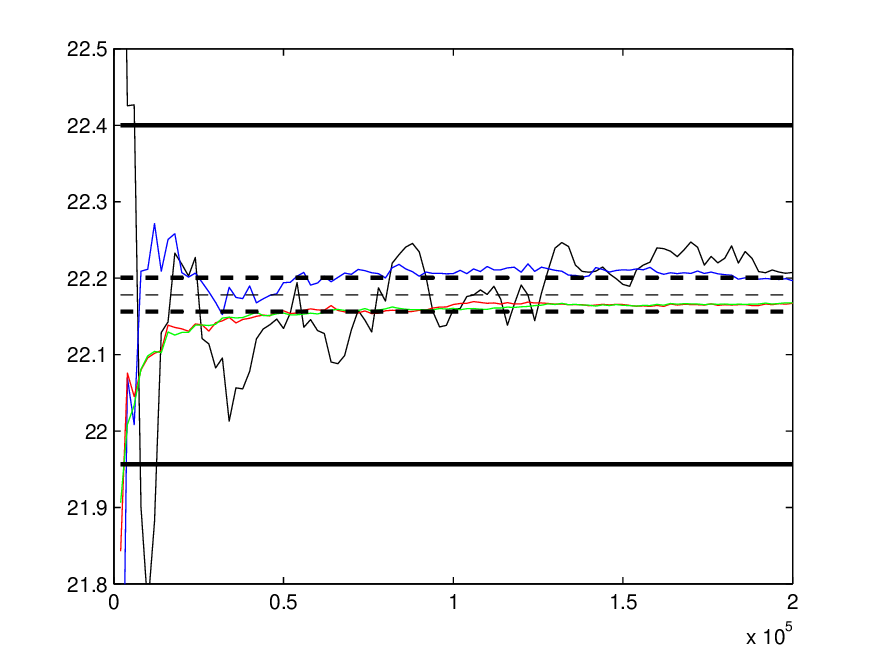}}
\caption{ Asian call option price $(a)$ and greeks $(b),(c),(d)$
convergence diagrams versus number of simulated paths for MC+SD (solid black line), QMC+SD (solid blue
line), QMC+BBD (solid red line) and QMC+PCA (solid green line). 1\% and 0.1\%
accuracy regions are marked by horizontal black solid and dashed
lines, respectively. Number of dimensions is $D=32$. Shift
parameter is $\epsilon = 10^{-3}$.} \label{fig:10}
\end{figure}
\begin{figure}[ht]
\centering
\subfigure[Price]{\includegraphics[width=3.1in,height=2.4in,keepaspectratio=false]{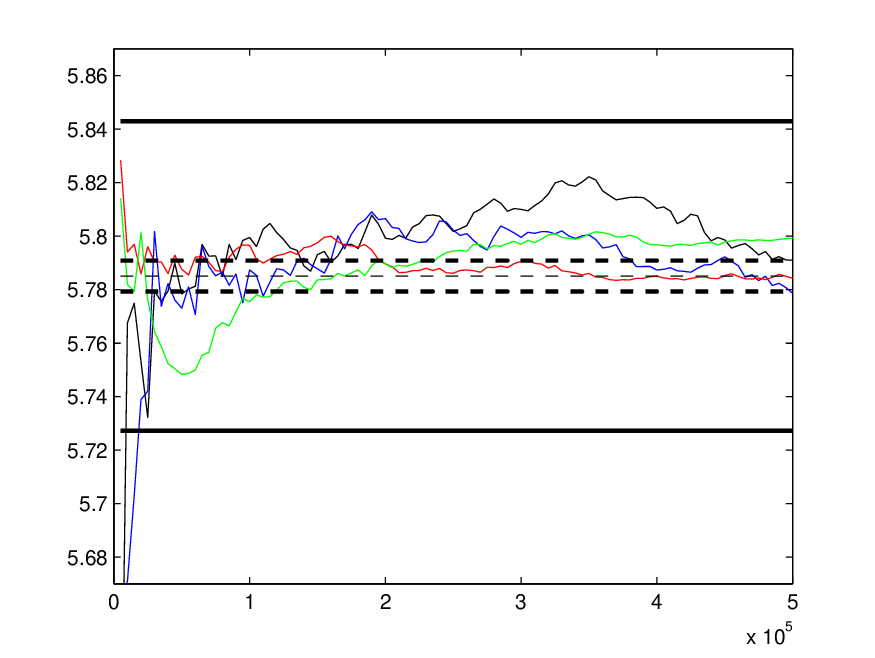}}
\subfigure[Delta]{\includegraphics[width=3.1in,height=2.4in,keepaspectratio=false]{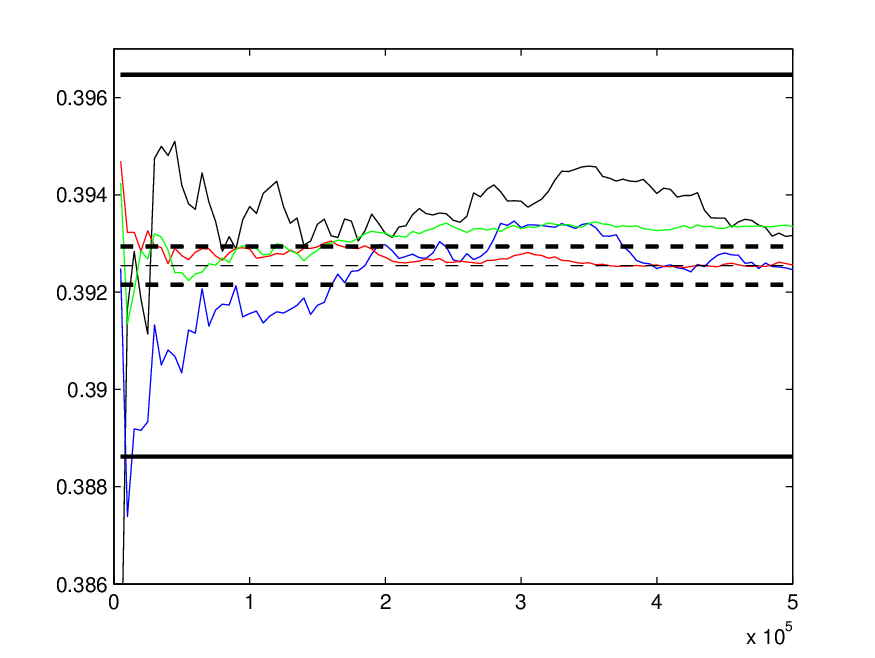}}
\subfigure[Gamma]{\includegraphics[width=3.1in,height=2.4in,keepaspectratio=false]{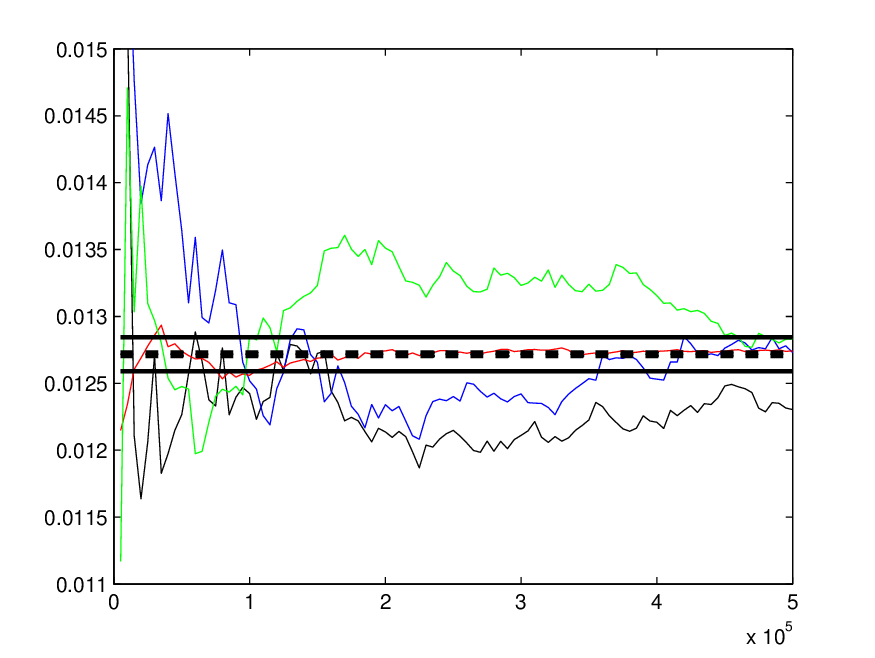}}
\subfigure[Vega] {\includegraphics[width=3.1in,height=2.4in,keepaspectratio=false]{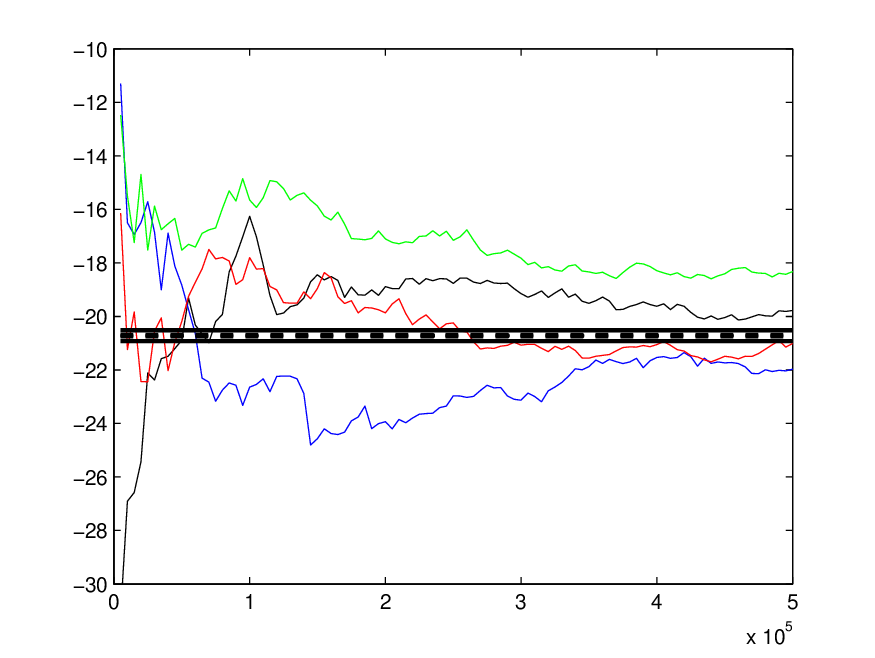}}
\caption{ Double Knock-out call option. Details as in Fig.
\ref{fig:10}.} \label{fig:11}
\end{figure}
\begin{figure}[ht]
\centering
\subfigure[Price]{\includegraphics[width=3.1in,height=2.4in,keepaspectratio=false]{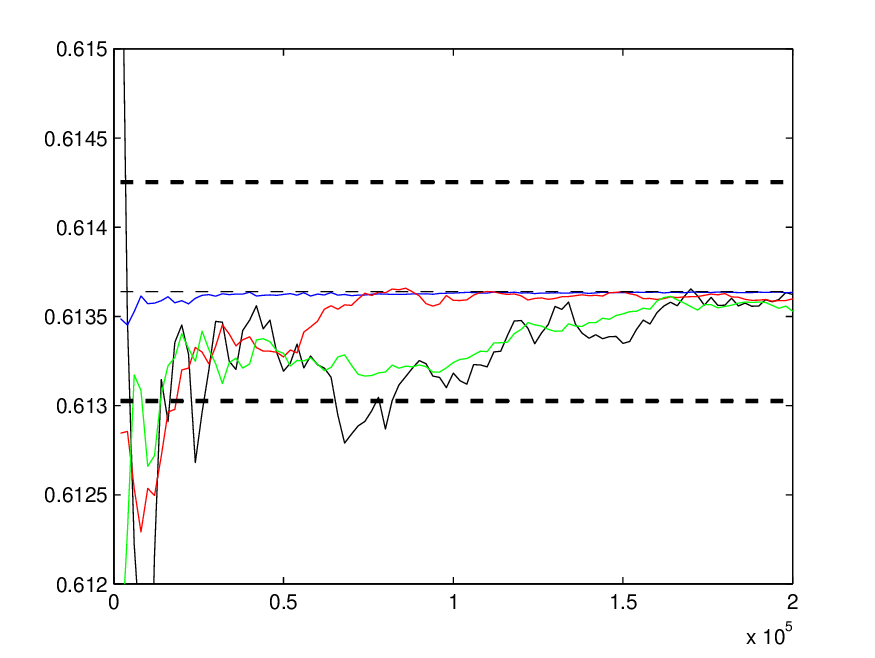}}
\subfigure[Vega] {\includegraphics[width=3.1in,height=2.4in,keepaspectratio=false]{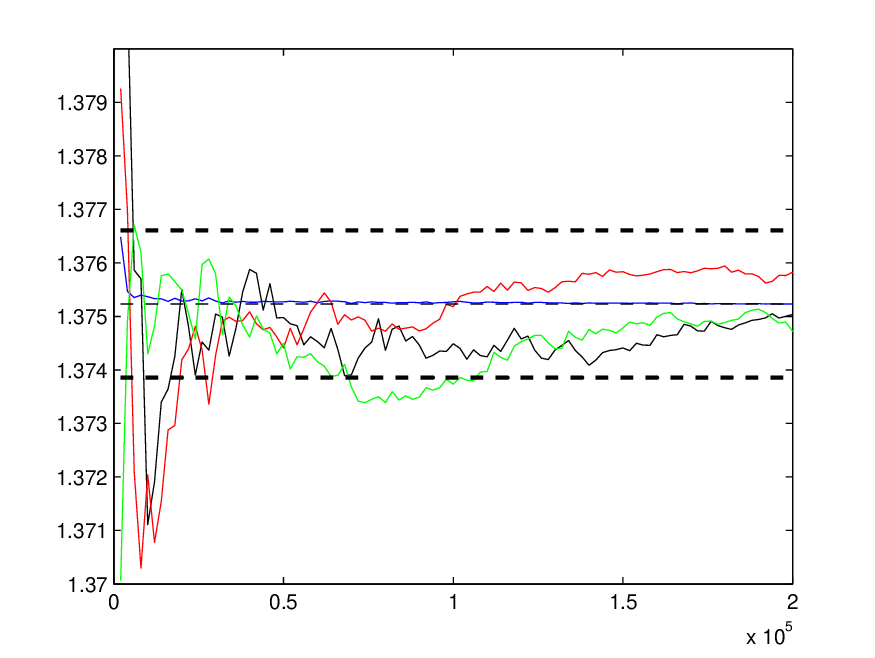}}
\caption{ Cliquet option. Details as in Fig. \ref{fig:10}.}
\label{fig:12}
\end{figure}
\begin{figure}[ht]
\centering
\subfigure[Price]{\includegraphics[width=3.1in,height=2.4in,keepaspectratio=false]{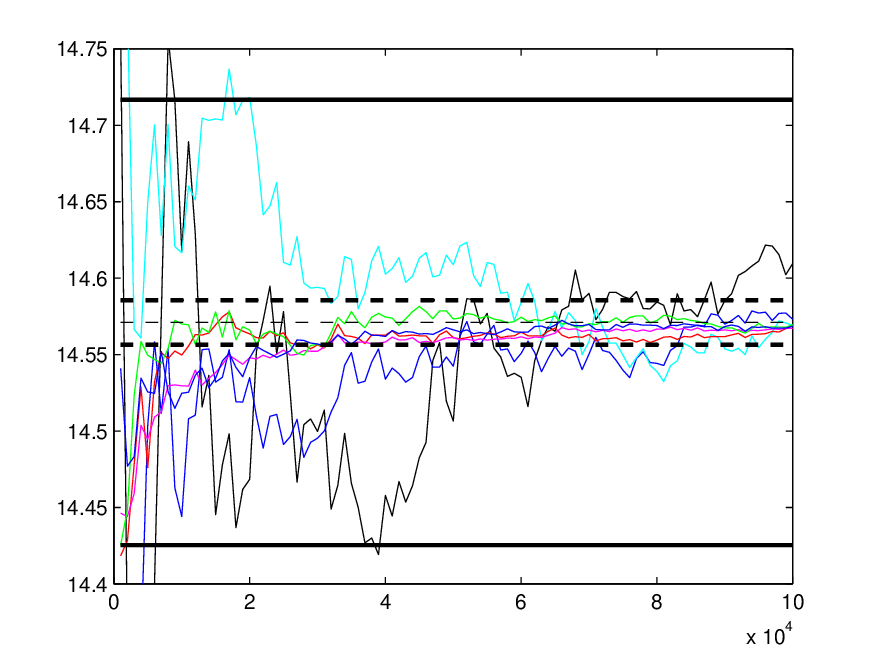}}
\subfigure[Delta \wrt 5th asset]{\includegraphics[width=3.1in,height=2.4in,keepaspectratio=false]{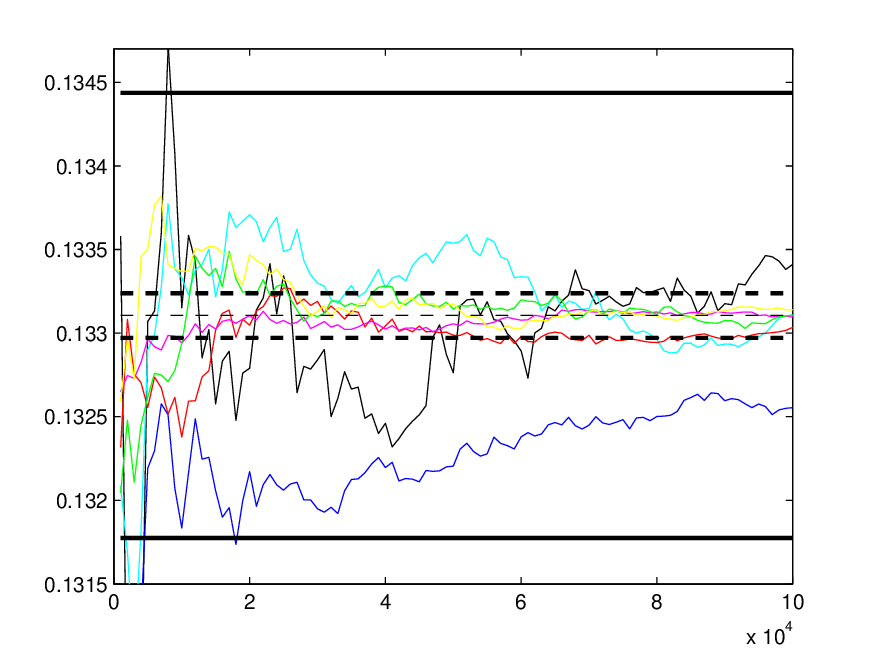}}
\subfigure[Vega \wrt 5th asset]{\includegraphics[width=3.1in,height=2.4in,keepaspectratio=false]{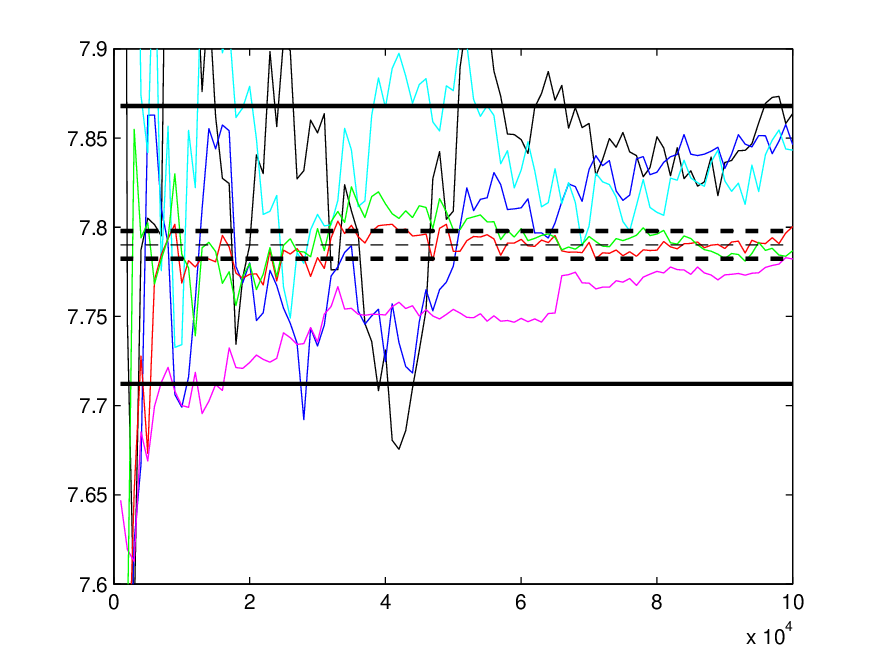}}
\caption{European basket call option price $(a)$ and selected greeks $(b),(c)$
convergence diagrams versus number of simulated paths for different combinations of sampling techniques: MC+SD+CHOL (black), QMC+SD+CHOL (blue), QMC+SD+PCA (cyan), QMC+BBD+CHOL (red), QMC+BBD+PCA (magenta), QMC+PCA+CHOL (green), QMC+PCA+PCA (yellow). 1\% and 0.1\%
accuracy regions are marked by horizontal black solid and dashed
lines, respectively. Number of dimensions is $D=80$, correlation is $\rho=0.6$ and shift
parameter is $\epsilon = 10^{-3}$.} \label{Fig:EuConv}
\end{figure}
\begin{figure}[ht]
\centering
\subfigure[Price]{\includegraphics[width=3.1in,height=2.4in,keepaspectratio=false]{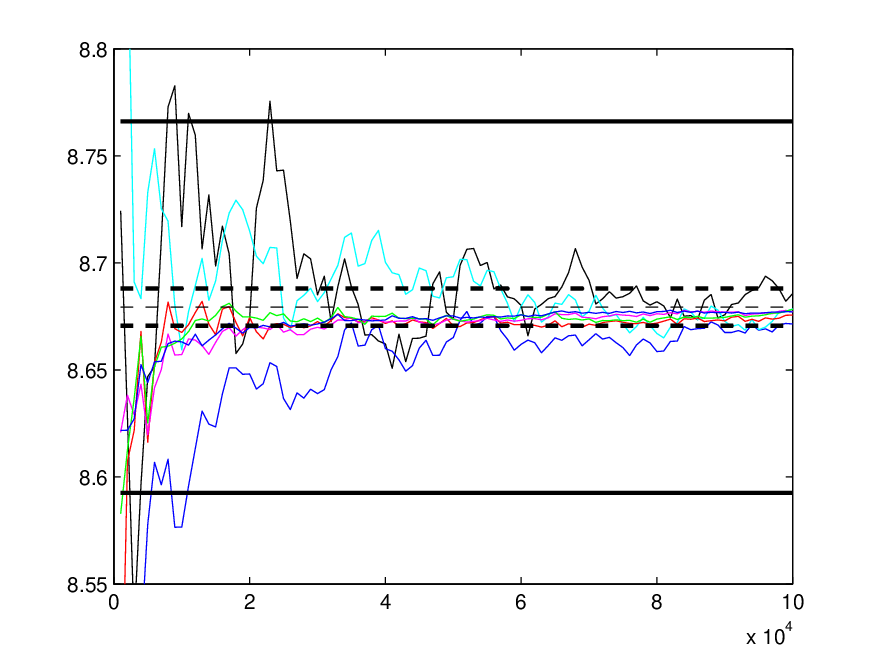}}
\subfigure[Delta \wrt 5th asset]{\includegraphics[width=3.1in,height=2.4in,keepaspectratio=false]{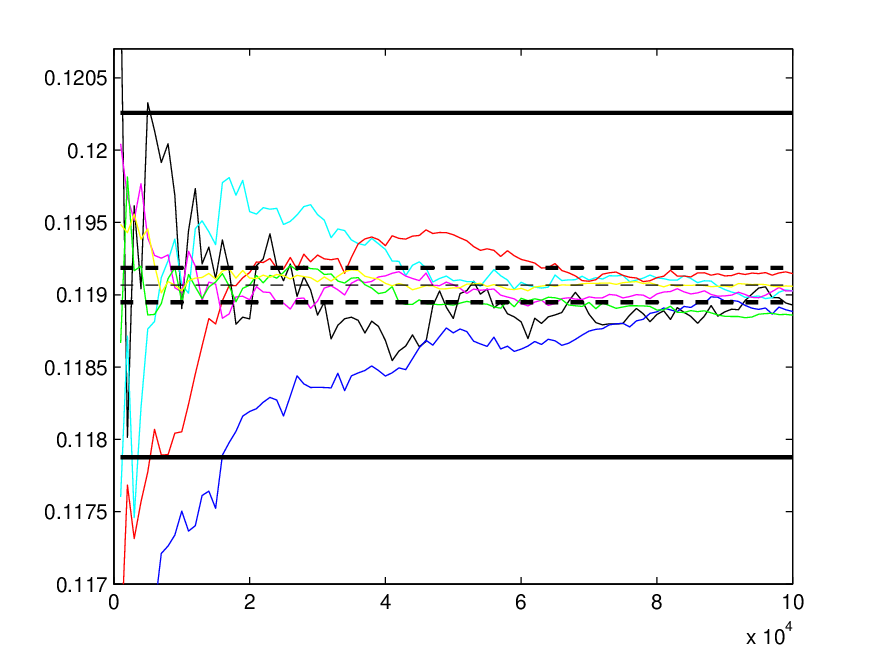}}
\subfigure[Vega \wrt 5th asset] {\includegraphics[width=3.1in,height=2.4in,keepaspectratio=false]{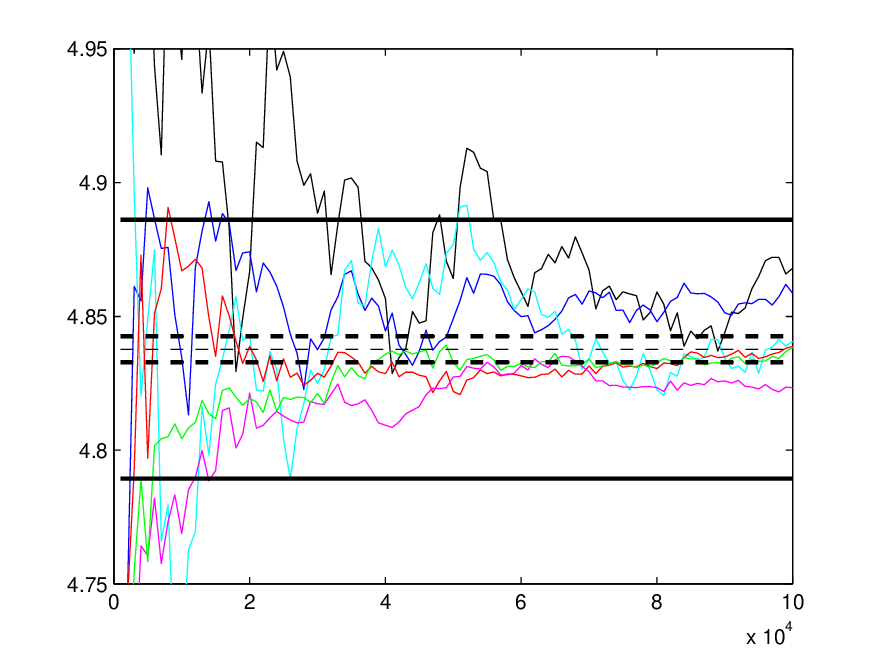}}
\caption{Asian basket call option. Details as in Fig. \ref{Fig:EuConv}.} \label{Fig:AsConv}
\end{figure}

Firstly, following the suggestion of \cite{Jac01}, Section 14.4,
we analyze convergence diagrams for prices and greeks, showing the
dependence of the MC simulation error upon the number of MC paths.
The results for the five payoffs are shown in Figs.
\ref{fig:10}-\ref{Fig:AsConv}. In the case of multi-asset options, we plot just delta and vega w.r.t. the fifth asset, for the case of correlation $\rho=0.6$. Similar results hold for other cases.

\begin{figure}[ht]
\centering
\subfigure[Price]{\includegraphics[width=3.1in,height=2.4in,keepaspectratio=false]{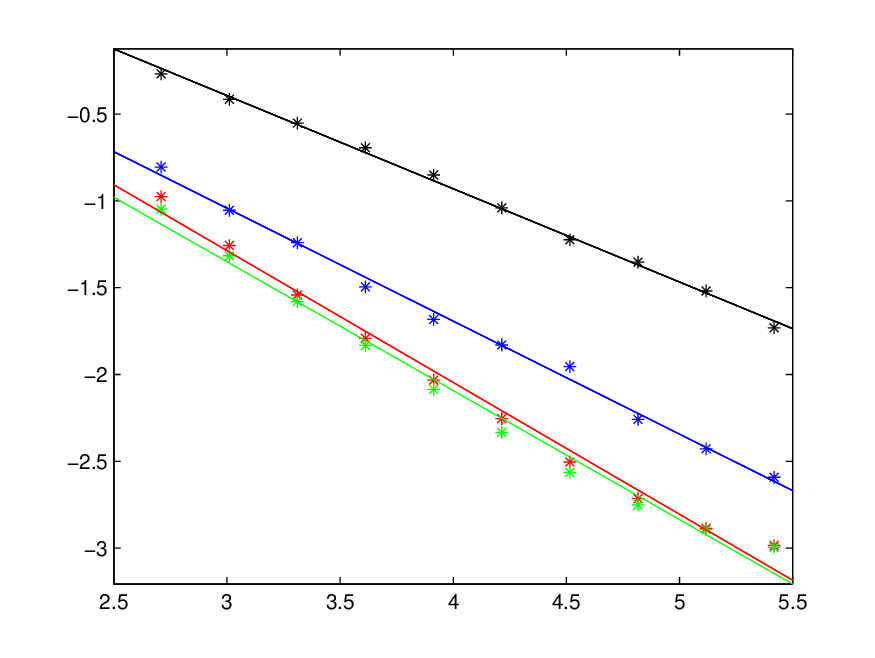}}
\subfigure[Delta]{\includegraphics[width=3.1in,height=2.4in,keepaspectratio=false]{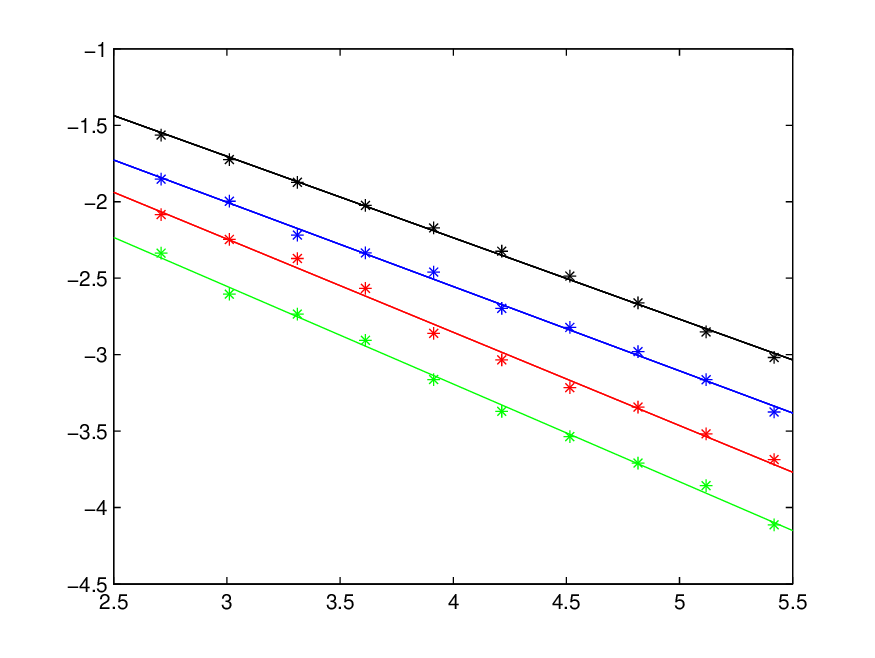}}
\subfigure[Gamma]{\includegraphics[width=3.1in,height=2.4in,keepaspectratio=false]{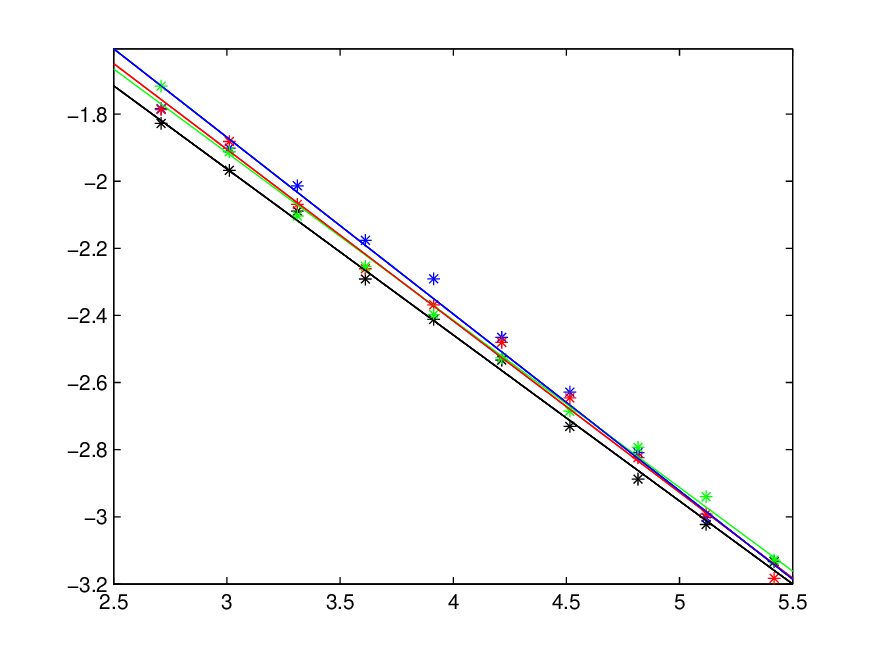}}
\subfigure[Vega] {\includegraphics[width=3.1in,height=2.4in,keepaspectratio=false]{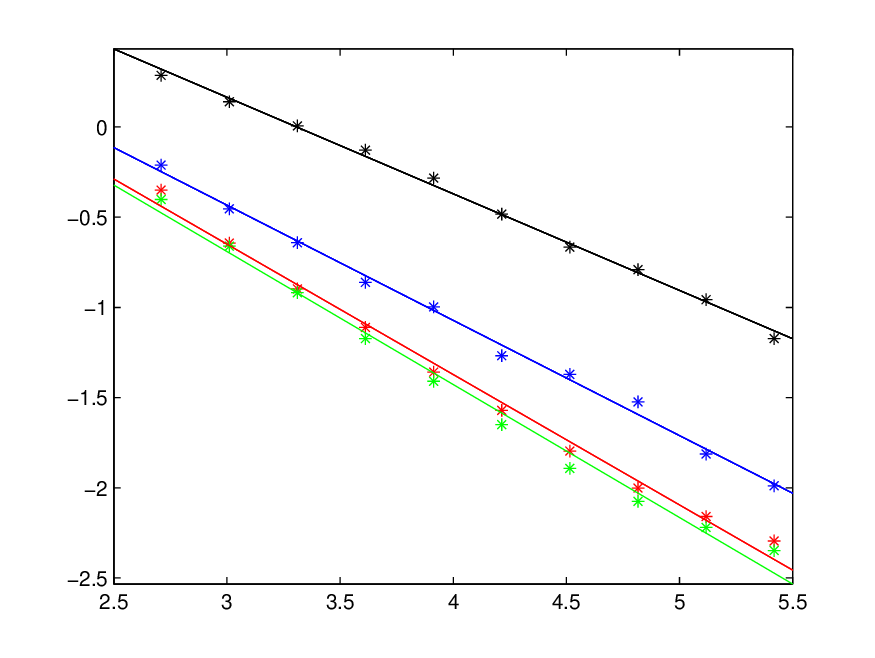}}
\caption{Asian call option price $(a)$ and greeks
$(b),(c),(d)$, Log-Log plots of error $\varepsilon_N$ versus
number of simulated paths $N=2^p,\;p=9,\ldots,18$, $D=32$,
$\epsilon=10^{-3}$, $L=100$ runs: MC+SD
(black), QMC+SD (blue), QMC+BBD (red), QMC+PCA (green). Linear regression lines
are also shown.} \label{fig:14}
\end{figure}

\begin{figure}[ht]
\centering
\subfigure[Price]{\includegraphics[width=3.1in,height=2.4in,keepaspectratio=false]{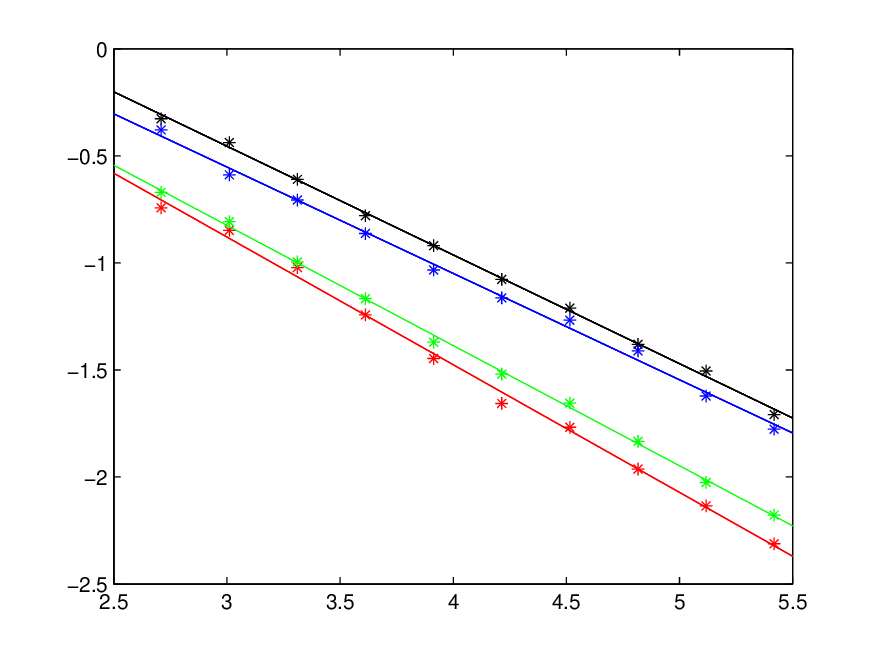}}
\subfigure[Delta]{\includegraphics[width=3.1in,height=2.4in,keepaspectratio=false]{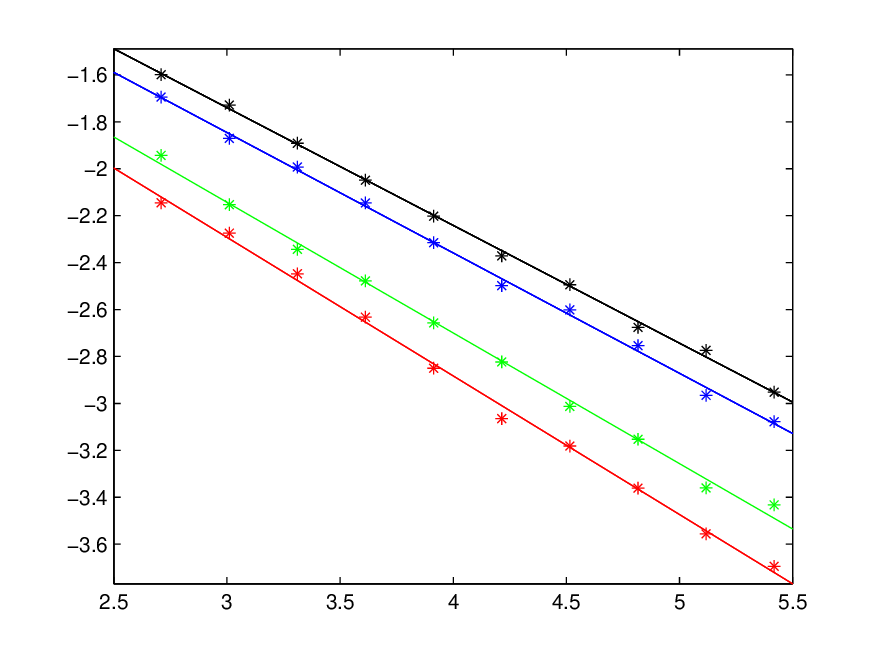}}
\subfigure[Gamma]{\includegraphics[width=3.1in,height=2.4in,keepaspectratio=false]{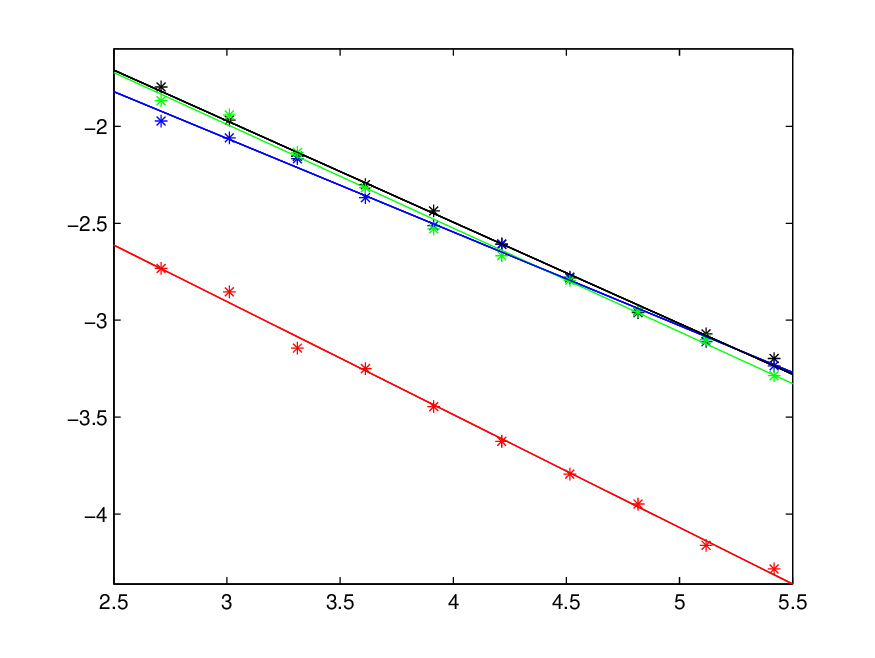}}
\subfigure[Vega] {\includegraphics[width=3.1in,height=2.4in,keepaspectratio=false]{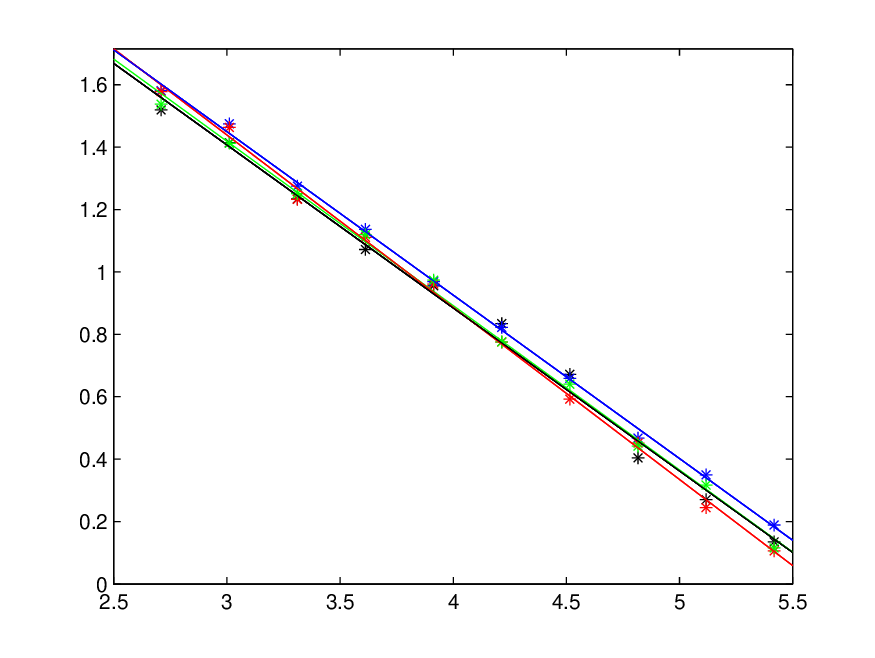}}
\caption{Double Knock-out call option. Details as in Fig.
\ref{fig:14}.} \label{fig:15}
\end{figure}

\begin{figure}[ht]
\centering
\subfigure[Price]{\includegraphics[width=3.1in,height=2.4in,keepaspectratio=false]{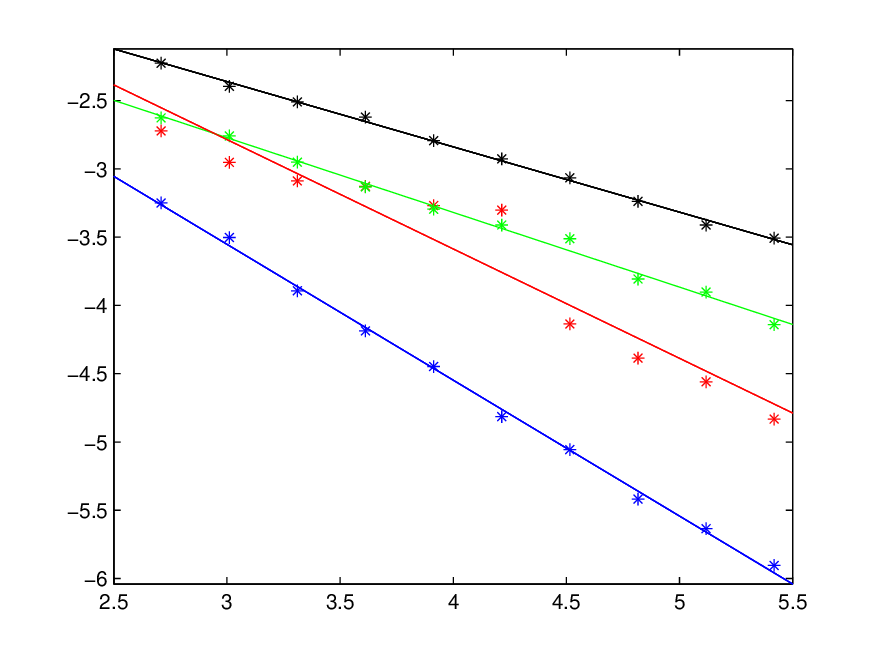}}
\subfigure[Vega] {\includegraphics[width=3.1in,height=2.4in,keepaspectratio=false]{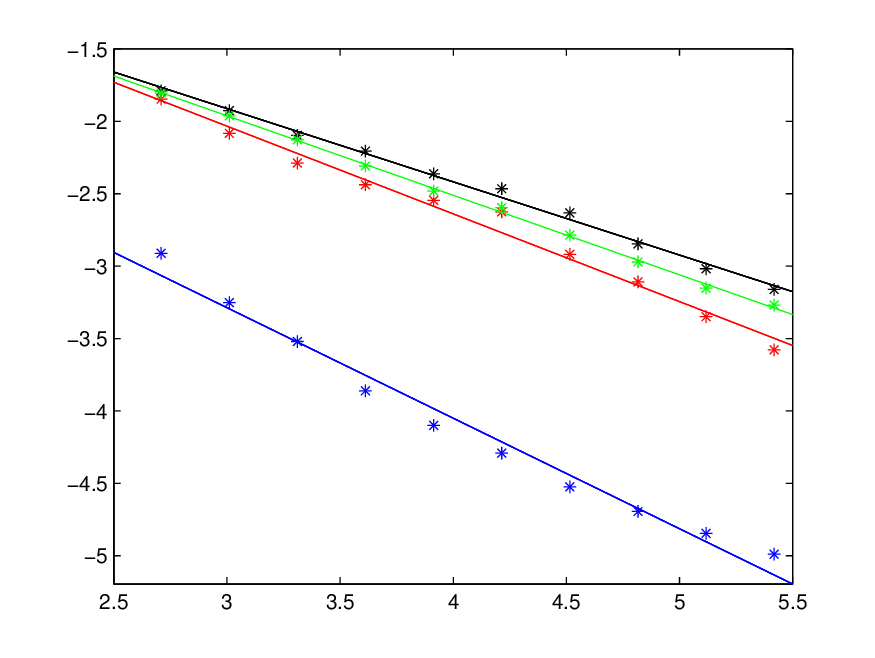}}
\caption{Cliquet option. Details as in Fig. \ref{fig:14}.}
\label{fig:16}
\end{figure}

\begin{figure}[ht]
\centering
\subfigure[Price]{\includegraphics[width=3.1in,height=2.4in,keepaspectratio=false]{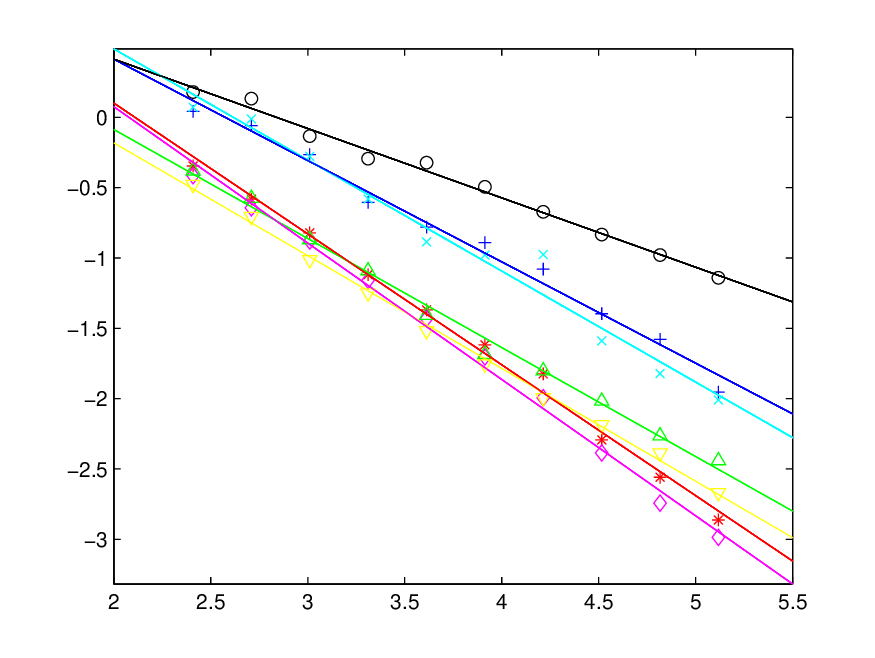}}
\subfigure[Delta asset 5]{\includegraphics[width=3.1in,height=2.4in,keepaspectratio=false]{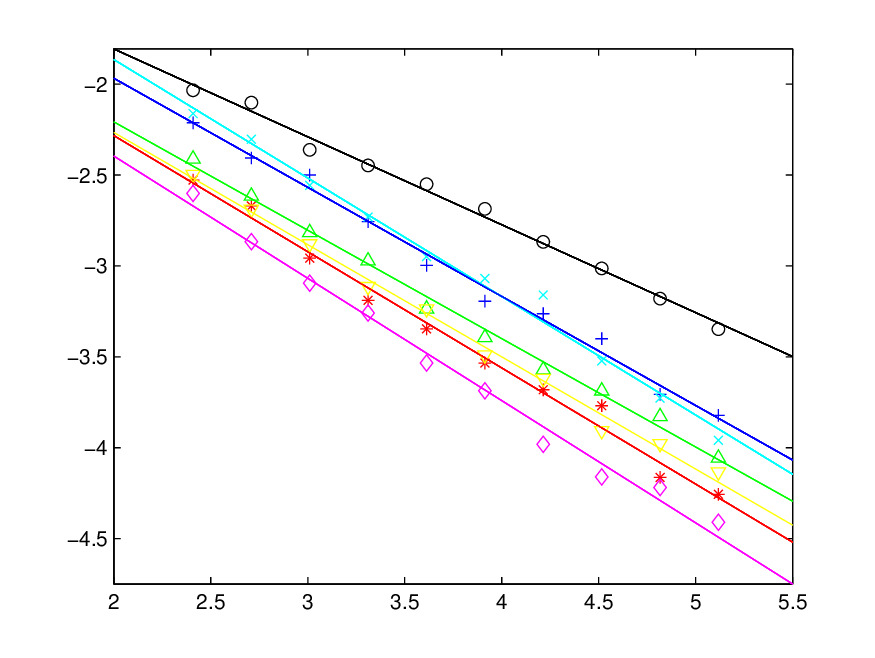}}
\subfigure[Vega asset 5]{\includegraphics[width=3.1in,height=2.4in,keepaspectratio=false]{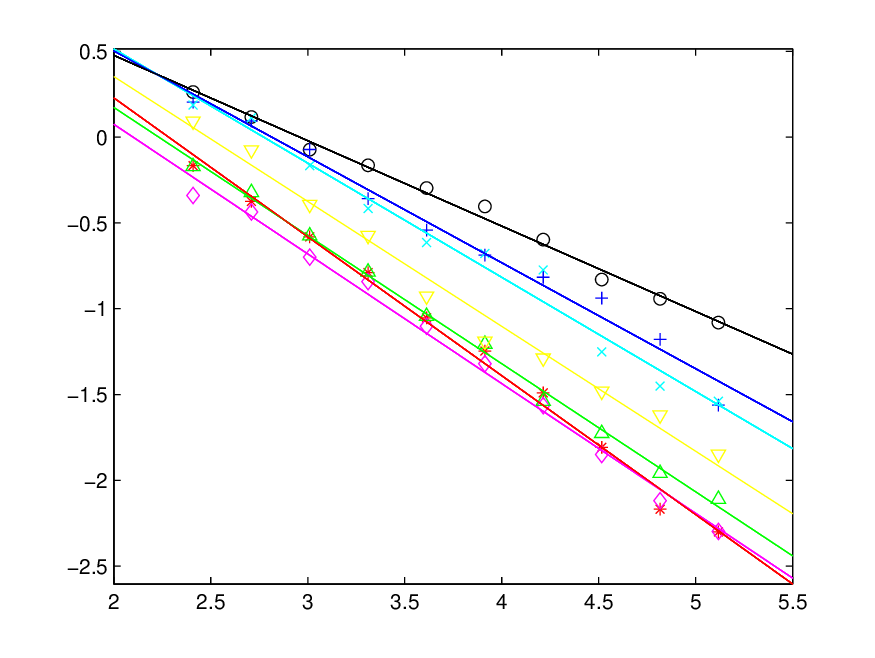}}
\caption{European basket call option price $(a)$ and selected greeks $(b),(c)$,
Log-Log plots of error $\varepsilon_N$ versus number of simulated paths, $N=2^p,\;p=8,\ldots,17$, $D=80$,
$\epsilon=10^{-3}$, $L=100$ runs:: MC+SD+CHOL (black), QMC+SD+CHOL (blue), QMC+SD+PCA (cyan), QMC+BBD+CHOL (red), QMC+BBD+PCA (magenta), QMC+PCA+CHOL (green), QMC+PCA+PCA (yellow). Linear regression lines are also
shown. Correlation is $\rho=0.6$.} \label{Fig:EuErr}
\end{figure}

\begin{figure}[ht]
\centering
\subfigure[Price]{\includegraphics[width=3.1in,height=2.4in,keepaspectratio=false]{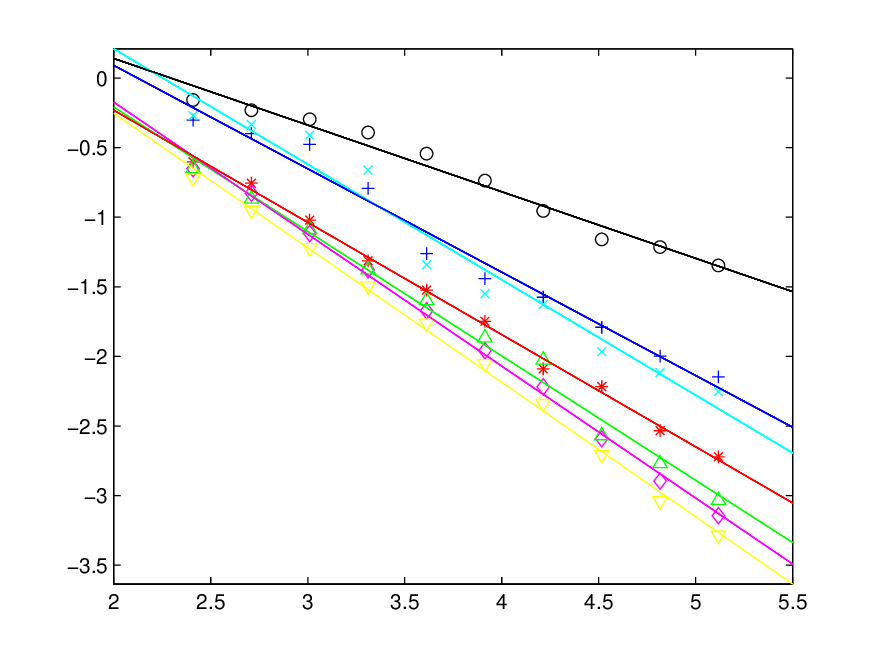}}
\subfigure[Delta asset 5]{\includegraphics[width=3.1in,height=2.4in,keepaspectratio=false]{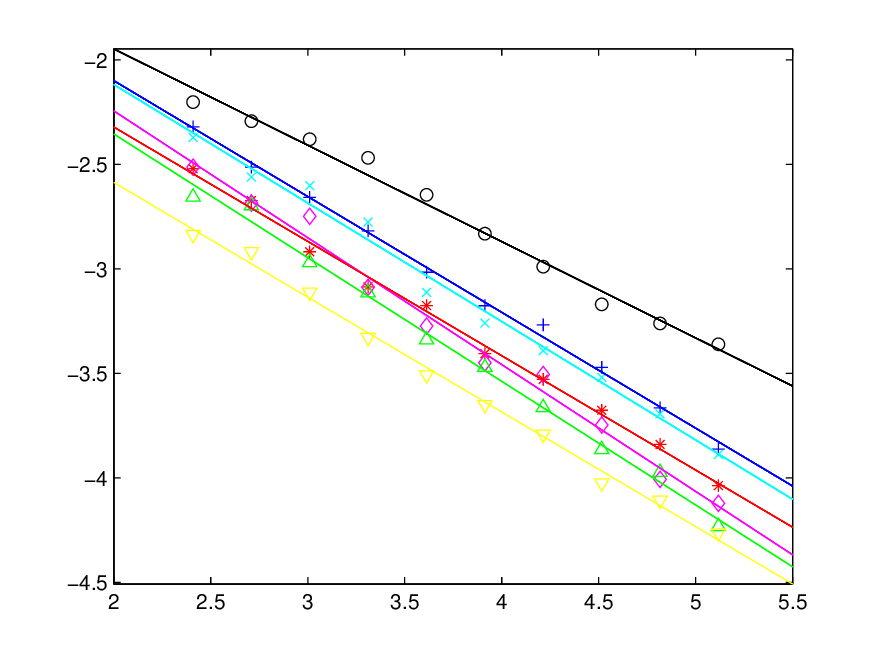}}
\subfigure[Vega asset 5]{\includegraphics[width=3.1in,height=2.4in,keepaspectratio=false]{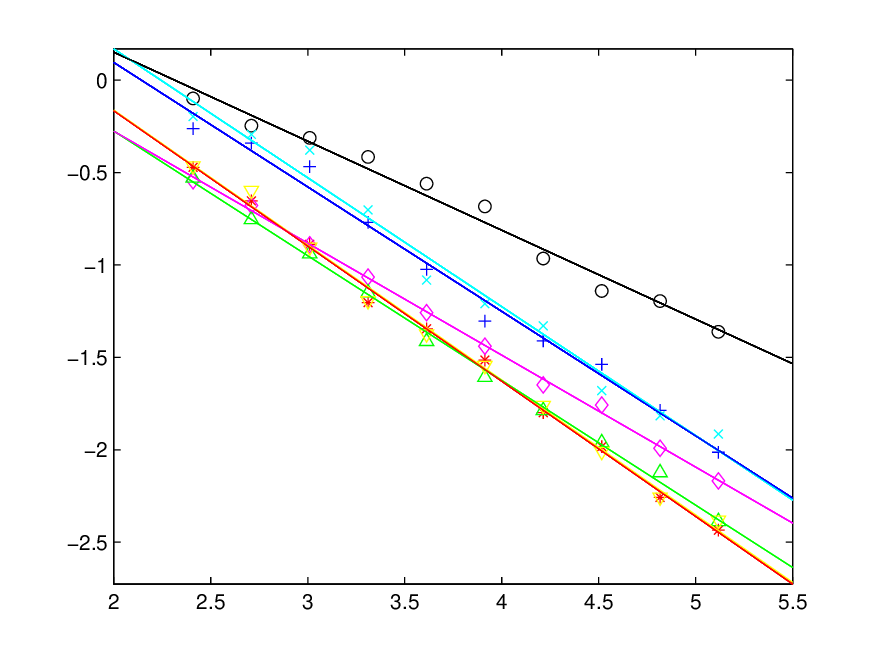}}
\caption{Asian basket call option. Details as in Fig. \ref{Fig:EuErr}.} \label{Fig:AsErr}
\end{figure}

Next, we analyze the relative performance of QMC vs MC in terms of convergence rate. We plot in Figs. \ref{fig:14}-\ref{Fig:AsErr} the root mean square error, eq. (\ref{error}), versus the number of MC scenarios $N$ in Log-Log scale. In all our tests we have chosen an appropriate range for $N$ such that, in the computation of greeks, the bias term is negligible with respect to the variance term (see Appendix \ref{App:GreekErr} for details). Hence, the observed relations are, with good accuracy, linear, therefore the power law (\ref{err:QMC}) is confirmed, and the convergence rates $\alpha$ can be extracted as the slopes of the regression lines.
Furthermore, also the intercepts of regression lines provide useful information about the efficiency of the QMC and MC methods: in fact, lower intercepts mean that the simulated value starts closer to the exact value. The resulting slopes and intercepts from linear regression are presented in Tables \ref{tab:3}-\ref{Tab:As} for all test cases.
\begin{table}[ht]
\small
\centering
\subtable{%
\begin{tabular}{c c c c c c}
  \toprule
  \textbf{Payoff} & \textbf{Function} & \textbf{MC+SD} & \textbf{QMC+SD} & \textbf{QMC+BBD} & \textbf{QMC+PCA}\\
  \midrule
  Asian    & Price & -0.3 & -0.9 & -1.1 & -1.1\\
           & Delta & -1.6 & -1.8 & -2.1 & -2.4\\
           & Gamma & -1.8 & -1.7 & -1.8 & -1.8\\
           & Vega & 0.2 & -0.2 & -0.4 & -0.5\\
  \hline
  Double KO& Price & -0.3 & -0.4 & -0.7 & -0.7\\
           & Delta & -1.6 & -1.7 & -2.1 & -2.0\\
           & Gamma & -1.8 & -1.9 & -2.7 & -1.8\\
           & Vega & 1.6 & 1.6 & 1.6 & 1.6\\
  \hline
  Cliquet  & Price & -2.2 & -3.3 & -2.6 & -2.6\\
           & Vega  & -1.8 & -3.1 & -1.9 & -1.8\\
  \bottomrule
\end{tabular}}
\caption{Intercepts from linear regression, for single-asset options with
MC+SD, QMC+SD, QMC+BBD and QMC+PCA, $L=100$ runs. Only significant digits are shown, standard error on the estimates is around 10\%. Results are shown for $N=512$ paths.} \label{tab:3}
\end{table}

\begin{table}[ht]
\small
\centering
\subtable{%
\begin{tabular}{c c c c c c}
  \toprule
  \textbf{Payoff} & \textbf{Function} & \textbf{MC+SD} & \textbf{QMC+SD} & \textbf{QMC+BBD}  & \textbf{QMC+PCA}\\
  \midrule
  Asian    & Price & -0.49 & -0.65 & -0.76 & -0.74\\
           & Delta & -0.48 & -0.55 & -0.61 & -0.64\\
           & Gamma & -0.48 & -0.53 & -0.51 & -0.50\\
           & Vega & -0.49 & -0.64 & -0.72 & -0.74\\
  \hline
  Double KO& Price & -0.51 & -0.50 & -0.60 & -0.56\\
           & Delta & -0.50 & -0.51 & -0.59 & -0.56\\
           & Gamma & -0.52 & -0.48 & -0.58 & -0.53\\
           & Vega & -0.52 & -0.52 & -0.55 & -0.53\\
  \hline
  Cliquet  & Price & -0.50 & -1.00 & -0.80 & -0.55\\
           & Vega & -0.51 & -0.76 & -0.61 & -0.55\\
  \bottomrule
\end{tabular}
}
\caption{Slopes from linear regression for single-asset options, as in Table \ref{tab:3}. Standard error on the estimates is around 5\% (20\% for the price of the Cliquet with QMC+BBD).}
\label{tab:4}
\end{table}

\begin{table}[htbp]
  \centering
    \begin{tabular}{rrrrrrrr}
    \toprule
    \multicolumn{1}{c}{\textbf{Correlation}} & \multicolumn{1}{c}{\textbf{Method}} & \multicolumn{3}{c}{\textbf{Slopes}} & \multicolumn{3}{c}{\textbf{Intercepts}} \\
    \textbf{} & \textbf{} & \multicolumn{1}{c}{\textbf{Price}} & \multicolumn{1}{c}{\textbf{Delta}} & \multicolumn{1}{c}{\textbf{Vega}} & \multicolumn{1}{c}{\textbf{Price}} & \multicolumn{1}{c}{\textbf{Delta}} & \multicolumn{1}{c}{\textbf{Vega}} \\
    \midrule
    $\rho=0$    & MC+STD+CHOL & -0.49 & -0.50 & -0.51 & -0.04 & -2.00 & 0.28 \\
         & QMC+STD+CHOL & -0.71 & -0.54 & -0.67 & -0.19 & -2.33 & 0.13 \\
         & QMC+STD+PCA & -0.75 & -0.64 & -0.74 & -0.07 & -2.14 & 0.32 \\
         & QMC+BB+CHOL & -0.91 & -0.63 & -0.76 & -0.47 & -2.49 & -0.17 \\
         & QMC+BB+PCA & -0.92 & -0.65 & -0.78 & -0.42 & -2.51 & -0.13 \\
         & QMC+PCA+CHOL & -0.80 & -0.58 & -0.74 & -0.55 & -2.45 & -0.15 \\
         & QMC+PCA+PCA & -0.75 & -0.56 & -0.61 & -0.51 & -2.47 & 0.07 \\
    \midrule
    $\rho=0.3$   & MC+STD+CHOL & -0.50 & -0.48 & -0.51 & 0.12  & -2.00 & 0.28 \\
       & QMC+STD+CHOL & -0.71 & -0.55 & -0.65 & -0.01 & -2.33 & 0.21 \\
       & QMC+STD+PCA & -0.75 & -0.60 & -0.62 & 0.02  & -2.22 & 0.30 \\
       & QMC+BB+CHOL & -0.92 & -0.63 & -0.80 & -0.36 & -2.49 & -0.09 \\
       & QMC+BB+PCA & -0.95 & -0.65 & -0.65 & -0.45 & -2.57 & -0.17 \\
       & QMC+PCA+CHOL & -0.78 & -0.54 & -0.76 & -0.46 & -2.49 & -0.11 \\
       & QMC+PCA+PCA & -0.82 & -0.59 & -0.64 & -0.53 & -2.50 & -0.01 \\
    \midrule
    $\rho=0.6$   & MC+STD+CHOL & -0.49 & -0.48 & -0.50 & 0.21  & -2.00 & 0.27 \\
       & QMC+STD+CHOL & -0.72 & -0.60 & -0.62 & 0.12  & -2.21 & 0.25 \\
       & QMC+STD+PCA & -0.79 & -0.65 & -0.67 & 0.16  & -2.13 & 0.24 \\
       & QMC+BB+CHOL & -0.93 & -0.64 & -0.81 & -0.28 & -2.54 & -0.10 \\
       & QMC+BB+PCA & -0.97 & -0.67 & -0.76 & -0.32 & -2.67 & -0.23 \\
       & QMC+PCA+CHOL & -0.78 & -0.60 & -0.75 & -0.40 & -2.45 & -0.13 \\
       & QMC+PCA+PCA & -0.80 & -0.62 & -0.73 & -0.51 & -2.52 & 0.06 \\
    \midrule
    $\rho=0.9$   & MC+STD+CHOL & -0.49 & -0.47 & -0.48 & 0.27  & -2.01 & 0.26 \\
       & QMC+STD+CHOL & -0.78 & -0.64 & -0.64 & 0.23  & -2.13 & 0.25 \\
       & QMC+STD+PCA & -0.83 & -0.64 & -0.74 & 0.28  & -2.16 & 0.30 \\
       & QMC+BB+CHOL & -0.97 & -0.64 & -0.83 & -0.22 & -2.64 & -0.09 \\
       & QMC+BB+PCA & -0.95 & -0.72 & -0.85 & -0.25 & -2.64 & -0.12 \\
       & QMC+PCA+CHOL & -0.86 & -0.56 & -0.78 & -0.29 & -2.57 & -0.10 \\
       & QMC+PCA+PCA & -0.83 & -0.58 & -0.74 & -0.46 & -2.61 & -0.02 \\
    \bottomrule
    \end{tabular}%
\caption{Slopes and intercepts from linear regression, for European basket option with different sampling techniques and correlation levels, $L=100$ runs. Only significant digits are shown, standard error on the estimates is around 10\% for intercepts and 5\% for slopes.
Results of intercepts are shown for $N=512$ paths.}\label{Tab:Eu}%
\end{table}

\begin{table}[htbp]
  \centering
    \begin{tabular}{rrrrrrrr}
    \toprule
    \multicolumn{1}{c}{\textbf{Correlation}} & \multicolumn{1}{c}{\textbf{Method}} & \multicolumn{3}{c}{\textbf{Slopes}} & \multicolumn{3}{c}{\textbf{Intercepts}} \\
    \textbf{} & \textbf{} & \multicolumn{1}{c}{\textbf{Price}} & \multicolumn{1}{c}{\textbf{Delta}} & \multicolumn{1}{c}{\textbf{Vega}} & \multicolumn{1}{c}{\textbf{Price}} & \multicolumn{1}{c}{\textbf{Delta}} & \multicolumn{1}{c}{\textbf{Vega}} \\
    \midrule
    $\rho=0$     & MC+STD+CHOL & -0.46 & -0.47 & -0.46 & -0.31 & -2.13 & -0.10 \\
         & QMC+STD+CHOL & -0.70 & -0.56 & -0.64 & -0.57 & -2.37 & -0.34 \\
         & QMC+STD+PCA & -0.76 & -0.56 & -0.80 & -0.40 & -2.31 & -0.02 \\
         & QMC+BB+CHOL & -0.76 & -0.58 & -0.69 & -0.75 & -2.44 & -0.46 \\
         & QMC+BB+PCA & -0.79 & -0.58 & -0.69 & -0.72 & -2.47 & -0.45 \\
         & QMC+PCA+CHOL & -0.87 & -0.65 & -0.72 & -0.79 & -2.43 & -0.53 \\
         & QMC+PCA+PCA & -0.84 & -0.57 & -0.64 & -0.74 & -2.62 & -0.39 \\
    \midrule
    $\rho=0.3$   & MC+STD+CHOL & -0.49 & -0.48 & -0.45 & -0.19 & -2.14 & -0.08 \\
       & QMC+STD+CHOL & -0.71 & -0.55 & -0.68 & -0.37 & -2.33 & -0.24 \\
       & QMC+STD+PCA & -0.82 & -0.59 & -0.64 & -0.23 & -2.32 & -0.01 \\
       & QMC+BB+CHOL & -0.76 & -0.56 & -0.70 & -0.66 & -2.52 & -0.48 \\
       & QMC+BB+PCA & -0.89 & -0.54 & -0.57 & -0.69 & -2.57 & -0.33 \\
       & QMC+PCA+CHOL & -0.86 & -0.57 & -0.75 & -0.68 & -2.53 & -0.46 \\
       & QMC+PCA+PCA & -0.93 & -0.59 & -0.62 & -0.81 & -2.69 & -0.60 \\
    \midrule
    $\rho=0.6$   & MC+STD+CHOL & -0.48 & -0.46 & -0.48 & -0.05 & -2.14 & -0.05 \\
       & QMC+STD+CHOL & -0.74 & -0.55 & -0.67 & -0.21 & -2.33 & -0.18 \\
       & QMC+STD+PCA & -0.83 & -0.57 & -0.70 & -0.13 & -2.35 & -0.12 \\
       & QMC+BB+CHOL & -0.81 & -0.55 & -0.73 & -0.56 & -2.54 & -0.46 \\
       & QMC+BB+PCA & -0.95 & -0.61 & -0.61 & -0.56 & -2.49 & -0.52 \\
       & QMC+PCA+CHOL & -0.89 & -0.59 & -0.67 & -0.57 & -2.60 & -0.55 \\
       & QMC+PCA+PCA & -0.97 & -0.55 & -0.73 & -0.65 & -2.81 & -0.46 \\
    \midrule
    $\rho=0.9$   & MC+STD+CHOL & -0.49 & -0.48 & -0.48 & 0.01  & -2.13 & -0.06 \\
       & QMC+STD+CHOL & -0.79 & -0.54 & -0.76 & -0.09 & -2.39 & -0.09 \\
       & QMC+STD+PCA & -0.84 & -0.56 & -0.82 & -0.05 & -2.37 & -0.05 \\
       & QMC+BB+CHOL & -0.89 & -0.58 & -0.77 & -0.48 & -2.52 & -0.50 \\
       & QMC+BB+PCA & -0.94 & -0.64 & -0.74 & -0.49 & -2.47 & -0.55 \\
       & QMC+PCA+CHOL & -0.92 & -0.53 & -0.82 & -0.52 & -2.71 & -0.49 \\
       & QMC+PCA+PCA & -0.95 & -0.63 & -0.74 & -0.54 & -2.79 & -0.53 \\
    \bottomrule
    \end{tabular}%
\caption{Slopes and intercepts for Asian basket option. Details as in Table \ref{Tab:Eu}.}\label{Tab:As}%
\end{table}%


We observe that, for the Asian call, QMC+PCA outperforms all other methods for price, delta and vega (for gamma all methods show similar convergence), with QMC+BBD being only marginally less efficient. Higher $\alpha$ and lower intercepts allow for faster convergence\footnote{We stress that slopes and intercepts shown in the previous Figs. \ref{fig:14}-\ref{Fig:AsErr} do not depend on the details of the simulations, in particular the MC seed or the LDS starting point, since we are averaging over $L=100$ runs.}. Moreover, convergence appears to be much smoother and more stable. Similar considerations hold for the double barrier option with QMC+BBD, the Cliquet with QMC+SD, the European basket option with QMC+BBD+PCA and the basket Asian option with QMC+PCA+PCA. These results can be explained by the fact that the above mentioned sampling strategies are optimal, in the sense that they are intrinsically designed to extract initial coordinates of the gaussian vector $\bm{Z}=(Z_1,\ldots,Z_D)$ to construct the most important coordinates of the underlying assets' vector $\bm{S}=\left(S_1(t_1),\ldots,S_{N_{rf}}(t_1),\ldots,S_1(t_{N_{ts}}),\ldots,S_{N_{rf}}(t_{N_{ts}})\right)$.

\subsection{Speed-Up analysis}
\label{SecSpeedUp} \noindent
A typical question with Monte Carlo simulation
is \QuoteDouble{how many scenarios are necessary to achieve a
given precision?}. When comparing two numerical simulation
methods, the typical question becomes \QuoteDouble{how many
scenarios may I save using method B instead of method A,
preserving the same precision?}.
\par
A useful measure of the relative computational performance of two
numerical methods is the so called speed-up $S_*(a)$
\cite{KreMer1998a,PapTrau1996}. It is defined as
\begin{equation}
\label{EqSpeedUp}
S_*^{(A,B)}(a) = \frac{N_*^{(A)}(a)}{N_*^{(B)}(a)}\, ,
\end{equation}
where $N_*^{(A)}(a)$ is the number of scenarios needed to computational scheme A (which can be any of the different combinations of random number generator, time discretization and covariance factorization algorithms) to reach and maintain a given accuracy $a$ \wrt exact or almost exact results\footnote{The thresholds $N_*$ could be evaluated through direct simulation, but this would be extremely computationally expensive. Thus we resort to a simpler algorithm: we identify the number of scenarios $N_*^{(A)}(a)$ in eq. (\ref{EqSpeedUp}) as the first number of simulated paths such that, for any $N>N_*$, $V-a\leq V_N\pm 3\, \varepsilon\leq V+a$, where $V$ and $V_N$ are respectively the exact and simulated
values of prices or greeks and $\varepsilon$ is the standard
error. Then, $N_*^{(A)}(a)$ can be estimated through linear regression results of Section \ref{SecPerformance}. Some concern should be given to extrapolation for finite differences, as discussed in \cite{BiaKuc15}.}.
Thus, the speed-up $S_*(a)$ quantifies the computational gain of method B \wrt method A.
\par
We show in table \ref{tab:5} the Speed-Up computation results for QMC method with optimal sampling strategies over standard MC, when accuracies of 1\% and 0.1\% are to be reached. The simulation methods chosen for the computation of QMC Speed-Up are those which achieved the highest performance for our test cases, as concluded from Section \ref{SecPerformance}.
\begin{table}[ht]
\small
\centering
\subtable{%
\begin{tabular}{c c c c}
\toprule
  \textbf{Payoff} & \textbf{Function} & \multicolumn{2}{c}{\textbf{QMCvsMC Speed-Up}}\\
   & & $a=1\%$ & $a=0.1\%$\\
  \midrule
  Asian     & Price & 80 & 300\\
            & Delta & 50 & 150\\
            & Gamma & 1 & -\\
            & Vega  & 60 & 300\\
  \hline
  Double KO & Price & 10 & 30\\
            & Delta & 20 & 30\\
            & Gamma & 150 & 300\\
            & Vega  & 3 & 5\\
  \hline
  Cliquet   & Price & 30 & 400\\
            & Vega  & 100 & 500\\
  \hline
  European basket   & Price & 110 & 800\\
                    & Delta  & 40 & 150\\
                    & Vega & 30 & 80\\
   \hline
  Asian basket   & Price & 80 & 700\\
                    & Delta  & 20 & 50\\
                    & Vega & 30 & 150\\
  \bottomrule
\end{tabular}
}
\caption{Speed-Up $S_*(a)$ of optimal QMC vs MC techniques, in order to achieve 1\% and 0.1\% accuracy in the computation of prices and FD greeks. The shift $\epsilon$ for finite differences and the correlations are the same as used in the previous sections. Missing values of $S_*$ mean that the required accuracy cannot be reached since it is smaller than the bias. Regarding the QMC simulation, the sampling strategies considered are: PCA for Asian call, BBD for double knock-out, SD for cliquet, BBD+PCA for European basket, PCA+PCA for Asian basket.}
\label{tab:5}
\end{table}
\par
In general, QMC with BBD or PCA largely outperforms the other methods, with a speed-up factor close to $10^3$ in some cases. QMC with SD is the best method for Cliquet. We notice in particular that, in most cases, a ten-fold increase of the accuracy $a$ results in a two-fold increase of speed-up $S_*(a)$. However, in a few cases (Cliquet and basket options), such an increase can result in up to ten-folds increase of $S_*(a)$.
\begin{figure}
\centering
\subfigure[Absolute CPU time]{\includegraphics[width=3.1in,height=2.4in,keepaspectratio=false]{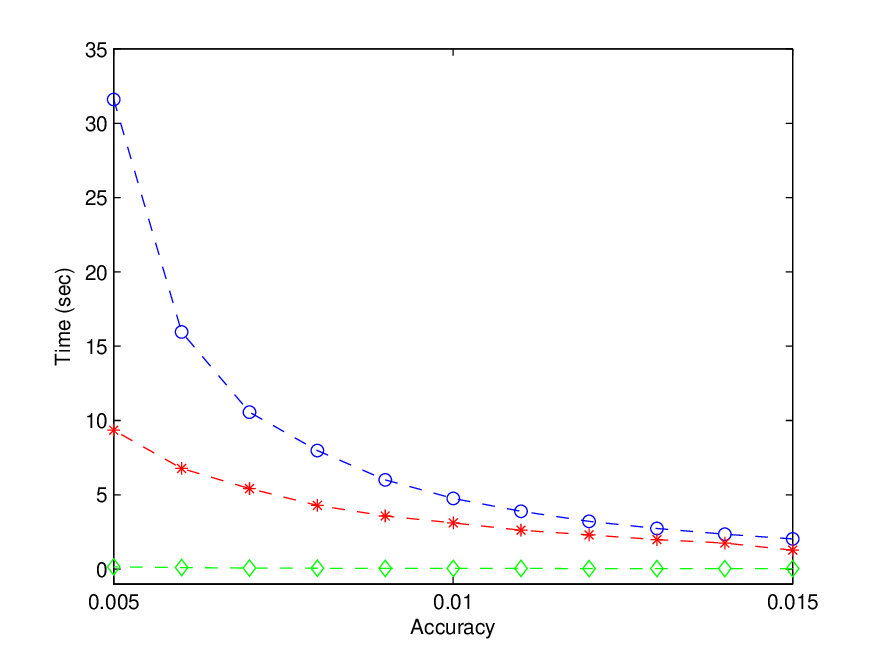}}
\subfigure[Log CPU time]{\includegraphics[width=3.1in,height=2.4in,keepaspectratio=false]{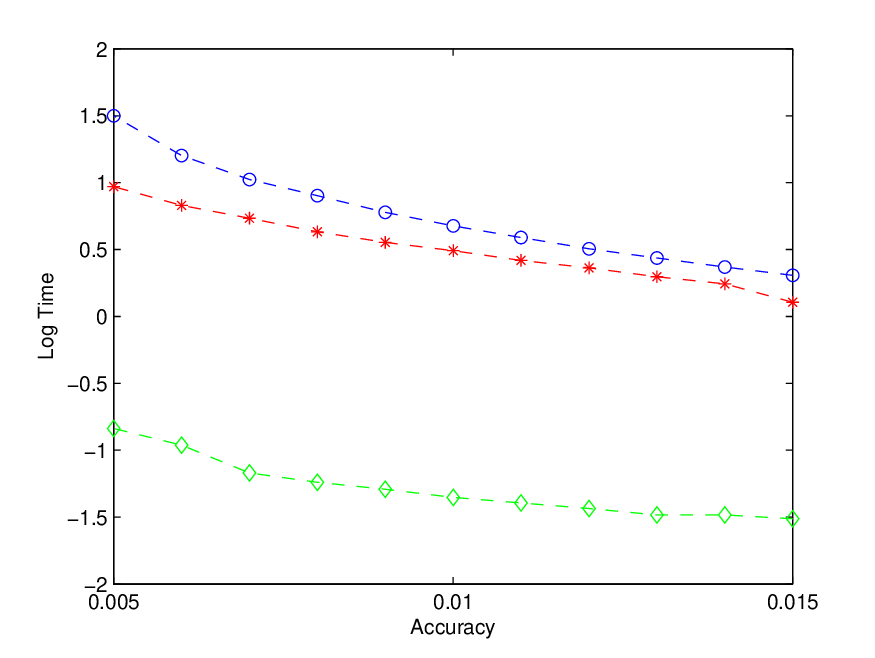}}
\caption{European basket option. Absolute CPU time $(a)$ and logarithmic CPU time $(b)$ needed to compute price and all greeks (deltas and vegas), for different target accuracies: MC+SD+CHOL with AAD (blue), QMC+BBD+PCA with FD (red), QMC+BBD+PCA with AAD (green). The number of underlyings is $N_{rf}=5$ and the number of time steps is $N_{ts}=16$. Correlation is $\rho=0.3$, the other parameters are as described in Section \ref{SecPayoffs}} \label{Fig:SpeedUpEu}
\end{figure}
\begin{figure}
\centering
\subfigure[Absolute CPU time]{\includegraphics[width=3.1in,height=2.4in,keepaspectratio=false]{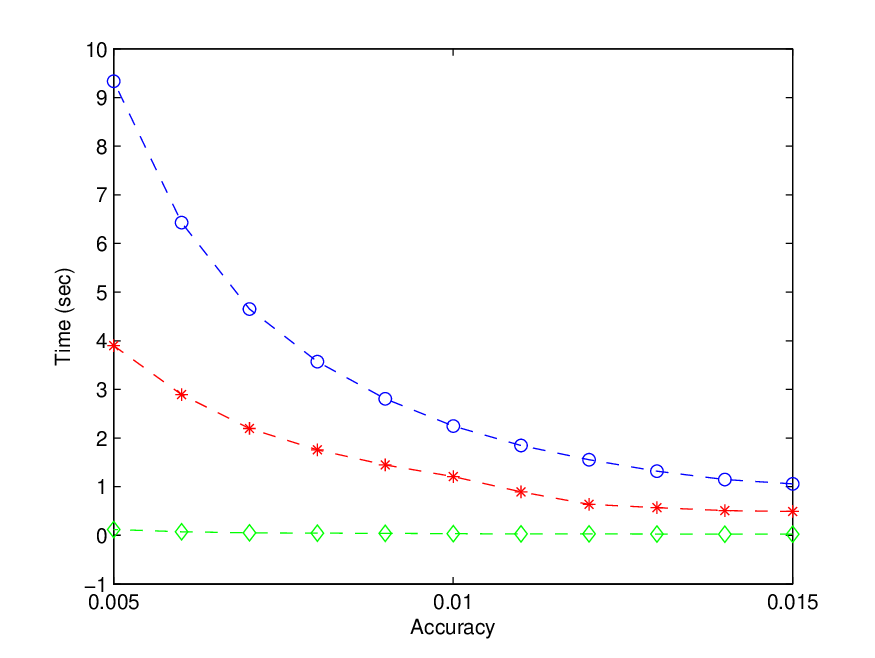}}
\subfigure[Log CPU time]{\includegraphics[width=3.1in,height=2.4in,keepaspectratio=false]{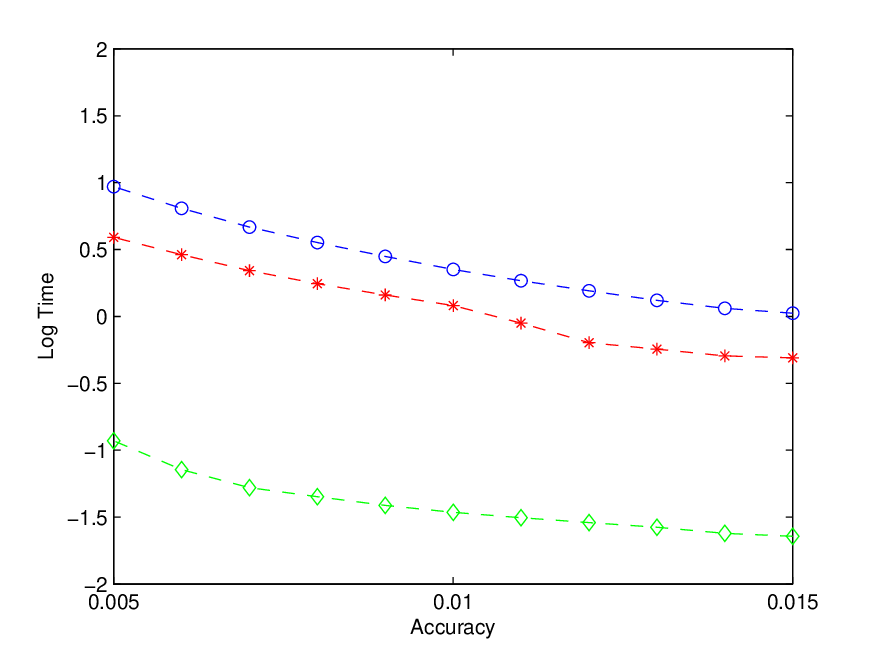}}
\caption{Asian basket option. Absolute CPU time $(a)$ and logarithmic CPU time $(b)$ needed to compute price and all greeks (deltas and vegas), for different target accuracies: MC+SD+CHOL with AAD (blue), QMC+PCA+PCA with FD (red), QMC+PCA+PCA with AAD (green). The number of underlyings is $N_{rf}=5$ and the number of time steps is $N_{ts}=16$. Correlation is $\rho=0.3$, the other parameters are as described in Section \ref{SecPayoffs}} \label{Fig:SpeedUpAs}
\end{figure}
\begin{figure}
\centering
\subfigure[Absolute CPU time]
{\includegraphics[width=3.1in,height=2.4in,keepaspectratio=false]{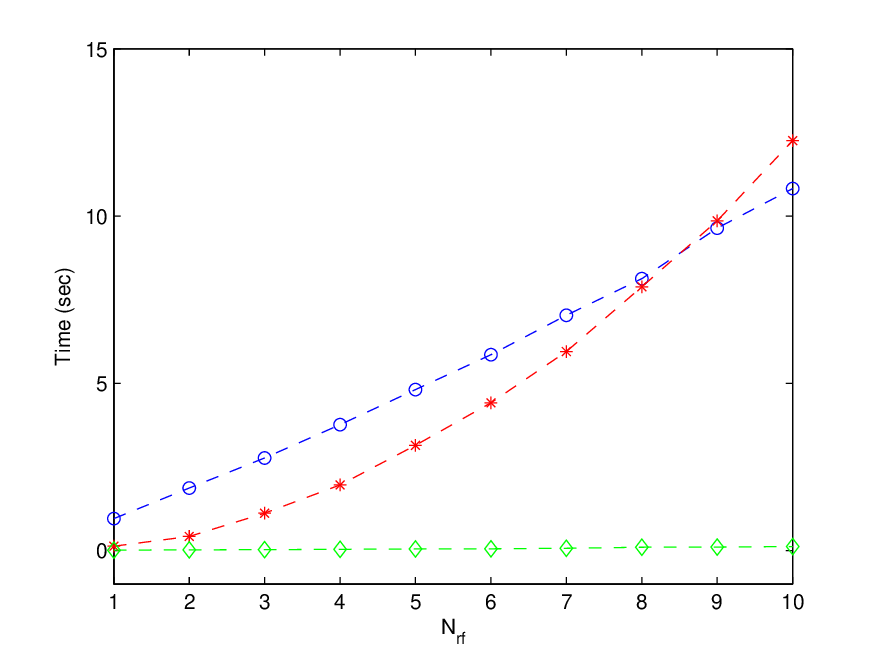}}
\subfigure[Log CPU time]{\includegraphics[width=3.1in,height=2.4in,keepaspectratio=false]{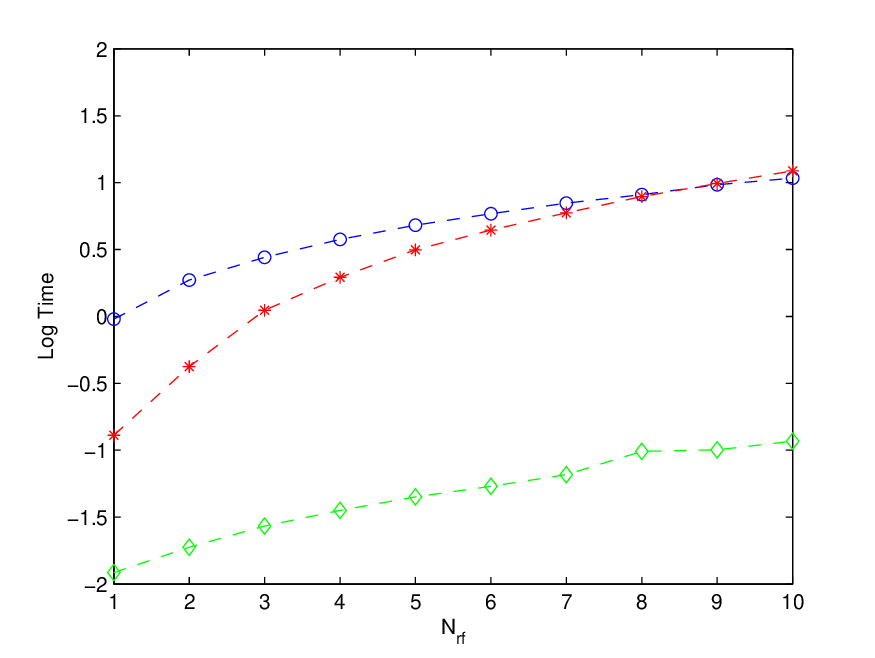}}
\caption{European basket option. Absolute CPU time $(a)$ and logarithmic CPU time $(b)$ needed to compute price and all greeks (deltas and vegas), for increasing number of underlying assets $N_{rf}$: MC+SD+CHOL with AAD (blue), QMC+BBD+PCA with FD (red), QMC+BBD+PCA with AAD (green). The target accuracy is fixed to 1\% and the number of time steps is always $N_{ts}=16$. Correlation is $\rho=0.3$, the other parameters are as described in Section \ref{SecPayoffs}} \label{Fig:SpeedUpIncrEu}
\end{figure}
\begin{figure}
\centering
\subfigure[Absolute CPU time]
{\includegraphics[width=3.1in,height=2.4in,keepaspectratio=false]{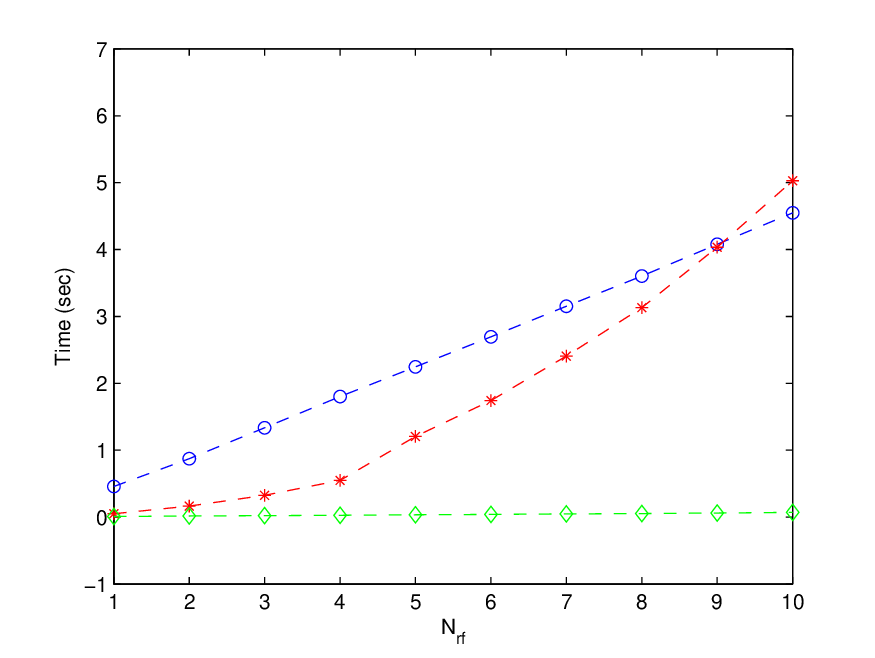}}
\subfigure[Log CPU time]{\includegraphics[width=3.1in,height=2.4in,keepaspectratio=false]{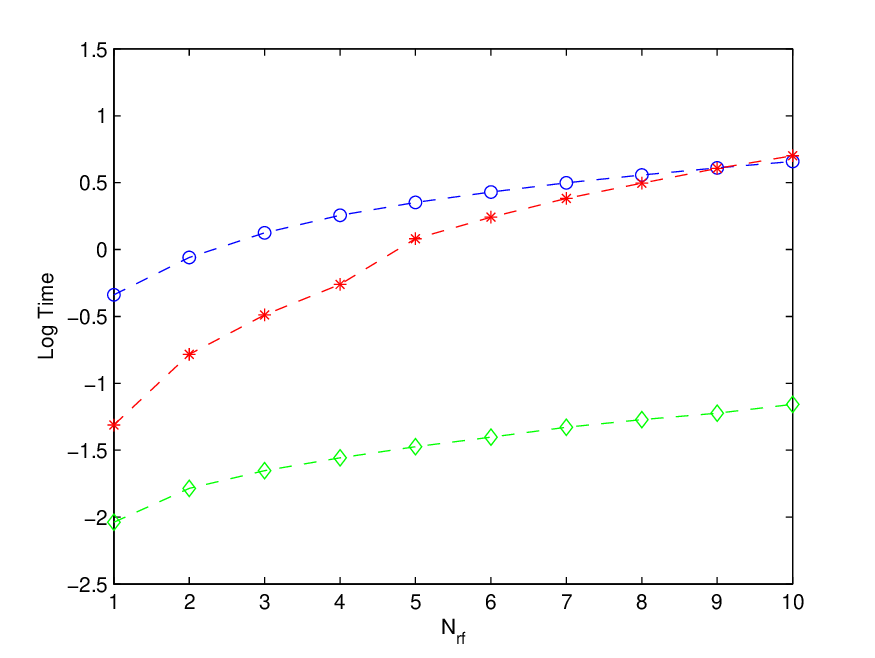}}
\caption{Asian basket option. Absolute CPU time $(a)$ and logarithmic CPU time $(b)$ needed to compute price and all greeks (deltas and vegas), for increasing number of underlying assets $N_{rf}$: MC+SD+CHOL with AAD (blue), QMC+BBD+PCA with FD (red), QMC+BBD+PCA with AAD (green). The target accuracy is fixed to 1\% and the number of time steps is always $N_{ts}=16$. Correlation is $\rho=0.3$, the other parameters are as described in Section \ref{SecPayoffs}}
\label{Fig:SpeedUpIncrAs}
\end{figure}
We notice that the Speed-Up measure actually makes no reference to the computational time but is rather defined as a ratio of number of simulations. This, in turn, is proportional to CPU time. Therefore, it is interesting to fix a given accuracy and compute the CPU time needed to reach it with various combinations of methods, including FD and AAD. Indeed, even though adjoints allow for big savings, in terms of computational time, w.r.t. finite differences, the accuracy of the computation is rather given by the simulation method: fixing a target accuracy $a$, QMC will reach it with much fewer scenarios than MC, as measured by $S_*(a)$. In Figs. \ref{Fig:SpeedUpEu} and \ref{Fig:SpeedUpAs} we show absolute CPU times necessary to evaluate price and all greeks at a given accuracy for MC and optimized QMC, with AAD or FD, for our test cases of European and Asian basket options with 5 correlated underlyings and 16 time steps, which is quite typical case in real financial applications.
\par
We observe that, while QMC with AAD is of course the best choice, QMC with FD runs in comparable times as MC with AAD for accuracies up to few percent and is actually faster for higher accuracies. When the number of underlyings is increased, AAD will become favourable w.r.t. FD in terms of computational time when the same accuracy is to be reached. Fixing target accuracy to 1\% and increasing the number of underlyings\footnote{For simplicity we assume that $S_*(a)$ is almost constant in the range $N_{rf}=1,\ldots,10$.}, we observe from Figs. \ref{Fig:SpeedUpIncrEu} and \ref{Fig:SpeedUpIncrAs} that AAD becomes faster than FD starting from $N_{rf}\simeq 10$. We recall that, since we are using central differences, FD computation of all deltas and vegas requires $4\cdot N_{rf}$ re-pricings.
\par
It follows from these simple experiments that AAD without QMC is not guaranteed to be faster than FD if accuracy is concerned. We further comment on this in the Conclusions.

\subsection{Stability analysis}
\label{SecStability}
\begin{figure}[t!]
\centering
\subfigure[Price]{\includegraphics[width=3.1in,height=2.9in,keepaspectratio=false]{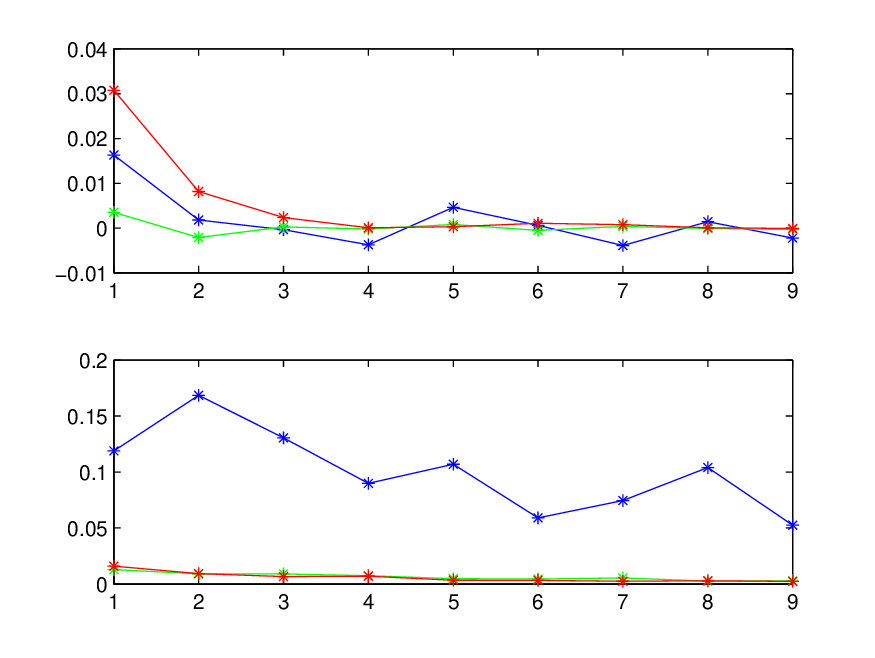}}
\subfigure[Delta]{\includegraphics[width=3.1in,height=2.9in,keepaspectratio=false]{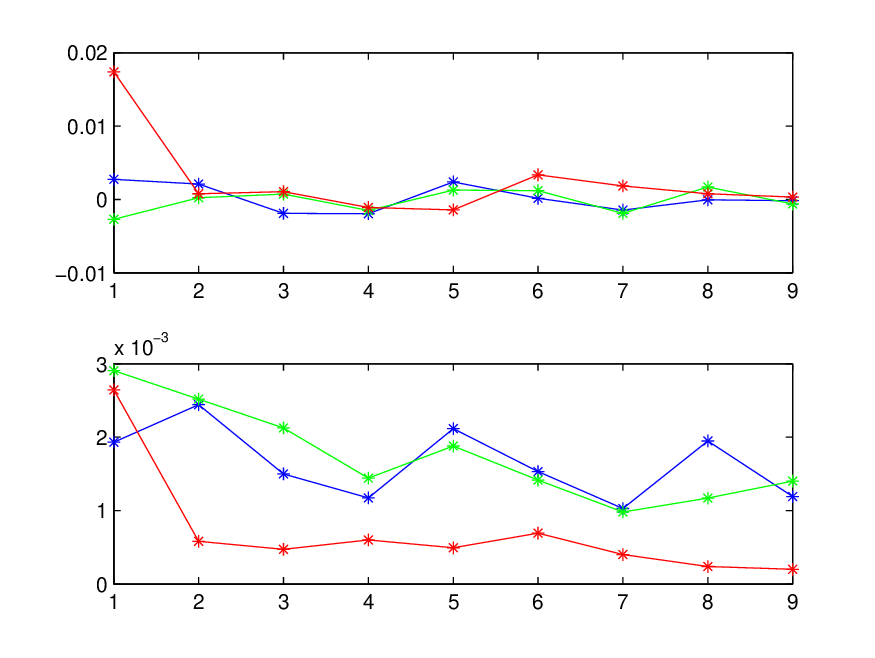}}
\subfigure[Gamma]{\includegraphics[width=3.1in,height=2.9in,keepaspectratio=false]{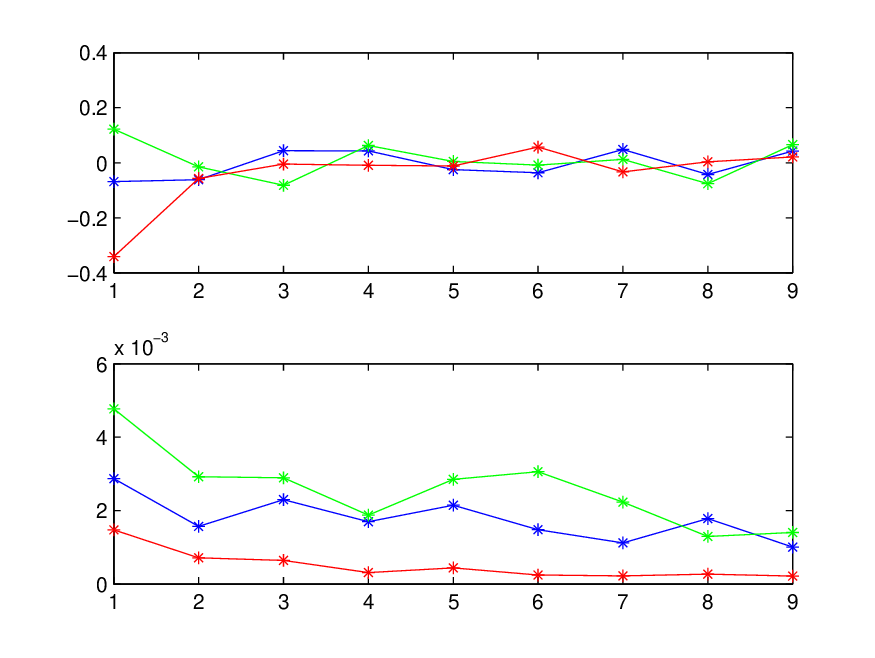}}
\subfigure[Vega] {\includegraphics[width=3.1in,height=2.9in,keepaspectratio=false]{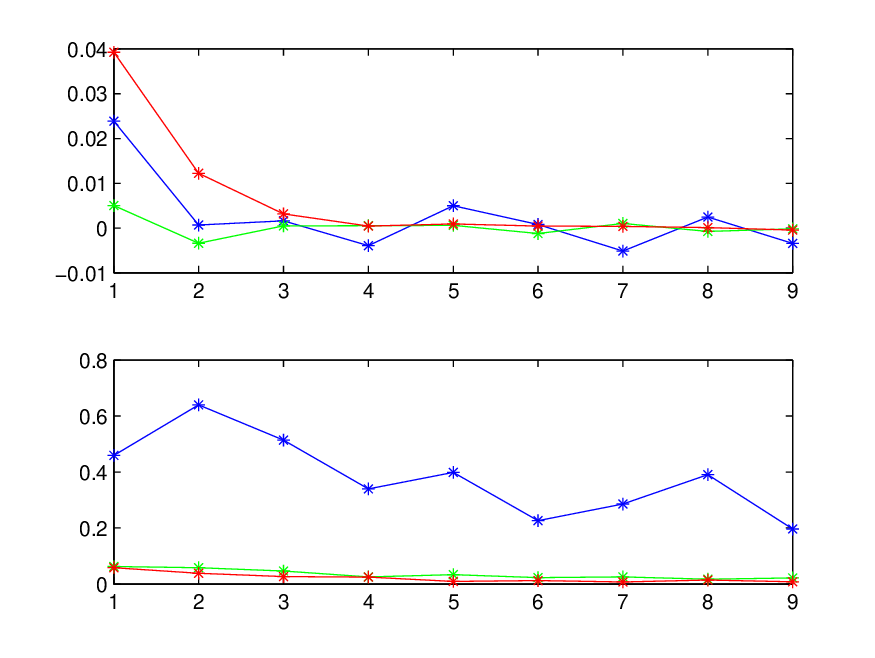}}
\caption{Log-returns (upper plots) and volatilities (lower plots)
of Asian call option price $(a)$ and greeks $(b),(c),(d)$, for
$D=32$, $\epsilon=10^{-3}$, MC+SD (blue), rQMC+BBD (green) and
pure QMC+BBD (red). The number of simulation paths ranges from 100
to 10,000 grouped in 10 windows each containing 10 samples
(x-axis).} \label{fig:18}
\end{figure}

 \noindent
We have already observed that QMC convergence
is often smoother than MC (see Figs. \ref{fig:10}-\ref{Fig:AsConv}):
such monotonicity and stability guarantee better convergence for a
given number of paths $N$. In order to quantify monotonicity and
stability of the various numerical techniques, the following
strategy is used: we divide the range of path simulations $N$ in
10 windows of equal length, and we compute the sample mean $m_i$
and the sample standard deviation (``volatility'') $s_i$ for each
window $i$. Then, log-returns $\log(m_i/m_{i-1})$ and volatilities
$s_i$, for $i=2,\ldots,10$, are used as measures of, respectively,
monotonicity and stability: ``monotonic'' convergence will show
non oscillating log-returns converging to zero, ``stable''
convergence will show low and almost flat volatility. We performed
stability analysis for MC and QMC methods. For QMC we used two
different generators: pure QMC with \texttt{BRODA} generator and
randomized Quasi Monte Carlo (rQMC) with \texttt{Matlab}
generator\footnote{Matlab Function \texttt{sobolset} with the
\texttt{MatousekAffineOwen} scrambling method was used.}.
As an example, we plot in Fig. \ref{fig:18} the results for the Asian option comparing the stability of QMC+BBD with MC+SD.

\begin{figure}[t!]
\centering
\subfigure[Price]{\includegraphics[width=3.1in,height=2.9in,keepaspectratio=false]{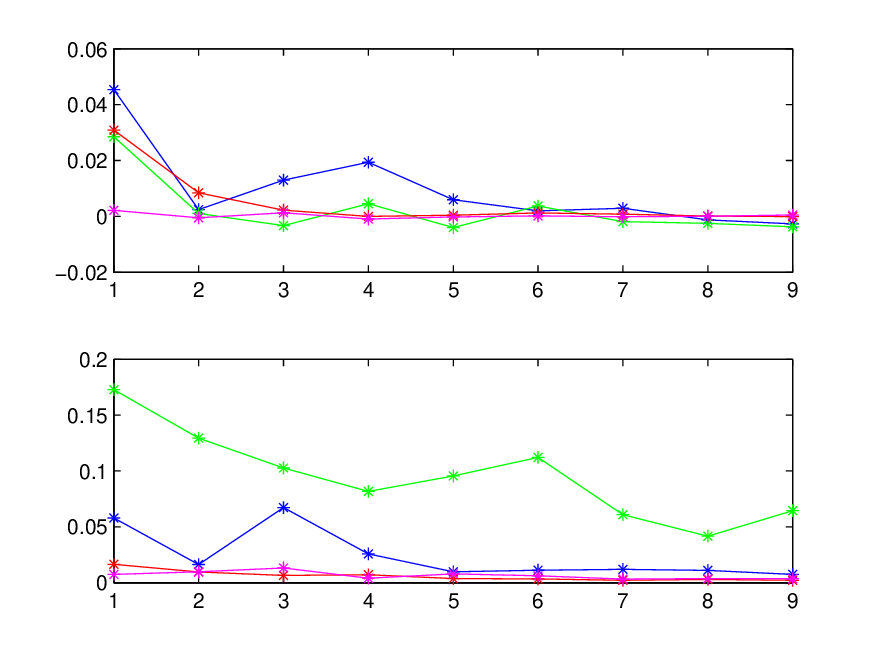}}
\subfigure[Delta]{\includegraphics[width=3.1in,height=2.9in,keepaspectratio=false]{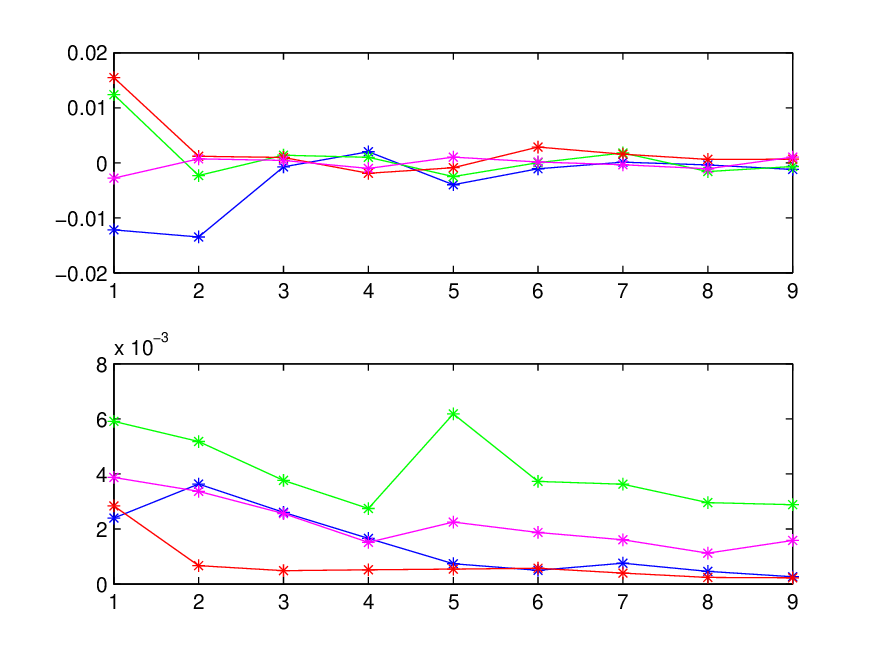}}
\subfigure[Gamma]{\includegraphics[width=3.1in,height=2.9in,keepaspectratio=false]{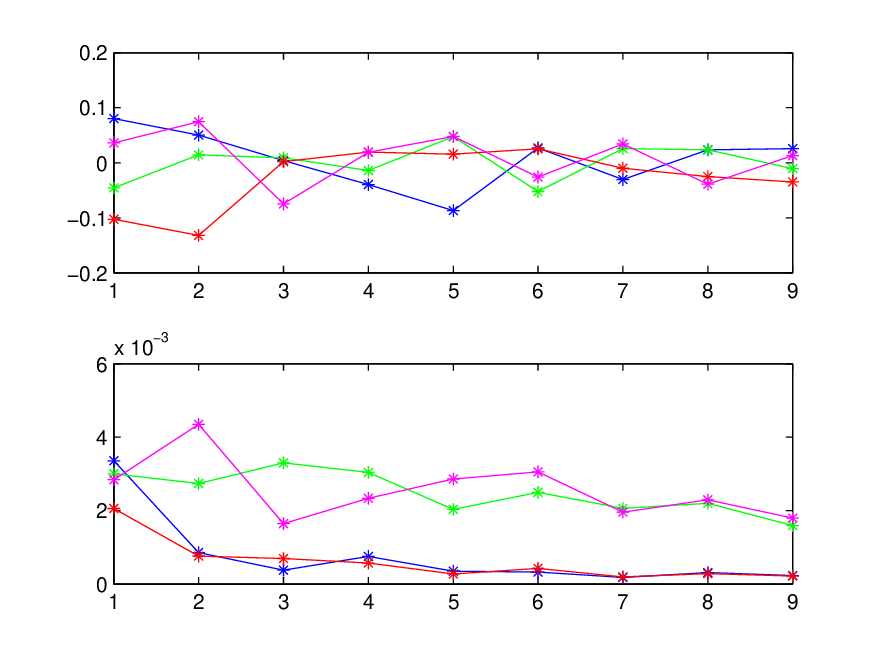}}
\subfigure[Vega] {\includegraphics[width=3.1in,height=2.9in,keepaspectratio=false]{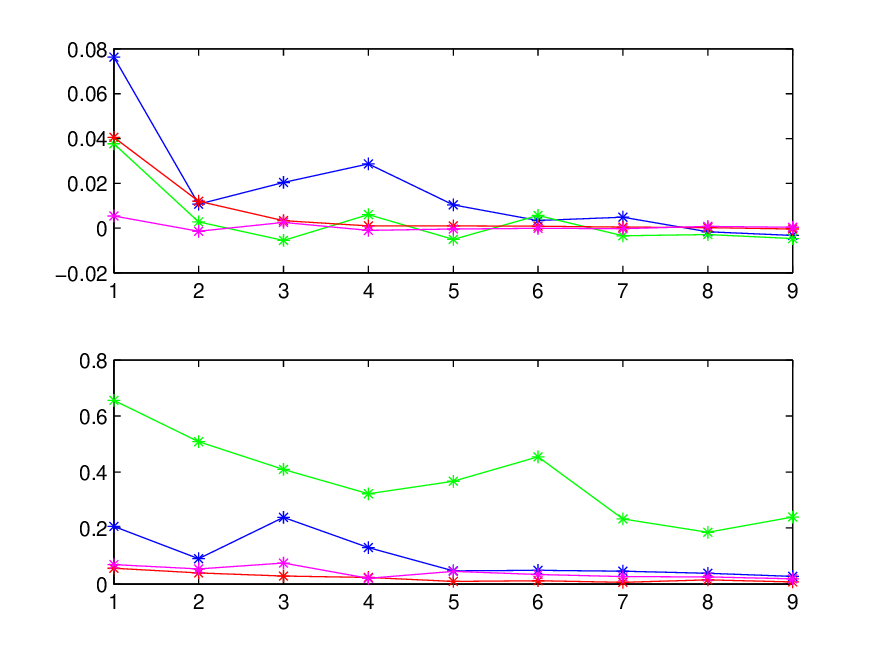}}
\caption{Asian call option with $D=252$, $\epsilon=10^{-3}$. Results are shown for rQMC+SD+\texttt{Matlab} (green)
and rQMC+BBD+\texttt{Matlab} (magenta), and QMC+SD+\texttt{BRODA}
(blue) and QMC+BBD+\texttt{Broda} (red).} \label{fig:21}
\end{figure}
\par
We observe that, in general, QMC+\texttt{BRODA} and rQMC+\texttt{Matlab} are more monotonic and stable than MC+SD. However, this fact is less evident for Asian delta and gamma, where QMC lacks monotonicity and stability \wrt MC, with QMC+BRODA being slightly more stable than rQMC+\texttt{Matlab}. As we know from the results of GSA for this case that higher order interactions are present and the effective dimensions are large (see Table \ref{tab:2}).
\par
In order to understand also the effect of dimension $D$ on monotonicity and stability, we run a similar experiment for an Asian option with $D=252$ fixing dates using both QMC and rQMC with SD and BBD. The results are shown in Fig. \ref{fig:21}. We observe that pure QMC with \texttt{BRODA} generator preserves monotonicity and stability much more than randomized QMC based on \texttt{Matlab} generator for all cases including delta and gamma, with QMC+BBD+\texttt{BRODA} showing the best stability.
It is also interesting to note that the increase in dimension resulted in the decrease in the effective dimensions for the case of the BBD but not for the SD.
\par
We conclude that good high-dimensional LDS generators are crucial to obtain a smooth monotonic and stable convergence of the MC Simulation in high effective dimensional problems.

\section{Conclusions}
\label{SecConclusions} \noindent
In this work we presented an updated overview of the application of QMC and GSA methods in finance, \wrt standard MC methods.
In particular, we considered prices and greeks (delta, gamma, vega) for selected payoffs with increasing degree of complexity and path-dependency (Arithmetic Asian Call, Double Knock-Out Barrier, Cliquet, European and Asian Basket options).
We compared standard discretization (SD) vs Brownian bridge discretization (BBD) and Principal Component Analysis (PCA) schemes of the underlying stochastic diffusion process, as well as Cholesky vs PCA factorization of the covariance matrix, and different sampling of the underlying distribution using pseudo random vs high dimensional Sobol' low discrepancy sequences.
We applied GSA and performed detailed and systematic analysis of convergence diagrams, error estimation, performance, speed-up and stability of the different MC and QMC simulations.
\par
The GSA results in Section \ref{SecGSAresults} revealed that effective dimensions associated with BBD and PCA simulations are generally lower than those associated with SD and Cholesky simulations, and showed how much such dimension reduction acts for different payoffs and greeks (Figures \ref{fig:2}-\ref{Fig:GSAmult} and Tables \ref{tab:1}-\ref{Tab:GSAmult}).
Effective dimensions, being linked with the structure of ANOVA decompositions (the number of important inputs, importance of high order interactions) fully explain the superior efficiency of QMC with BBD and PCA due to the specifics of Sobol' sequences. An exception is represented by Cliquet options.
\par
The performance analysis results presented in Section \ref{SecPerformance} confirmed that QMC with BBD or PCA outperforms MC in most cases, showing faster and more stable convergence to exact or almost exact results (Figs. \ref{fig:10}-\ref{Fig:AsConv}, \ref{fig:14}-\ref{Fig:AsErr}, and Tables \ref{tab:3}-\ref{Tab:As}).
\par
The speed-up analysis results in Section \ref{SecSpeedUp} shows that QMC with optimal sampling strategies allows significative reduction of the number of scenarios to achieve a given accuracy, leading to significative reduction of computational efforts (Table \ref{tab:5}). The size of the reduction costs scales up to almost $10^3$. This fact is very important when the computation of a large number of price sensitivities (greeks) has to be performed, as in the case of multi-asset options, because the computational time increases linearly with the number of underlyings if standard (finite differences, or FD) techniques are employed. We compared the computational efforts needed by FD and AAD in computing price and all first order greeks with and without QMC, fixing the desired accuracy. Remarkably, we obtained that QMC with FD runs in comparable times as MC with AAD for medium sized baskets, while the best choice is clearly QMC with AAD, which allows for very fast and efficient results as shown in Figs. \ref{Fig:SpeedUpEu}-\ref{Fig:SpeedUpIncrAs}. It means that, taking into account the accuracy of the computation, AAD is not guaranteed to be faster than FD if it is implemented with standard MC rather than QMC (at least for a modest number of derivatives to be computed). Since, as discussed in appendix \ref{SecAAD}, AAD requires a considerable implementation effort, especially in industrial applications, our results suggest that, if a financial institution doesn't have AAD implemented, the use of FD coupled with QMC (which is much easier to implement) remains competitive in many realistic applications. Moreover, if a financial institution already has  AAD, it should use QMC instead of MC: this allows for huge savings in computational time and achieves high accuracy, in contrast to standard MC.
\par
Finally, the stability analysis results presented in Section \ref{SecStability} confirmed that QMC simulations with optimal sampling strategies are generally more stable and monotonic than MC (Figures \ref{fig:18}, \ref{fig:21}).
\par
The methodology presented in this paper can be used
for more complex problems in finance, in particular, fair value adjustments
(XVAs) and market and counterparty risk
measures\footnote{Some of these metrics, such as EPE/ENE or
expected shortfall, are defined as means or conditional means,
while some other metrics, such as VaR or PFE, are defined as
quantiles of appropriate distributions.}, based on
multi-dimensional, multi-step Monte Carlo simulations of large
portfolios of trades. Such simulations can require, in typical real
cases, $\sim 10^2$ time simulation steps, $\sim 10^3$ (possibly
correlated) risk factors, $\sim 10^3-10^4$ MC scenarios, $\sim
10^4-10^5$ trades, $60$ years maturity, leading to a nominal
dimensionality of the order $D \sim 10^5$, and to a total of
$10^9-10^{11}$ evaluations. Moreover, a fraction $\sim 1\%$
of exotic trades may require distinct MC simulations for their
evaluation, nesting another set of $\sim 10^3-10^5$ MC scenarios,
thus leading up to $10^{14}$ evaluations. Finally, hedging
CVA/DVA/FVA/KVA/MVA valuation adjustments \wrt to their underlying
risk factors (typically credit/funding curves) also requires the
computation of their corresponding greeks \wrt each term structure
node, adding another $\sim 10^2$ simulations. This is the reason
why the industry is continuously looking for advanced techniques
to reduce computational times: grid computing, GPU computing, AAD, etc. (see \eg
\cite{She15a}).
\par
We argue that, using QMC sampling (instead of MC) to generate the
scenarios of the underlying risk factors and to price exotic
trades may significantly improve the accuracy, the performance and
the stability of such monster-simulations, as shown by preliminary
results on real portfolios in \cite{BiaKuc14}. Furthermore, GSA
should suggest how to order the risk factors according to their
relative importance, thus reducing the effective dimensionality.
Such applications will need further research.

\begin{appendices}
\section{FD vs AAD greeks}
\label{SecAAD}
\label{App:GreekErr} \noindent
Complex financial instruments can be priced only resorting to Monte Carlo simulation. The main drawback of this approach is that it is generally computationally expensive to reach an acceptable degree of accuracy. This problem becomes even more striking when the computation of Greeks is concerned: indeed, due to its simplicity, the most widely used technique is to form finite difference (FD) approximations and then re-price the instrument on bumped scenarios (\ie with the relevant parameter shifted by a predefined finite amount). The FD estimator of a generic greek (say, the sensitivity w.r.t. to parameter $\theta_i$) is defined as\footnote{For simplicity, here the one-sided forward difference is shown. As it will be discussed, generally central (two-sided) differences are preferable \cite{Gla03}.}
\begin{equation}\label{FDestimator}
\frac{\partial{V_0}}{\partial{\theta_i}} \doteq \frac{Y(\theta_i+h)-Y(\theta_i)}{h}
\end{equation}
where $Y$ is the discounted payoff and $h$ the increment on $\theta_i$, which is chosen to be $h=\epsilon S_i(0)$, for deltas and gammas, and $h=\epsilon$, for vegas, for a given\footnote{Of course, the shift parameters need not to be the same for delta and vega. They can also be different for each component greek.} ``shift parameter'' $\epsilon$. Clearly, the value of the greek is obtained by averaging (\ref{FDestimator}) on many Monte Carlo paths: all that is required for the computation of the price is thus sufficient for the computation of FD greeks as well, no additional implementation effort is required. However, this approach has two disadvantages. The first one concerns with the accuracy of the computation: finite differences are subject to truncation errors, which can be mitigated by the use of central differences so that the bias of the greek estimator (\ref{FDestimator}) is of second order in the increment. Bias can be decreased by choosing a small increment, however this would also increase the variance of the estimator, even if path recycling is adopted, so that a fine tuning is needed. It is often hard to find the optimal increment and, in concrete applications, the same increment usually has to be applied to many different situations\footnote{The choice of the appropriate increment
is guided by the following considerations. The MC/QMC root mean
square error estimate of finite differences is given by
\cite{Gla03}
\begin{equation}\label{RMSEGreeks}
\varepsilon=\sqrt{\frac{c}{N^{2\alpha}h^{\beta}}+b^2h^4}\, .
\end{equation}
The first term in the square root is a ``statistical'' error
related to the variance $c$. It depends on $N$ as well as on
$\epsilon$. $\alpha=0.5$ for MC and, usually, $0.5<\alpha<1$ for
QMC, while $\beta=1$ for first derivatives and $\beta=3$ for
second derivatives. The second term is the systematic error due to
the bias of finite differences: it is independent of $N$ but it
depends on $\epsilon$. The constant $b$ is given by
$b=\frac{1}{6}\frac{\partial^3 V}{\partial \theta^3}(\theta)$ for
central differences of the first order (delta and vega) and
$b=\frac{1}{12}\frac{\partial^4 V}{\partial \theta^4}(\theta)$ for
central differences of the second order (gamma). The optimal value of $h$
doesn't vary too much for reasonable ranges of $N$ and it can be chosen in such a way that the bias term remains negligible so that (\ref{RMSEGreeks})
follows approximately a power law.}.The second disadvantage of FD techniques is concerned with computational efforts: indeed, the instrument has to be re-priced as many times as the number of derivatives to compute (actually twice as many, if central differences are used). Therefore, the computational cost of evaluating the price and all greeks, \ie the gradient of the price function w.r.t. all relevant parameters, increases linearly with the number of required sensitivities. This becomes particularly expensive \eg in the case of options on multiple underlyings, where at least deltas and vegas w.r.t. each underlying are usually needed.
\par
Both disadvantages of FD techniques can be eliminated by applying Adjoint Algorithmic Differentiation (AAD), which we briefly review in the following. This method was introduced in finance in \cite{GilGla06} and further developed in \cite{LecLia09,CapGil10,CapGil11,Cap11,CapLee12}. Instead of the FD estimator, let's introduce the ``Pathwise Derivative'' estimator of the greeks \cite{BroGla1996,Gla03},
\begin{equation}\label{PDestimator}
\frac{\partial{V_0}}{\partial{\theta_i}} \doteq \frac{\partial{Y}}{\partial\theta_i}\, ,
\end{equation}
which is simply the derivative of the discounted payoff. If the pathwise derivative (\ref{PDestimator}) exists with probability 1 and if the payoff function is regular enough (\eg it is Lipschitz continuous, see \cite{Gla03} for other sufficient conditions), then (\ref{PDestimator}) provides an unbiased estimator of the greeks since expectation and differentiation can be safely interchanged:
\begin{equation}
\frac{\partial{V_0}}{\partial{\theta_i}} = \frac{\partial{\mathbb{E}[Y]}}{\partial\theta_i} = \mathbb{E}\left[\frac{\partial{Y}}{\partial\theta_i}\right].
\end{equation}
In other words, we just need to differentiate the discounted payoff path by path and the value of the greek is then recovered by a Monte Carlo average as usual. Notice that the pathwise derivatives have to be computed explicitly in order to compute greeks according to this approach, so that extra implementation effort is required. This can be tedious for complex payoffs. Moreover, in the limit $h\to 0$, both the FD and Pathwise Derivative estimators provide the same estimates with the same Monte Carlo variance, so that the implementation effort required by the latter seems to be hardly justifiable at first sight. However, the major benefit comes from the fact that, if one has to compute the gradient of a single output w.r.t. many variables (as in the case of greeks of multi-asset options), the adjoint mode of algorithmic differentiation dramatically increases the efficiency of pathwise differentiation.
\par
We now describe the basics ideas behind this methodology. Algorithmic differentiation (AD) is a set of programming techniques aimed at calculating \emph{exact} (\ie free of truncation errors) derivatives of functions given as computer codes \cite{Gri08,Nau12}. Let $f:\mathds{R}^n\rightarrow\mathds{R}\, ,f(X)=Y$ be a scalar function of $n$ variables $X=(x_1,\ldots,x_n)$, such as the discounted payoff regarded as a function of parameters $\bm{\theta}$. No matter how complicated $f$ might be, it is always given as the composition of elementary functions and/or basic arithmetic operations. AD exploits the information on the structure of the code and on the dependencies between its various parts, in order to optimize the calculation of the derivatives. The main tool which makes AD work is the chain rule, which is repeatedly used on the arcs connecting the nodes of the computational graph representing the instructions needed to execute $f$. Here we discuss only the adjoint mode of algorithmic differentiation (AAD), which is the most efficient when the derivatives of few outputs w.r.t. many inputs are needed\footnote{The discussion of the tangent mode, useful when the derivatives of many outputs w.r.t. few inputs have to be computed, is left to the references \cite{Gri08,Nau12,Cap11}.}. First of all, a \emph{forward sweep} is performed where, starting from the values of the inputs, the value of the output is computed recording all necessary information in intermediate steps. After that, a \emph{backward sweep} is performed, where the derivatives w.r.t all the intermediate variables, \ie the adjoints, are propagated from the output to the inputs until the whole gradient is obtained in a single run.
\begin{align*}
&X \rightarrow \cdots \rightarrow U_i \rightarrow V_j \rightarrow \cdots \rightarrow Y\\
&\phantom{X \rightarrow \cdots \rightarrow U_i \rightarrow V_j \rightarrow \cdots \rightarrow}\downarrow\\
&\bar{X} \leftarrow \cdots \leftarrow \bar{U}_i \leftarrow \bar{V}_j \leftarrow \cdots \leftarrow \bar{Y}
\end{align*}
In more detail: any intermediate instruction is an intrinsic (unary or binary) operation of the form
\begin{equation}
V_j = V_j(\{U_i\}_{i\prec j})
\end{equation}
whose derivative is known from calculus. Here, the notation means that $U_i$ are the variables on which $V_j$ explicitly depends. The adjoints are then defined as
\begin{equation}
\bar{U}_i=\sum_{j\succ i}\,\frac{\partial V_j}{\partial U_i}\, \bar{V}_j\, .
\end{equation}
Initializing $\bar{Y}=1$ and propagating the adjoints backward through each intermediate step, at the end of the computation the adjoints of the inputs (\ie the gradient) are obtained.
\par
\begin{figure}[ht]
\centering
\includegraphics[width=3.1in,height=2.4in,keepaspectratio=false]{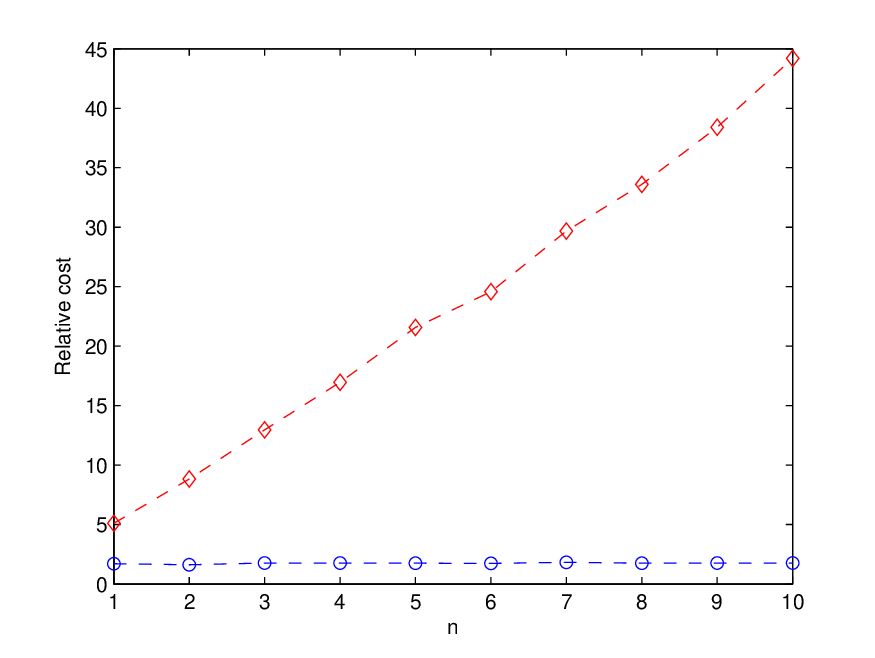}
\caption{CPU time required to compute price and all greeks (deltas and vegas) divided by CPU time required to compute only price of an Asian option on a basket of $n$ underlyings, for increasing $n$. Monte Carlo simulation with 200,000 scenarios is used. The red curve refers to central finite differences computation, while the blue curve refers to AAD computation.}
\label{Fig:RelCost}
\end{figure}

It is easy to appreciate that the cost for the propagation of the chain rule back to the inputs is of the same order as the cost of evaluating the function $f$ itself. Indeed, there's a precise result stating that, in the adjoint mode, AD provides the full gradient of $f$ at a cost which is up to 4 times the computational cost of evaluating the function $f$ itself, independently of the number of variables. This explains the power of AAD, enabling an extremely fast computation of an arbitrary number of greeks\footnote{AAD, being mechanical in nature, can be automated. Indeed, several AD tools have been developed which automatically implement the adjoint counterpart of a given computer code. These tools typically make use of source code transformations or operator overloading. The latter are well suited to the object oriented programming paradigm. The drawback is that the computation of derivatives is slower, since a preliminary step is needed where all the information of the original code is recorded in a kind of tape, which is a representation of the computational graph necessary to run the forward and backward sweeps. A lot of memory is also necessary since all intermediate variables cannot be overwritten. We refer to the literature for further details on AD tools \cite{Gri08,Nau12}.}. In Fig. \ref{Fig:RelCost} the relative cost of evaluating the price and all deltas and vegas of an Asian basket option w.r.t the cost of evaluating just the price is shown for an increasing number of underlyings and for both FD and AAD techniques. It is easily observed that, while for FD the relative cost increases linearly with the number of underlyings (or greeks to be computed), for AAD it remains constant.
However, AAD also has some disadvantages w.r.t FD: first of all, it is harder to implement than finite differences and the implementation effort within large architectures would be challenging without the availability of automatic tools. Secondly, it is not always applicable: in particular it cannot handle discontinuous payoffs. One must regularize the payoff by explicitly smoothing the discontinuity, \eg approximating a digital call with a call spread or something smoother, or using conditional expectations, \eg smoothing payoffs with barriers \cite{Gla03}. However this introduces a bias and the use of automatic tools is not straightforward, so that extra effort is needed. Finally, second order Greeks do not have the benefits of the adjoints for multi-asset options. One usually is forced to use a mixed approach AAD+FD. Recently, a mixed approach AAD + Likelihood Ratio Method has been proposed \cite{Cap15}. Furthermore, we want to stress that the results such as those shown in Fig. \ref{Fig:RelCost}, typically presented in many works on adjoint methods in finance, solely refer to the speed of the computation and by no means are indicative of the accuracy of the computation. The latter is rather given by the details of the simulation method, such as the number of scenarios, the random number generator, the use of variance reduction techniques, and so forth. Of course, an optimal use of the simulation technique will increase accuracy. In other words, the same accuracy can be reached with less scenarios if the simulation details are accurately chosen (\eg QMC with BBD): this is, therefore, another source of speed-up of the whole computation, besides the mere use of adjoints instead of finite differences in a MC simulation. These two sources of speed-up are compared in the present work.
\end{appendices}

\bibliographystyle{alpha}
\bibliography{FinanceBibliography}

\end{document}